\documentclass[fleqn,usenatbib]{mnras}

\usepackage{newtxtext,newtxmath}
\usepackage[T1]{fontenc}
\DeclareRobustCommand{\VAN}[3]{#2}
\let\VANthebibliography\thebibliography
\def\thebibliography{\DeclareRobustCommand{\VAN}[3]{##3}\VANthebibliography}

\usepackage{graphicx}
\usepackage{amsmath}
\usepackage{pdflscape}
\usepackage{subfig}
\usepackage{soul}

\makeatletter
\newcommand\footnoteref[1]{\protected@xdef\@thefnmark{\ref{#1}}\@footnotemark}
\makeatother

\usepackage{scalerel,tikz}
\usetikzlibrary{svg.path}
\definecolor{orcidlogocol}{HTML}{A6CE39}
\tikzset{orcidlogo/.pic={
 \fill[orcidlogocol] svg{M256,128c0,70.7-57.3,128-128,128C57.3,256,0,198.7,0,128C0,57.3,57.3,0,128,0C198.7,0,256,57.3,256,128z};
 \fill[white] svg{M86.3,186.2H70.9V79.1h15.4v48.4V186.2z}
 svg{M108.9,79.1h41.6c39.6,0,57,28.3,57,53.6c0,27.5-21.5,53.6-56.8,53.6h-41.8V79.1z M124.3,172.4h24.5c34.9,0,42.9-26.5,42.9-39.7c0-21.5-13.7-39.7-43.7-39.7h-23.7V172.4z}
 svg{M88.7,56.8c0,5.5-4.5,10.1-10.1,10.1c-5.6,0-10.1-4.6-10.1-10.1c0-5.6,4.5-10.1,10.1-10.1C84.2,46.7,88.7,51.3,88.7,56.8z};
}}
\newcommand\orcidicon[1]{\href{https://orcid.org/#1}{\mbox{\scalerel*{
\begin{tikzpicture}[yscale=-1,transform shape]
\pic{orcidlogo};
\end{tikzpicture}
}{|}}}}

\title[Galactic Plane Faraday Complexity]{A new window into the sub-parsec scale magnetic field in the Milky Way? Unveiling small-scale magneto-ionic structures with Faraday complexity}

\author[Y.~K.~Ma et al.]{Yik~Ki~Ma,$^{\orcidicon{0000-0003-0742-2006}\,1}$\thanks{E-mail: \href{mailto:yikki.ma@anu.edu.au}{yikki.ma@anu.edu.au}}
Amit~Seta,$^{\orcidicon{0000-0001-9708-0286}\,1}$
N.~M.~McClure-Griffiths,$^{\orcidicon{0000-0003-2730-957X}\,1}$
C.~L.~Van Eck$^{\orcidicon{0000-0002-7641-9946}\,1}$
S.~A.~Mao,$^{\orcidicon{0000-0001-8906-7866}\,2}$
\newauthor
A.~Ordog,$^{\orcidicon{0000-0002-2465-8937}\,3,4}$
J.~C.~Brown,$^{\orcidicon{0000-0003-4781-5701}\,5}$
T.\ O.\ Kovacs,$^{\orcidicon{0000-0001-6649-8559}\,6}$
Takuya~Akahori,$^{\orcidicon{0000-0001-9399-5331}\,7}$
K.~Kurahara,$^{\orcidicon{0000-0003-2955-1239}\,7}$
\newauthor
L.~Oberhelman,$^{\orcidicon{0000-0003-3672-146X}\,1}$
and
C.~S.~Anderson$^{\orcidicon{0000-0002-6243-7879}\,1}$
\\
$^{1}$Research School of Astronomy \& Astrophysics, Australian National University, Canberra, ACT 2611, Australia\\
$^{2}$Max-Planck-Institut f\"ur Radioastronomie, Auf dem H\"ugel 69, 53121 Bonn, Germany\\
$^{3}$Dominion Radio Astrophysical Observatory, Herzberg Research Centre for Astronomy and Astrophysics, National Research Council Canada, PO Box 248,\\
\phantom{$^{3}$}Penticton, BC V2A 6J9, Canada\\
$^{4}$Department of Computer Science, Math, Physics, and Statistics, University of British Columbia, Okanagan Campus, 3187 University Way, Kelowna,\\
\phantom{$^{4}$}BC V1V 1V7, Canada\\
$^{5}$Department of Physics and Astronomy, University of Calgary, Calgary, AB T2N 1N4, Canada\\
$^{6}$Max-Planck-Institut f\"ur Astronomie, K\"onigstuhl 17, 69117 Heidelberg, Germany\\
$^{7}$Mizusawa VLBI Observatory, National Astronomical Observatory Japan, 2-21-1 Osawa, Mitaka, Tokyo 181-8588, Japan}

\date{Accepted 2025 June 12. Received 2025 May 16; in original form 2024 September 27}

\pubyear{2025}

\begin{document}
\label{firstpage}
\pagerange{\pageref{firstpage}--\pageref{lastpage}}
\maketitle

\begin{abstract}
Radio broadband spectro-polarimetric observations are sensitive to the spatial fluctuations of the Faraday depth (FD) within the telescope beam. Such FD fluctuations are referred to as ``Faraday complexity'', and can unveil small-scale magneto-ionic structures in both the synchrotron-emitting and the foreground volumes. We explore the astrophysical origin of the Faraday complexity exhibited by 191 polarised extragalactic radio sources (EGSs) within $5^\circ$ from the Galactic plane in the longitude range of $20^\circ$--$52^\circ$, using broadband data from the Karl G.\ Jansky Very Large Array presented by a previous work. A new parameter called the FD spread is devised to quantify the spatial FD fluctuations. We find that the FD spread of the EGSs (i) demonstrates an enhancement near the Galactic mid-plane, most notable within Galactic latitude of $\pm 3^\circ$, (ii) exhibits hints of modulations across Galactic longitude, (iii) does not vary with the source size across the entire range of $2\farcs5$--$300^{\prime\prime}$, and (iv) has an amplitude higher than expected from magneto-ionic structures of extragalactic origin. All these suggest that the primary cause of the Faraday complexity exhibited by our target EGSs is $<2\farcs5$-scale magneto-ionic structures in the Milky Way. We argue that the anisotropic turbulent magnetic field generated by galactic-scale shocks and shears, or the stellar feedback-driven isotropic turbulent magnetic field, are the most likely candidates. Our work highlights the use of broadband radio polarimetric observations of EGSs as a powerful probe of multi-scale magnetic structures in the Milky Way.
\end{abstract}


%%%%%%%%%%%%%%%%%%%%%%%%%%%%%%%%%%%%%%%%%%%%%%%%%%

\section{Introduction} \label{sec:intro}

The magnetic field is an integral part of the interstellar medium (ISM) of galaxies. This is reflected by its energy density, which is comparable to that of thermal gas, turbulence, and cosmic rays \citep{beck16}. At the galactic scale ($\gtrsim {\rm kpc}$), magnetic pressure can have significant effects on the gas dynamics \citep[e.g.][]{beck05,pakmor14,chan22,khademi23}. Meanwhile at small-scales ($\lesssim 10\,{\rm pc}$), magnetic fields can modulate star formation \citep[e.g.][]{federrath12,krumholz19} and restrain cosmic ray particles \citep[e.g.][]{seta18}. It is therefore crucial to obtain a detailed map of the galactic magnetic fields of the Milky Way \citep{boulanger18} and other galaxies near \citep{beck19} and far \citep{mao17,geach23,chen24}.

The Faraday rotation effect is a useful tool to unveil the magnetic field strength and structure in galaxies \citep[e.g.][]{haverkorn13,beck16}. As linearly polarised emission propagates through a magneto-ionic medium, the birefringence of circular polarisation leads to a rotation of the linear polarisation plane, with the observed polarisation position angle (PA) given by
\begin{equation}
{\rm PA} (\lambda^2) = {\rm PA}_0 + \left[ 0.81 \int_L^0 n_e (s) B_\parallel (s)\,{\rm d}s \right] \cdot \lambda^2 \equiv {\rm PA}_0 + \phi \cdot \lambda^2{\rm ,} \label{eq:fd}
\end{equation}
where $\lambda$ [m] is the observed wavelength, ${\rm PA}_0$ [rad] is the intrinsic PA (determined by the magnetic field orientation in the emission volume, for the case of synchrotron emission), $L$ [pc] is the distance to the emission volume from the observer, $n_e$ [${\rm cm}^{-3}$] is the electron number density, $B_\parallel$ [$\mu{\rm G}$] is the magnetic field strength along the line-of-sight ($s$), and $\phi$ [${\rm rad\,m}^{-2}$] is the Faraday depth (FD) of the polarised emission \citep[e.g.][]{ferriere21}. In the simplest case\footnote{Historically, the FD in such case is often referred to as the rotation measure (RM).}, where all of the observed polarised emission have the same FD (``Faraday simple''), the PA will follow a linear relationship with $\lambda^2$, meaning that the FD can be obtained as the slope of PA against $\lambda^2$. The more general case is where the observed polarised emission is composed of emission with different FD values (``Faraday complex''), which can be caused by, e.g.\ spatial fluctuations of the foreground magneto-ionic medium across the plane-of-sky. In such a situation, the constituent polarised emission can be decomposed and recovered by applying algorithms such as RM-Synthesis \citep{brentjens05} and Stokes \textit{QU}-fitting \citep{farnsworth11,osullivan12} to broadband spectro-polarimetric data in radio wavelengths. In other words, the spatial FD fluctuations within the telescope beam volume are encrypted in the broadband radio data, which can be unlocked with modern polarisation analysis algorithms. Such analysis allows us to peer into the intricate magneto-ionic structures of the intervening medium even if we do not have the normally required spatial resolution. Indeed, in the past decade, there have been numerous studies exploiting the broadband capabilities of radio telescopes to observe background polarised extragalactic radio sources (EGSs) as a means to understand the small-scale magnetic structures in the Milky Way \citep[e.g.][]{anderson15,livingston21,ranchod24} as well as beyond our Galaxy \citep[e.g.][]{mao17,osullivan17,ma19a}.

The magnetic fields in galaxies can be broadly separated into three components \citep[e.g.][]{jaffe10,haverkorn15,beck16}~-- the regular magnetic field (also commonly called the coherent magnetic field), the isotropic turbulent magnetic field (also called the random magnetic field), and the anisotropic turbulent magnetic field \citep[also referred to as the striated magnetic field in][and as the ordered magnetic field in \citealt{jaffe10}]{jansson12,jansson12b}. These three magnetic components have different astrophysical origins, geometrical properties, and observational signatures:
\begin{itemize}
\item The \textit{regular magnetic field} is believed to have been generated by a galactic-scale $\alpha$-$\Omega$ dynamo that amplifies and orders the magnetic field on $\sim {\rm Gyr}$ time-scale \citep[e.g.][]{ruzmaikin88,beck96,arshakian09,ntormousi20,brandenburg23}. The dominant mode of the resulting magnetic field is coherent on the galactic scale (coherence length $\gtrsim {\rm kpc}$), and follow a single direction without frequent reversals. This field component contributes to the diffuse synchrotron emission (in both total intensity and linear polarisation), and can also be observed through the spatially coherent FD of both the background EGSs \citep[e.g.][]{ma20,livingston22} as well as the diffuse synchrotron emission of the galaxies themselves \citep[e.g.][]{kierdorf20,dickey22}.
\item The \textit{isotropic turbulent magnetic field} can primarily be formed by the compression and tangling of the magnetic fields caused by energetic events such as supernova explosions \citep[e.g.][]{ferriere91,maclow04,haverkorn08}, or by the turbulent dynamo \citep[e.g.][]{seta22}. Such magnetic fields have no preferred directions (i.e.\ isotropic random), and can have a coherence length of about $10$--$100\,{\rm pc}$ when driven by supernova explosions \citep{haverkorn08}. We can identify this magnetic component through its resulting spatial fluctuations of FD to background EGSs \citep[e.g.][]{simonetti84,stil11,seta23}, depolarisation of background emission \citep[e.g.][]{stil07,thomson19}, or contributions to the synchrotron emission in total intensity \citep[e.g.][]{hassani22}.
\item The \textit{anisotropic turbulent magnetic field}, first identified in the Milky Way by \cite{brown01}, is presumed to have been formed by shock compression (by, e.g.\ spiral arms) and shearing force (by the galactic differential rotation) exerted on an originally isotropic turbulent magnetic field \citep[e.g.][]{laing80}. Both these actions can diminish the magnetic field component parallel (perpendicular) to the shock (shear) motion, resulting in magnetic fields that follow a single orientation (i.e.\ with possible flips in the direction by $180^\circ$ when projected onto the plane-of-sky). Observationally, the most common way to probe this class of galactic magnetic fields is through its contributions to the diffuse synchrotron emission (both in total intensity and linear polarisation). Specifically, the differences in the magnetic field strength estimates from the polarised synchrotron emission (tracing both the regular and the anisotropic turbulent magnetic fields) and from the spatially coherent FD (tracing the regular magnetic fields only) have been routinely used to constrain the magnetic field strength of the anisotropic turbulent component for the Milky Way \citep{jansson12b} and other nearby galaxies \citep{han99,beck05,beck07,fletcher11,beck15,mulcahy17,beck20,kierdorf20}. In principle, this magnetic field component can also lead to small-scale FD fluctuations that can be identified by high-angular-resolution observations, but this has not yet been achieved in practice \citep{beck19}. This is hindered by the immense challenge of involving sensitive polarimetric observations at high angular resolution and/or across a broad frequency range, in addition to algorithms that can accurately separate this galactic magnetic field component from the two counterparts. Thus, the anisotropic turbulent magnetic field is the least understood of the three. However, an improved understanding of this magnetic field component is critical, especially for galaxies with strong spiral shocks \citep[e.g.\ M~51;][]{fletcher11}.
\end{itemize}

We point out here a possible confusion in the terminology -- the terms ``isotropic'' or ``anisotropic'' discussed above refer to the possible orientations of the magnetic fields, with the former spanning full $180^\circ$ in projection and the latter confined to a single orientation (with possible flips in direction by $180^\circ$). This is distinct from the fundamental anisotropy (in coherence length) in magneto-hydrodynamics (MHD) turbulence \citep[e.g.][]{goldreich95,cho02,cho03,ferriere20}. For example, in \cite{houde13} the ``anisotropic random magnetic field'' of M~51 was studied by quantifying the spatial fluctuations of the PA of synchrotron emission from the \cite{fletcher11} data. From our definitions above \citep[following, e.g.][]{jaffe10,haverkorn15,beck16}, the anisotropic turbulent magnetic field cannot change in orientation, which the synchrotron PA measures. Therefore, in our terminology, the \cite{houde13} work studied the isotropic turbulent magnetic field, and their use of the term ``anisotropic'' referred to the anisotropy of the turbulence coherence lengths (Houde \textit{priv.\ comm.}).

\begin{figure*}
\includegraphics[width=0.99\textwidth]{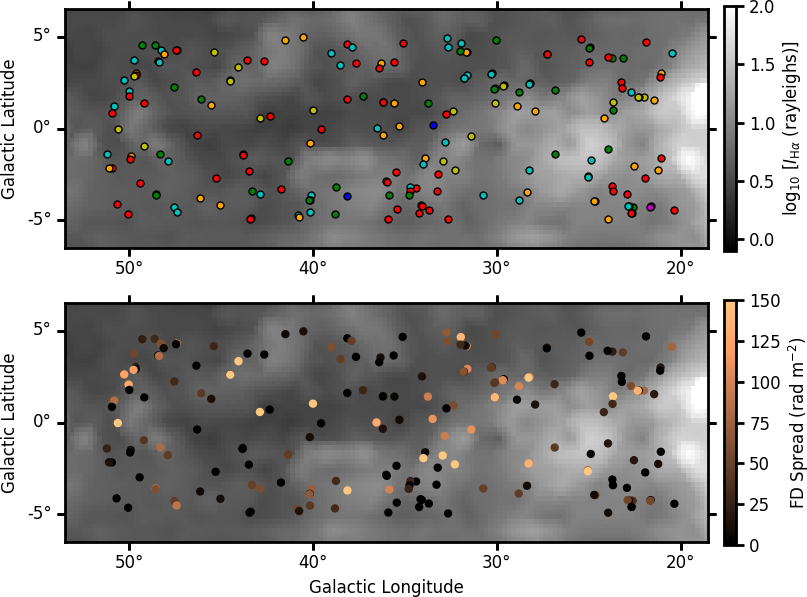}
\caption{\textbf{(Top)} Spatial distribution of the best-fit Stokes \textit{QU}-fitting model (Section~\ref{sec:qufit}) of the 191 polarised EGSs used in this study: 1T (red), 2T (cyan), 1Ed (orange), 2Ed-c (yellow), 1S (green), 2S (blue), and 1Id (magenta). \textbf{(Bottom)} Correspondingly, the spatial distribution of the spatial FD spread (Section~\ref{sec:fdspread}) measured from the polarised EGSs. In both panels, the background grey-scale image is the Wisconsin H-Alpha Mapper Sky Survey (WHAMSS) H$\alpha$ map \citep{haffner03,haffner10}.}
\label{fig:map_fdspread}
\end{figure*}

The small-scale magnetic structures of galaxies can be studied at the highest details in the Milky Way, owing to our natural vantage point from within the Galaxy. There have been a few studies in the literature using the Faraday complexity of background EGSs to probe the small-scale Galactic magneto-ionic structures within the telescope beam. \cite{anderson15} used data from the Australia Telescope Compact Array (ATCA) at $1.3$--$2.0\,{\rm GHz}$ to study the polarisation behaviour of 563 EGSs at high Galactic latitude ($b \approx -55^\circ$), and found that the 19 Faraday complex sources are preferentially situated at sky regions with Galactic H\textsc{i} column density within an intermediate range of $(1.4$--$1.65) \times 10^{20}\,{\rm cm}^{-2}$. They interpreted the observed Faraday complexity as being induced by the interstellar medium (ISM) interface where phase transitions are occurring, though it is also possible to have been caused by other extragalactic effects (see Section~\ref{sec:exgal}). More recently, \cite{livingston21} observed 62 EGSs within a $12^\circ \times 12^\circ$ field in the Galactic Centre region using the ATCA at $1.1$--$3.1\,{\rm GHz}$, and found evidence of enhanced Faraday complexity most likely caused by stellar feedback around the Galactic Centre area. And finally, \cite{ranchod24} studied the Faraday complexity of EGSs in the southern Galactic plane using also the ATCA at $1.1$--$3.1\,{\rm GHz}$, and found an enhanced density of Faraday complex sources near spiral arm tangents. The goals of our work here are (1) to investigate if the magneto-ionic medium in the Galactic plane can lead to detectable Faraday complexity of the background EGSs and, if it is found to be the case, (2) to quantify the Faraday complexity to help identify the responsible component of the magneto-ionic medium. These are achieved by repurposing the \cite{ma20} data of 194 polarised EGSs in the northern Galactic plane (specifically, within the longitude range of $20^\circ$--$52^\circ$; see Section~\ref{sec:vladata}), which was used to study the $\gtrsim 1^\circ$-scale FD distribution that traces the Galactic-scale regular magnetic field.

Our paper is structured as follows. In Section~\ref{sec:data}, we describe the data and their processing for this study; in Section~\ref{sec:results} we introduce a parameter called the FD spread, and present our findings based on this parameter; in Section~\ref{sec:discussion} we explore the astrophysical origin of the FD spread observed from our EGS sample and discuss the implications; and in Section~\ref{sec:conclusions} we make concluding remarks of this work. 

\section{Data and Data Processing} \label{sec:data}

For our work, we primarily focus on using the \cite{ma20} broadband spectro-polarimetric data obtained from the Karl G.\ Jansky Very Large Array (VLA). The flux densities of the EGSs in Stokes \textit{I}, \textit{Q}, and \textit{U} are calculated from the data, and subsequently analysed using Stokes \textit{QU}-fitting. We further supplement these with data from the Australian Square Kilometre Array Pathfinder's \citep[ASKAP's;][]{hotan21} Rapid ASKAP Continuum Survey \citep[RACS;][]{mcconnell20}, as well as the VLA Sky Survey \citep[VLASS;][]{lacy20}. We describe these data and the methods in detail below.

\subsection{Broadband spectro-polarimetric data from the VLA} \label{sec:vladata}

We use the \cite{ma20} broadband spectro-polarimetric data obtained from the VLA in L-band (1--2\,GHz). This frequency range is suitable for our study of Faraday complexity here, as it provides a good balance between the FD precision and the maximum scale in FD space that the data are sensitive to (see Section~\ref{sec:qufit}). In short, on-axis snapshot observations of 176 EGSs near the Galactic mid-plane ($|b| \leq 5^\circ$) within the Galactic longitude range of $20^\circ$ to $52^\circ$ were conducted in the D-array configuration. Standard flagging and calibration procedures for spectro-polarimetric datasets were followed, and Stokes \textit{I}, \textit{Q}, and \textit{U} image cubes were formed for each on-axis target at a frequency interval of 4\,MHz. The resulting images have a typical angular resolution of $50^{\prime\prime} \times 42^{\prime\prime}$ at 1.5\,GHz, and a typical per-channel noise level of $4.3$, $1.4$, and $1.5\,{\rm mJy\,beam}^{-1}$ for Stokes \textit{I}, \textit{Q}, and \textit{U} respectively.

From the image cubes, \cite{ma20} proceeded to extract the flux densities in the three Stokes parameters across frequency for all 176 on-axis EGSs, in addition to 43 off-axis sources within $5^\prime$ from the pointing centres of the observations. After discarding 15 spurious sources in total, they have analysed the remaining 204 sources (171 on-axis and 33 off-axis) using RM-Synthesis \citep{brentjens05,purcell20} for their study of the Galactic-scale magnetic field, with 10 sources found to be unpolarised. For this work, we start from the \cite{ma20} Stokes \textit{I}, \textit{Q}, and \textit{U} image cubes, and proceed to re-evaluate the flux density values across frequency for those 194 polarised sources (Section~\ref{sec:fg_sub}). Subsequently, we apply Stokes \textit{QU}-fitting to identify and investigate the Faraday complexity exhibited by these sources (Section~\ref{sec:qufit}).

\subsection{Flux density determination} \label{sec:fg_sub}

We choose to extract from the image cubes the Stokes \textit{I}, \textit{Q}, and \textit{U} values of the 194 polarised sources from \cite{ma20}, instead of directly using their obtained flux density values. The underlying reason is that the diffuse polarised emission from the Milky Way can contaminate the obtained flux densities of the polarised background sources if the former is not taken into account. This is especially an issue for observations using radio interferometric arrays in compact configurations \citep[see][]{ranchod24}, which was indeed the case for the \cite{ma20} observations using the VLA in its most compact (D-array) configuration. 

We therefore extract the flux densities of the polarised sources largely following the steps described in section~3 of \cite{ma20}, but with a foreground subtraction routine implemented in order to mitigate the effects of the potential Galactic polarised emission. The Common Astronomy Software Application (\textsc{casa}) package \cite[version 5.3.0;][]{mcmullin07,casa22} is used for such operations. Specifically, sources that are spatially unresolved are processed using the task \texttt{IMFIT} to obtain the integrated flux densities, with the foreground contamination removed by allowing for an amplitude offset whilst performing the 2D Gaussian fitting. Such amplitude offset can be different between sources, frequency channels, and Stokes parameters, but is otherwise a constant value spatially. Meanwhile, the flux densities of sources that are spatially extended are obtained by using the task \texttt{IMSTAT}, with the foreground contamination determined by using an annulus with inner/outer radii of $70^{\prime\prime}$/$170^{\prime\prime}$ centred on the target source \citep[following, e.g.][]{oberhelman24}. The inner radius of $70^{\prime\prime}$ is chosen with respect to the typical angular resolution of the \cite{ma20} observations ($\approx 50^{\prime\prime}$ full-width) to minimise the chance that the target sources can be included within the annulus. The outer radius of $170^{\prime\prime}$ has been decided to ensure that the annulus is large enough (with the annulus area being about 40 times the beam area) to be robust against measurement uncertainties, while still being small enough to avoid spatial fluctuations of the Galactic emission that will reduce the accuracy of the foreground subtraction. Within the annulus, we determine the median value of the intensity, which is then converted to the amount of Galactic diffuse emission contained by the target source extraction region. We decide to use the median instead of the mean intensity, since the former is more robust against cases where EGSs with large angular extents (larger than twice the inner radius; five sources out of the total of 194) can have part of their emission being included in the annulus regions. The foreground emission is then subtracted from the flux density directly obtained from the source extraction region, with the uncertainties added in quadrature.

The foreground-subtracted flux densities of the 194 sources in Stokes \textit{I}, \textit{Q}, and \textit{U} across frequency are used for the analysis in the remainder of the paper. In Appendix~\ref{sec:fg_pi} we briefly assess the polarised intensity of the foreground emission, and conclude that most of our target EGSs are not obscured by significant foreground polarised emission. We further explore in Appendix~\ref{sec:fg_test} whether the omission of the foreground subtraction can alter our scientific results. We find that while the qualitative conclusions of our work will remain unchanged, the quantitative aspects of the results can change significantly. We further show that the \cite{ma20} conclusions are not impacted by not having a foreground subtraction routine in place.

\subsection{Stokes \textit{QU}-fitting} \label{sec:qufit}

We apply the Stokes \textit{QU}-fitting algorithm \citep{farnsworth11,osullivan12} to the Stokes \textit{I}, \textit{Q}, and \textit{U} values across frequency for the 194 EGSs after foreground emission subtraction (Section~\ref{sec:fg_sub}) to extract information of the synchrotron-emitting volume, as well as the foreground magneto-ionic medium causing Faraday rotation. Previous studies have shown that for sources well represented by two Faraday thin components (see below), Stokes \textit{QU}-fitting can have higher accuracy in identifying the two components than, for example, RM-Synthesis \citep{sun15}, motivating our choice here for this study. In particular, we use the \textsc{RM-Tools} package \citep[][Van Eck et al.\ in prep.]{purcell20}\footnote{Available on \href{https://github.com/CIRADA-Tools/RM-Tools}{https://github.com/CIRADA-Tools/RM-Tools}.} coded in \textsc{Python}. For our case, we do not follow the standard approach of using \textsc{RM-Tools}' in-built Stokes \textit{I} fitting routine to model the total intensity spectrum for the division of the Stokes \textit{Q} and \textit{U} values. Instead, we perform a direct channel-wise division using the measured values of Stokes \textit{I} in order to obtain the fractional Stokes $q = Q/I$ and $u = U/I$ values, and used them as the inputs to \textsc{RM-Tools}. Since the flux density scales in Stokes $I$, $Q$, and $U$ are all tied together on a per-channel basis, the $q$ and $u$ values are independent of the bandpass calibration solution. Therefore, our procedure is expected to be more robust against potential bandpass errors. Furthermore, this method will not be significantly impacted by Stokes \textit{I} uncertainties as long as the target sources are sufficiently bright in the total intensity, which indeed is the case since $88$\,per cent of the 194 sources are brighter than $50\,{\rm mJy}$ at $1.4\,{\rm GHz}$ (with a signal-to-noise of higher than 12 per frequency channel), and the dimmest source in the sample (NVSS J190832$-$005319) still has a flux density of $18.0 \pm 0.3\,{\rm mJy}$.

A total of eight astrophysically motivated models are considered and applied to each of our target sources to determine the best-fit model and the associated parameters. All these models are described below \citep[see also][]{burn66,sokoloff98,osullivan12,ma19a}:
\begin{enumerate}
\item \textit{Single Faraday thin (1T)}: the synchrotron emission from the purely emitting volume traverses through a foreground homogeneous magneto-ionic medium that causes Faraday rotation. This is the only model referred to as being Faraday simple. The complex polarisation ($\mathbf{p}$) is given by
\begin{equation}
\mathbf{p}(\lambda^2) = p_0 {\rm e}^{2{\rm i} ({\rm PA}_0 + \phi \lambda^2)}{\rm ,}
\end{equation}
where $p_0$ denotes the intrinsic polarisation fraction.
\item \textit{Double Faraday thin (2T)}: the sight-lines towards the purely synchrotron-emitting volume are separated into two parts subjected to different amount of Faraday rotation. The complex polarisation is given by
\begin{equation}
\mathbf{p}(\lambda^2) = p_{0,1} {\rm e}^{2{\rm i} ({\rm PA}_{0,1} + \phi_1 \lambda^2)} + p_{0,2} {\rm e}^{2{\rm i} ({\rm PA}_{0,2} + \phi_2 \lambda^2)} {\rm .}
\end{equation}
\item \textit{Single external Faraday dispersion (1Ed)}: the emission from the purely emitting volume passes through a foreground turbulent magneto-ionic medium, leading to a dispersion in the observed FD characterised by $\sigma_\phi$. The complex polarisation is given by
\begin{equation}
\mathbf{p}(\lambda^2) = p_0 {\rm e}^{-2 \sigma_\phi^2 \lambda^4} {\rm e}^{2{\rm i} ({\rm PA}_0 + \phi \lambda^2)}{\rm .}
\end{equation}
\item \textit{Double external Faraday dispersion with common depolarisation (2Ed-c)}: the sight-lines towards the synchrotron-emitting volume are divided into two parts, experiencing different amounts of Faraday rotation from the regular magnetic field, but the same amount of dispersion in FD from the turbulent magneto-ionic medium. The complex polarisation is given by
\begin{equation}
\mathbf{p}(\lambda^2) = \left( p_{0,1} {\rm e}^{2{\rm i} ({\rm PA}_{0,1} + \phi_1 \lambda^2)} + p_{0,2} {\rm e}^{2{\rm i} ({\rm PA}_{0,2} + \phi_2 \lambda^2)} \right) {\rm e}^{-2 \sigma_\phi^2 \lambda^4}{\rm .}
\end{equation}
\item \textit{Double external Faraday dispersion with separate depolarisation (2Ed-s)}: similar to the 2Ed-c model above, but instead the two parts of emission experience different amounts of Faraday dispersion. The complex polarisation is given by
\begin{align}
\mathbf{p}(\lambda^2) &= p_{0,1} {\rm e}^{-2 \sigma_{\phi,1}^2 \lambda^4} {\rm e}^{2{\rm i} ({\rm PA}_{0,1} + \phi_1 \lambda^2)} \nonumber \\
&+ p_{0,2} {\rm e}^{-2 \sigma_{\phi,2}^2 \lambda^4} {\rm e}^{2{\rm i} ({\rm PA}_{0,2} + \phi_2 \lambda^2)}{\rm .}
\end{align}
\item \textit{Single sinc component (1S)}: named after its $\mathbf{p}(\lambda^2)$ modulation. This can either be caused by the synchrotron-emitting volume embedded within a homogeneous magneto-ionic medium that causes Faraday rotation \citep[commonly referred to as the Burn slab;][equation 3]{sokoloff98}, or by a gradient of FD in the plane-of-sky \citep[][equation 41]{sokoloff98}. The differential Faraday rotation, along the line-of-sight in the former case and in the plane-of-sky in the latter case, is characterised by $\Delta\phi$. The overall emission is further subjected to a uniform Faraday rotation. The complex polarisation is given by
\begin{equation}
\mathbf{p}(\lambda^2) = p_0 \frac{\sin (\Delta \phi \lambda^2)}{\Delta \phi \lambda^2} {\rm e}^{2{\rm i} \left[ {\rm PA}_0 + \left( \frac{1}{2} \Delta \phi + \phi \right) \lambda^2\right]} {\rm .} \label{eq:burn}
\end{equation}
\item \textit{Double sinc component (2S)}: the superposition of two sinc components, each subjected to different amounts of differential Faraday rotation as well as foreground Faraday rotation. The complex polarisation is given by
\begin{align}
\mathbf{p}(\lambda^2) &= p_{0,1} \frac{\sin (\Delta \phi_1 \lambda^2)}{\Delta \phi_1 \lambda^2} {\rm e}^{2{\rm i} \left[ {\rm PA}_{0,1} + \left( \frac{1}{2} \Delta \phi_1 + \phi_1 \right) \lambda^2\right]} \nonumber \\
&+ p_{0,2} \frac{\sin (\Delta \phi_2 \lambda^2)}{\Delta \phi_2 \lambda^2} {\rm e}^{2{\rm i} \left[ {\rm PA}_{0,2} + \left( \frac{1}{2} \Delta \phi_2 + \phi_2 \right) \lambda^2\right]} {\rm .}
\end{align}
\item \textit{Single internal Faraday dispersion (1Id)}: similar to the 1S model above, but the embedded magneto-ionic medium has a turbulent component. The complex polarisation is given by
\begin{equation}
\mathbf{p} (\lambda^2) = p_0 {\rm e}^{2 {\rm i} ({\rm PA}_0 + \phi \lambda^2)} \left( \frac{1 - {\rm e}^{{\rm i} \Delta \phi \lambda^2 - 2 \sigma_\phi^2 \lambda^4}}{2 \sigma_\phi^2 \lambda^4 - {\rm i} \Delta \phi \lambda^2} \right) {\rm .}
\end{equation}
\end{enumerate}

\begin{figure*}
\includegraphics[width=0.95\textwidth]{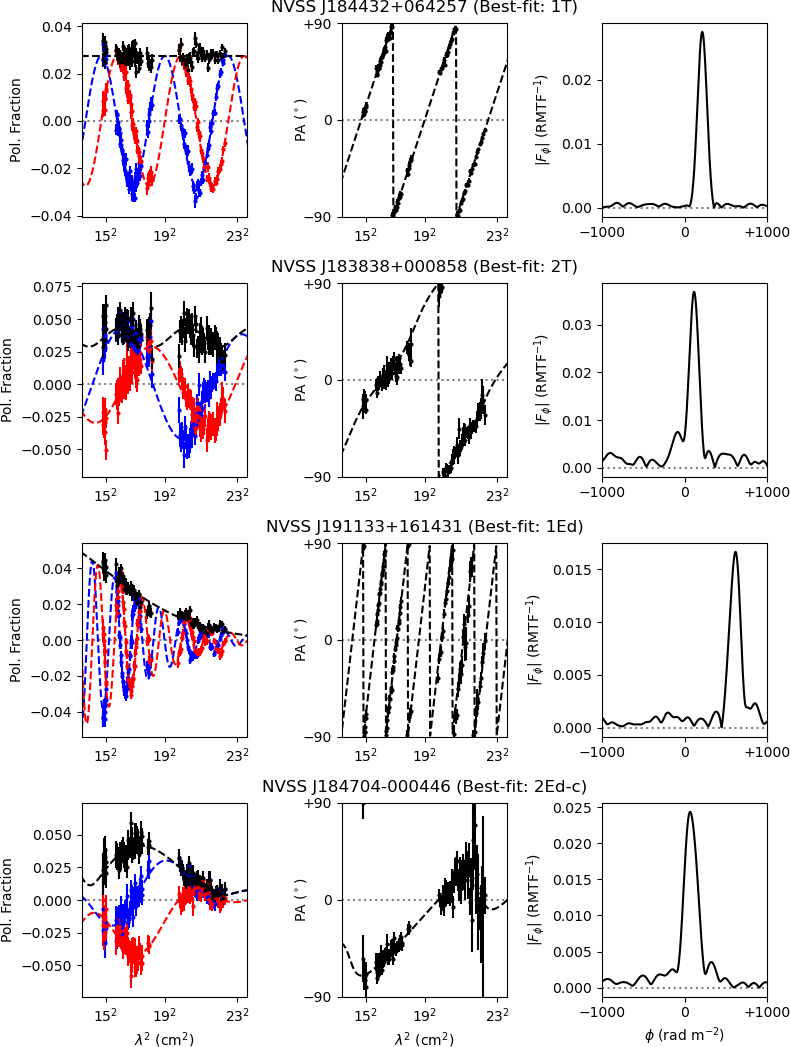}
\caption{Example polarisation spectra of our target EGSs. \textbf{(Left panels)} The measured polarisation fraction, Stokes $q = Q/I$, and Stokes $u = U/I$ across $\lambda^2$ are shown as the black, blue, and red data points, respectively. The dashed lines depict the corresponding best-fit results from Stokes \textit{QU}-fitting (Section~\ref{sec:qufit}). \textbf{(Middle panels)} The measured PA values across $\lambda^2$ are shown as the black data points, with the dashed lines tracing the PA predicted by the best-fit Stokes \textit{QU}-fitting models. \textbf{(Right panels)} The Faraday spectra obtained from RM-Synthesis with RM-Clean.}
\label{fig:queg}
\end{figure*}

\begin{figure*}
\ContinuedFloat
\includegraphics[width=0.95\textwidth]{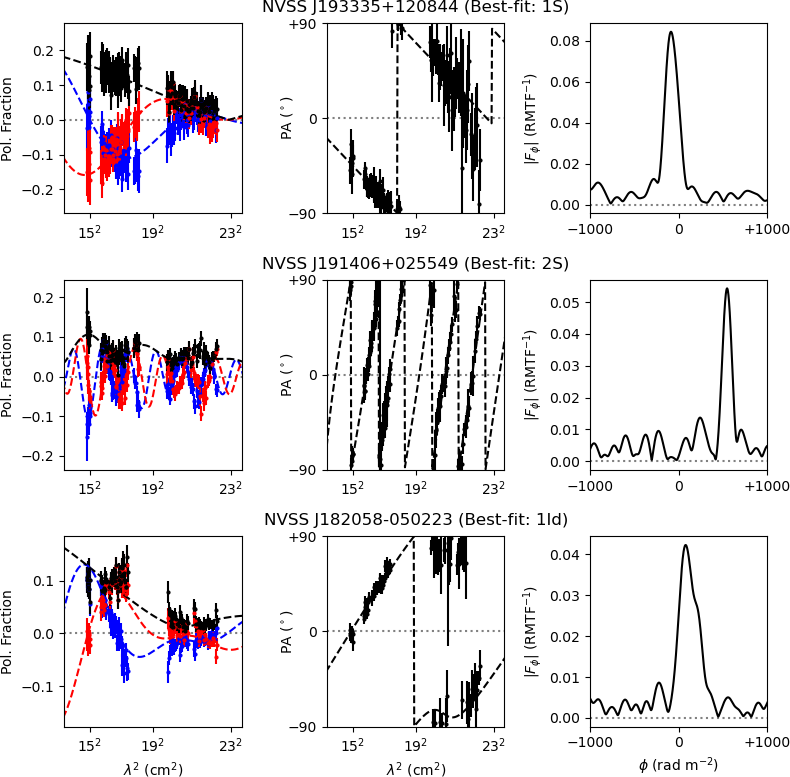}
\caption{\textit{(Continued)} Example polarisation spectra of our target EGSs.}
\end{figure*}

We used the \textsc{Nestle}\footnote{Available on \href{https://github.com/kbarbary/nestle}{https://github.com/kbarbary/nestle}.} nested sampling algorithm \citep{skilling04} through the \textsc{Bilby}\footnote{Available on \href{https://git.ligo.org/lscsoft/bilby}{https://git.ligo.org/lscsoft/bilby}.} library \citep{ashton19} to determine the best-fit parameters for each source and each model. The priors are set as uniform distribution within the ranges listed below:
\begin{itemize}
\item $\displaystyle -1000 \leq \phi, \phi_1, \phi_2 [{\rm rad\,m}^{-2}] \leq +1000$,
\item $\displaystyle +10 \leq (\phi_1 - \phi_2) [{\rm rad\,m}^{-2}] \leq +2000$,
\item $\displaystyle 1 \leq \sigma_\phi, \Delta\phi [{\rm rad\,m}^{-2}] \leq 150$,
\item $\displaystyle 0.1 \leq p, p_1, p_2 [\%] \leq 100$,
\item $\displaystyle 0.1 \leq (p_1 + p_2) [\%] \leq 100$, and
\item $\displaystyle 0 \leq {\rm PA}_0, {\rm PA}_{0,1}, {\rm PA}_{0,2} [^\circ] \leq 180$.
\end{itemize}
These chosen ranges are motivated by the absence of obvious polarisation components with $|\phi| > 1000\,{\rm rad\,m}^{-2}$ upon inspection of the \cite{ma20} Faraday spectra from RM-Synthesis, as well as the Faraday spectrum resolution ($\delta\phi$) and the maximum detectable scale in FD space (max-scale) for these observations \citep[][numbers taken from \citealt{ma20}]{brentjens05}:
\begin{gather}
\delta\phi \approx \frac{2 \sqrt{3}}{\Delta \lambda^2} \approx 123\,{\rm rad\,m}^{-2}\text{, and} \\
\text{max-scale} \approx \frac{\pi}{\lambda_{\rm min}^2} \approx 144\,{\rm rad\,m}^{-2}{\rm .}
\end{gather}
Here, we assume that RM-Synthesis and Stokes \textit{QU}-fitting can uncover polarised emission across a continuous range of FD up to a similar max-scale. To ensure adequate sampling of the posterior, we have set the number of active points (\texttt{nlive} in \textsc{bilby}) as 128.

Before proceeding with identifying the best-fit model for each of the target sources, we first reject models that are deemed as bad for each source \citep[similar to][]{ma19a}. These rejected models have best-fit parameters that (i) have $p_0$ signal-to-noise ratio of less than five (or six for the single Faraday thin model\footnote{This is to align with the \cite{ma20} polarisation fraction signal-to-noise cutoff of six, applied to their RM-Synthesis peak-fitting results.}), (ii) have uncertainty in $\phi$ of more than $20\,{\rm rad\,m}^{-2}$, (iii) have $\Delta \phi$ signal-to-noise ratio of less than three, or (iv) have $\sigma_\phi$ signal-to-noise ratio of less than three. The rationales behind these are to remove models for which the added extra polarisation component or the depolarisation introduced by $\Delta \phi$ or $\sigma_\phi$ may not be necessary.

Finally, we sort the remaining models by their Bayesian information criterion (BIC) values, and identify the best-fit model as the one with the lowest BIC. The difference in BIC values between the second-best model and the best model ($\Delta {\rm BIC}$) is also calculated, with $\Delta {\rm BIC} > 10$ indicating a strong preference for the best-fit model over the second-best, and $\Delta {\rm BIC} < 2$ indicating a weak preference \citep[e.g.][]{osullivan12,schnitzeler18}. The $\Delta{\rm BIC}$ value is excluded if only one model remains after the filtering step above. The best-fit models, their associated best-fit parameters, and the BIC and $\Delta{\rm BIC}$ values are all listed in Table~\ref{table:qufit} in Appendix~\ref{sec:qufit_results}. We remark that none of the 194 sources are best represented by the 2Ed-s model, and therefore it is omitted in the rest of this paper. In addition, we find that three sources (NVSS~J183931$-$101336, NVSS~J184415$-$041757, and NVSS J184547$-$093821) are now classified as being unpolarised, as they have no remaining models after filtering. We believe that the linear polarisation signal from these sources reported by \cite{ma20} is from the Galactic diffuse polarised emission, which is now removed by our implemented foreground emissions subtraction steps (Section~\ref{sec:fg_sub}). The spatial distribution of the remaining 191 polarised EGSs is shown in Figure~\ref{fig:map_fdspread}, coloured by their best-fit models. In Figure~\ref{fig:queg}, we show the polarisation spectra of one example source for each best-fit model.
 
\subsection{High resolution total intensity images} \label{sec:highres_images}

The $\approx 50^{\prime\prime}$ angular resolution of the \cite{ma20} observations were not optimal for obtaining the angular scale and morphology of the target sources. We therefore supplement our study with higher angular resolution data from radio surveys at similar frequency bands, as described below\footnote{Cutout images from both surveys have been obtained from the CIRADA Image Cutout Web Service: \href{http://cutouts.cirada.ca/}{http://cutouts.cirada.ca}.}. For completeness, this is done for all of the 194 EGSs that we begin this study with. Note that as of the time of writing, only the total intensity information is available from these surveys, with the linear polarisation data products still being processed by the respective survey teams \citep[e.g.][]{thomson23}.

\begin{figure*}
\includegraphics[height=\textwidth]{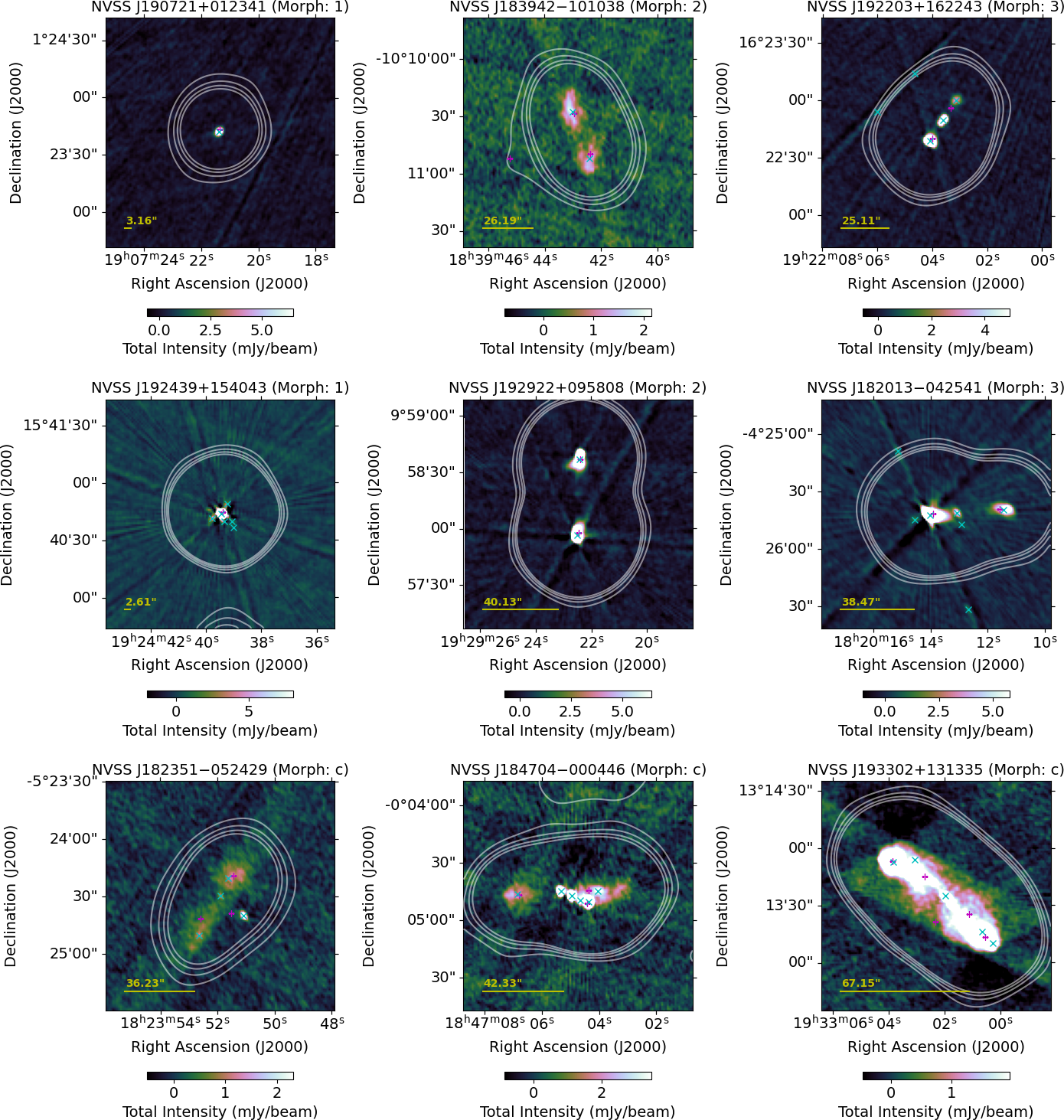}
\caption{Examples of our EGS sample with various morphologies (see panel title suffixes). The background maps in cubehelix colour scheme \citep{green11} are from VLASS \citep{lacy20} epoch 2 with $2\farcs5$ beam, while the contours are from RACS-low1 \citep{mcconnell20} at [2.5, 5.0, 7.5, 10.0] ${\rm mJy\,beam}^{-1}$ with $25^{\prime\prime}$ beam. The cyan crosses and magenta plus signs mark the positions of source components from the VLASS \citep{gordon21} and the RACS-low1 \citep{hale21} source catalogues, respectively. The scale bars to the lower left of each panel represent the angular scale of the source (Section~\ref{sec:highres_images}). Finally, note that the polarisation data that we use for this work have an angular resolution of about $50^{\prime\prime}$ \citep[not shown here;][]{ma20}. The full sample of 194 sources is shown in the Online Supplementary Materials.}
\label{fig:sample_src}
\end{figure*}

First, we obtain the data from the RACS survey \citep{mcconnell20} of ASKAP \citep{hotan21}. Specifically, we use the first epoch of the low-frequency component ($887.5\,{\rm MHz}$) of the survey (RACS-low1). Cutout images of $2^\prime \times 2^\prime$ (for most of the 194 sources; larger for sources with large angular extents) with angular resolution of $25^{\prime\prime}$ are acquired. Furthermore, we obtain the sky positions of the total intensity source components within the cutout images from the RACS-low1 source catalogue \citep{hale21}.

Second, we obtain the data from VLASS \citep{lacy20} at $3\,{\rm GHz}$ with angular resolution of $\approx 2\farcs5$. Cutouts of the VLASS epoch 2 Quick Look images are made for our target sources with identical image sizes as RACS-low1 above, and we similarly extract the source positions in total intensity from the VLASS source catalogue \citep{gordon21}. We note that our study here is largely unaffected by the potential limited accuracy in absolute astrometry ($\approx 1^{\prime\prime}$) and flux density ($\approx 10\,\%$) of the VLASS data, since we are mainly concerned about the angular scale and morphology here. 

Finally, we visually inspect each of the images (nine sample sources shown in Figure~\ref{fig:sample_src} in the main text, and all 194 sources shown in the Online Supplementary Materials) to determine the angular scales and morphologies of our target sources. Given the factor of $\approx 10$ difference in angular resolution between the two datasets above, we find that most of our conclusions are drawn from the VLASS data alone. However, note that since the largest angular scale that the VLASS observations are sensitive to is $\approx 58^{\prime\prime}$, the RACS-low1 data are invaluable for the evaluation of our target sources with spatial extents larger than this scale (e.g.\ NVSS J182351$-$052429 shown in Figure~\ref{fig:sample_src}). We determine the angular scale by three different methods depending on the EGS source morphology:
\begin{enumerate}
\item Wherever possible, we use the angular distance between two source components that reasonably describe the size of our target source;
\item For point-like sources, we use the size of the semi-major axis of the fitted Gaussian component as reported by the VLASS catalogue; and
\item For sources with highly complicated morphologies\footnote{Specifically, NVSS J182058$-$050223, NVSS J184919$+$063211, NVSS J185857$+$000727, NVSS J190832$-$005319, and NVSS J191725$+$044236 are sources with highly complicated morphologies that we have to manually measure the angular extents.}, we measure the spatial scale by manual inspection of the VLASS and RACS-low1 images.
\end{enumerate}
The morphology is decided by visual inspection of the images, and classified into four categories: single-component (1), double-components (2), triple-components (3), and complicated (c). These results are listed in Table~\ref{table:qufit} in Appendix~\ref{sec:qufit_results}, and also shown as the title suffix of each subplot of Figures~\ref{fig:sample_src} as well as in the Online Supplementary Materials.

\begin{figure*}
\includegraphics[width=0.47\textwidth]{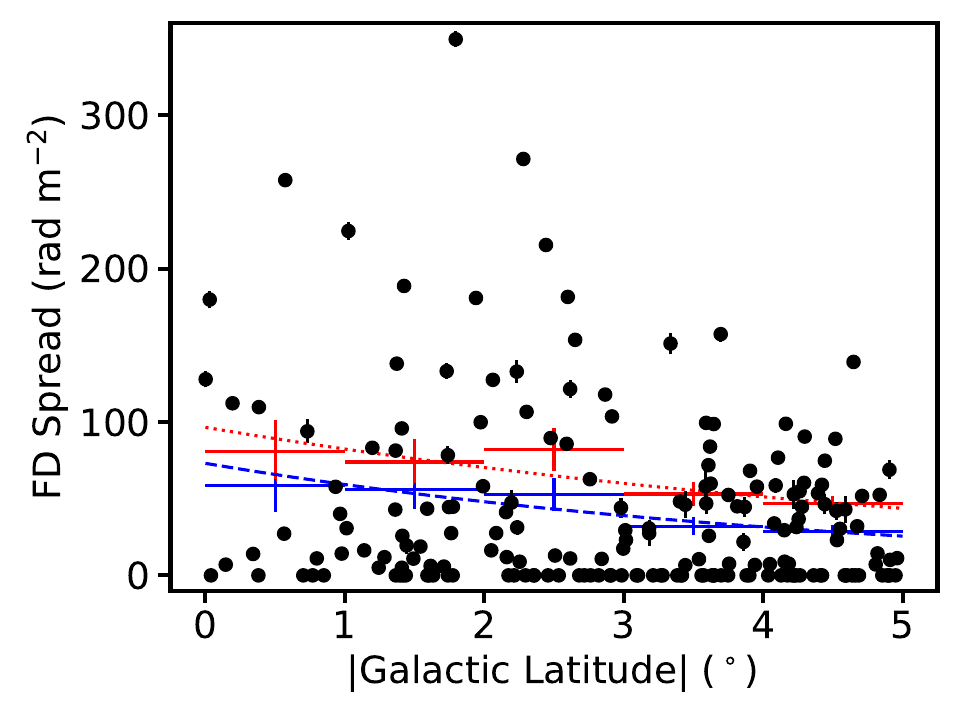}
\includegraphics[width=0.47\textwidth]{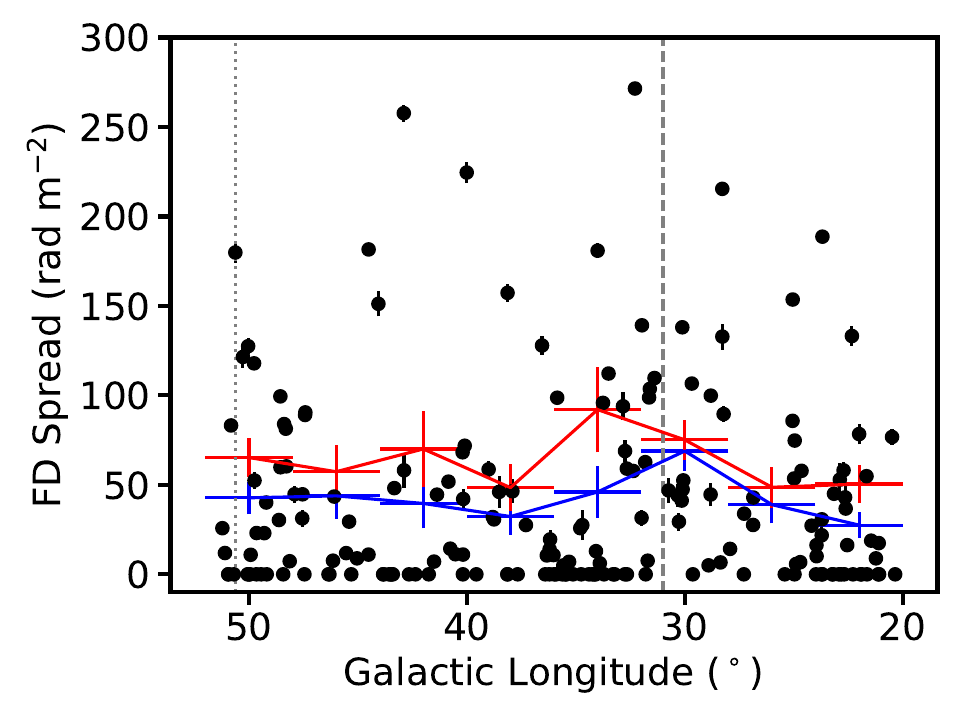}
\caption{The Galactic latitude and longitude dependences of the spatial FD spread. The 191 polarised EGSs are individually plotted as the black data points in both panels, with error bars included but in most cases are too small to be visible. The data points are further binned \textbf{(Left)} along the absolute value of Galactic latitude at a bin width of $1^\circ$, and \textbf{(Right)} along Galactic longitude at a bin width of $4^\circ$. The binned values are shown as the blue and red data points for the cases of binning all data (i.e.\ including data points with FD spread of zero) and binning non-zero data only, respectively. For the Galactic latitude dependence, we fit the binned data points with exponential functions as shown by the dashed / dotted lines (see Section~\ref{sec:latitude}). Meanwhile, for the Galactic longitude dependence, we mark the Scutum and Sagittarius spiral arm tangents as inferred from H\textsc{i} observations \citep{englmaier99,hou15,nakanishi16,vallee22} with the vertical dashed and dotted lines, respectively.}
\label{fig:latitude}
\end{figure*}

\section{Results} \label{sec:results}

The two most significant contributions from this work are the creation of a parameter called the FD spread that we use to quantify the fluctuation of FD across each of our target sources, and the detailed investigation of the best-fit Stokes \textit{QU}-fitting models for these sources. For the former, we examine how FD spread varies with Galactic latitude, longitude, source angular size, and spectral index. The potential influences of Galactic ionised structures are further considered. For the latter, we examine the relationship of the best-fit models with spatial distribution, source morphology, and source angular size. We expand on our results below.

\subsection{The spatial Faraday depth spread} \label{sec:fdspread}
\subsubsection{Definition} \label{sec:fds_definition}

To quantify the spatial FD fluctuations across each of our polarised target sources, we devise a parameter called the FD spread (listed in Table~\ref{table:qufit} in Appendix~\ref{sec:qufit_results}). The FD spread is calculated by combining the parameters from the best-fit Stokes \textit{QU}-fitting model following:
\begin{equation}
\text{FD spread} = \sqrt{\frac{1}{N} \sum_i (\phi_i - \overline{\phi})^2 + \sum_i (\sigma_{\phi,i}^2) + \sum_i (\Delta \phi_{i}^2)}{\rm ,} \label{eq:fdspread}
\end{equation}
where subscript $i$ runs through the $N$ polarisation components of the corresponding Stokes \textit{QU}-fitting model, and $\overline\phi$ is the mean $\phi$ of the $N$ components. In particular, the first term of $(1/N)\sum_i(\phi_i - \overline\phi)^2$ is defined in this manner such that, for example, sources comprised of two compact spatial components and situated behind an external Faraday dispersion medium characterised by $\sigma_\phi$ (and thus best represented by the 2T model) will yield $\text{FD spread} \approx \sigma_\phi$. Each of the terms in Equation~\ref{eq:fdspread} is added to the sum only when appropriate and is set to zero otherwise. For example, sources that are best-fit by the 1Ed model will have both the $(1/N) \sum_i (\phi_i - \overline\phi)^2$ and $\sum_i (\Delta \phi_i^2)$ terms set as zero, reducing the equation to $\text{FD spread} = \sigma_\phi$; sources best-fit by the 2S model will have the $\sum_i (\sigma_{\phi,i}^2)$ term set as zero, yielding $\text{FD spread} = \sqrt{[(\phi_1 - \phi_2)/2]^2 + \Delta \phi_1^2 + \Delta \phi_2^2}$; and sources best-fit by the 1T model simply have $\text{FD spread} = 0\,{\rm rad\,m}^{-2}$. The spatial distribution of FD spread is shown in the bottom panel of Figure~\ref{fig:map_fdspread}.

The FD spread defined here is, in some ways, similar to the second moment of the Faraday spectrum from RM-Synthesis \citep[$M_2$; e.g.][]{anderson15,dickey19,livingston21}. The key difference between the two is that, the calculation of $M_2$ involves a weighting of each FD value by the corresponding polarisation fraction or PI, which corresponds to the magnetic properties of the synchrotron-emitting region (for our case here, within the EGSs themselves). We argue that such a weighting scheme should not be applied for our case here if the spatial FD fluctuations are dominated by the Galactic contributions (see Section~\ref{sec:origin}), and therefore we have decided to devise the FD spread above that instead performs a uniform weighting between different FD values.

We particularly point out that models involving the $\Delta\phi$ parameter (i.e.\ 1S, 2S, and 1Id), most often used in the literature to represent the Burn slab \citep{burn66,sokoloff98}, can possibly have the $\Delta\phi$ parameter representing the foreground ISM (instead of strictly being a property intrinsic to the background source). As derived by \cite{sokoloff98} in equation 41, an FD gradient in the plane-of-sky caused by, for example, the Galactic magneto-ionic medium, can cause the observed $\mathbf{p}(\lambda^2)$ to be modulated by a sinc function. This behaviour is identical to that of the Burn slab. We therefore elect to adopt a more general terminology here -- "sinc component" instead of "Burn slab" -- to highlight that $\Delta\phi$ can reflect the magneto-ionic medium either intrinsic to the EGSs, or in the Galactic foreground. Nonetheless, in Appendix~\ref{sec:remove_deltaphi} we repeat our analysis presented below but with the sinc-component models (i.e.\ 1S, 2S, and 1Id) excluded, under the hypothesis that the $\Delta\phi$ parameter of these sinc-component models is not associated with the foreground ISM. The results of this careful test show that our results below still hold true.

For our investigations of the FD spread trends below, we first perform the analysis by using all of the 191 selected polarised sources, followed by repeating the same analysis by excluding the 66 sources best-fit by the 1T model (which by definition always gives $\text{FD spread} = 0\,{\rm rad\,m}^{-2}$). This is because the 1T sources can either be reflections of truly homogeneous magneto-ionic medium in the foreground (for which case the former method of including them will be more appropriate), or they can be spatially compact sources that individually only probe small spatial scales in the plane-of-sky ($\ll$ coherence length of the magneto-ionic medium) and therefore do not represent the structural changes of the magneto-ionic medium at larger spatial scales (for which case the latter method of discarding them is favoured). As we will show in Section~\ref{sec:qumodel_size}, the use of the full data set is favoured.

\subsubsection{Overall statistics} \label{sec:fds_stats}

First, we report the statistics of the FD spread of our EGSs. Out of the full 191 samples (i.e.\ including sources with $\text{FD spread} = 0$), we find that the mean FD spread is $42\,{\rm rad\,m}^{-2}$, the median is $19\,{\rm rad\,m}^{-2}$, and the standard deviation is $58\,{\rm rad\,m}^{-2}$. Meanwhile, if we consider the 125 Faraday complex sources only (i.e.\ excluding the 1T sources that have $\text{FD spread} = 0$), the mean FD spread is $65\,{\rm rad\,m}^{-2}$, the median is $46\,{\rm rad\,m}^{-2}$, while the standard deviation is $61\,{\rm rad\,m}^{-2}$.

We further explore here the uncertainty values of FD spread. Out of the 125 Faraday complex sources, the mean (maximum) FD spread uncertainty is $3.2\,{\rm rad\,m}^{-2}$ ($9.9\,{\rm rad\,m}^{-2}$). In addition, the mean signal-to-noise of FD spread of the sample is $28$, with 108 out of the 125 EGSs ($86\,\%$) having signal-to-noise of six or above. Therefore, we consider our investigation of the EGS FD spread to be robust against measurement uncertainty.

\subsubsection{As a function of Galactic latitude} \label{sec:latitude}

Next, we look into the dependence of the FD spread on the magnitude of Galactic latitude ($|b|$), as shown in the left panel of Figure~\ref{fig:latitude}. There appears to be signs of decreasing FD spread with increasing $|b|$, with a Pearson correlation coefficient of $-0.21$ ($-0.21$) and a corresponding $p$-value of $3.6 \times 10^{-3}$ ($1.8 \times 10^{-2}$) when considering all 191 EGSs (only the 125 sources with non-zero FD spread). We further perform spatial averaging of the data in independent bins of $1^\circ$ along $|b|$, for both cases of using all data (including data points with zero FD spread; blue) and using non-zero data only (red). The plotted values are the mean values within the $|b|$ bins, the error bars along the $x$-axis being the bin size, and those along the $y$-axis represent the standard error of mean (SEM):
\begin{equation}
{\rm SEM} = \frac{\sigma}{\sqrt{N}}{\rm ,} \label{eq:sem}
\end{equation}
where $\sigma$ here is the standard deviation of FD spread in each bin, and $N$ is the number of FD spread measurements within the same bin. The individual uncertainty values of the FD spread (median of $3\,{\rm rad\,m}^{-2}$) are not considered here, as they are negligible when compared with $\sigma$ ($\approx 30$--$80\,{\rm rad\,m}^{-2}$).

We note that the FD spread distribution at $|b| < 3^\circ$ appears distinct from the $|b| \geq 3^\circ$ counterpart. When considering all data, the mean FD spread values are $55 \pm 7\,{\rm rad\,m}^{-2}$ and $30 \pm 4\,{\rm rad\,m}^{-2}$ for EGSs at $|b| < 3^\circ$ and at $|b| \geq 3^\circ$, respectively. By performing a Kolmogorov-Smirnov test (KS-test) on these two samples, the resulting $p$-value is $1.9 \times 10^{-2}$. Meanwhile, when considering non-zero FD spread data only, the mean FD spread values for $|b| < 3^\circ$ and $|b| \geq 3^\circ$ sources are $78 \pm 9\,{\rm rad\,m}^{-2}$ and $50 \pm 4\,{\rm rad\,m}^{-2}$, respectively. The KS-test $p$-value for such a case is $8.9 \times 10^{-3}$.

We further proceed to fit an exponential function to each of the two sets of the latitude-binned FD spread values:
\begin{equation}
\text{FD Spread} = A \cdot {\rm exp}\left( \frac{-|b|}{C}\right){\rm .} \label{eq:expo}
\end{equation}
Using all FD spread data, we obtain
\begin{equation}
\text{FD Spread} = (70 \pm 10\,{\rm rad\,m}^{-2}) \cdot {\rm exp}\left( \frac{-|b|}{4.8 \pm 0.9\,{\rm deg}}\right){\rm ;} \label{eq:fds_all}
\end{equation}
and by using non-zero FD spread data only, we obtain
\begin{equation}
\text{FD Spread} = (100 \pm 10\,{\rm rad\,m}^{-2}) \cdot {\rm exp}\left( \frac{-|b|}{6.3 \pm 1.6\,{\rm deg}}\right){\rm .} \label{eq:fds_nonzero}
\end{equation}
Finally, we attempt to fit the exponential function to the unbinned data points instead, from which we obtain agreeing results as above.

Overall, from the investigations above, we find an enhanced FD spread near the Galactic mid-plane, most noticeably within Galactic latitude of $\pm 3^\circ$. Finally, we test the robustness of our results above by excluding sources with low $\Delta {\rm BIC}$ values ($< 10$), to see if similar results can be recovered by only using sources with highly reliable best-fit Stokes \textit{QU}-fitting model (see Section~\ref{sec:qufit}). The results are presented in Appendix~\ref{sec:bictest}, from which we conclude that our results are indeed robust against the effects of $\Delta{\rm BIC}$.

\subsubsection{As a function of Galactic longitude} \label{sec:longitude}

We next turn to FD spread as a function of the Galactic longitude ($\ell$), as shown in the right panel of Figure~\ref{fig:latitude}. We further mark by vertical lines the Galactic longitudes of the spiral arm tangents covered by our studied region -- the Scutum arm at $31\fdg0$, and the Sagittarius arm at $50\fdg6$ \citep{vallee22}. Both these values were determined by summarising the H\textsc{i} studies in the literature \citep{englmaier99,hou15,nakanishi16}. From this, we see hints of modulations of FD spread across Galactic longitude, especially for the heightened FD spread in the $28^\circ$--$32^\circ$ bin that matches with the Scutum arm tangent. However, a similar increase in FD spread cannot be found towards the Sagittarius arm tangent in the $48^\circ$--$52^\circ$ bin.

We further evaluate the statistical significance of the modulation with the two-sample KS-test. The FD spread values are separated into eight independent bins, each with a width of $4^\circ$ in $\ell$. We then apply the KS-test to each of the 28 permutations of the $\ell$ bins, again for both cases of using all data and non-zero data only. The resulting $p$-values are listed in Table~\ref{table:longitude_ks}. From this we note that, from considering the full sample, the longitude bin of $28^\circ$--$32^\circ$ appears to stand out, with four out of its seven $\ell$-bin combinations from using all data having $p$-values of less than 0.05. Otherwise, we do not notice any other $\ell$ bins that are obviously statistically different from the others. From considering the EGSs with non-zero FD spread only, we do not see any bin that stand out from the others with statistical significance. We note that Faraday simple sources (which may be reflecting the lack of small-scale Galactic magneto-ionic structures in the foreground; Sections~\ref{sec:qumodel_size} and \ref{sec:gal}) are found in abundance outside of the Galactic longitude range of $28^\circ$--$32^\circ$ (see Figure~\ref{fig:latitude}; to be discussed further in Section~\ref{sec:qumodel_spatial}). Therefore, the removal of these sources with zero FD spread can have significantly enhanced the average FD spread of all longitude bins except for that of $28^\circ$--$32^\circ$, and washing out the statistical differences.

Finally, following the results from Section~\ref{sec:latitude}, we use only EGSs situated at $|b| < 3^\circ$ and repeat the same investigation. From this, we find similar modulations of FD spread across Galactic longitude, but with weaker statistical significance than with the full $|b| \leq 5^\circ$ dataset.

\begin{figure}
\includegraphics[width=0.47\textwidth]{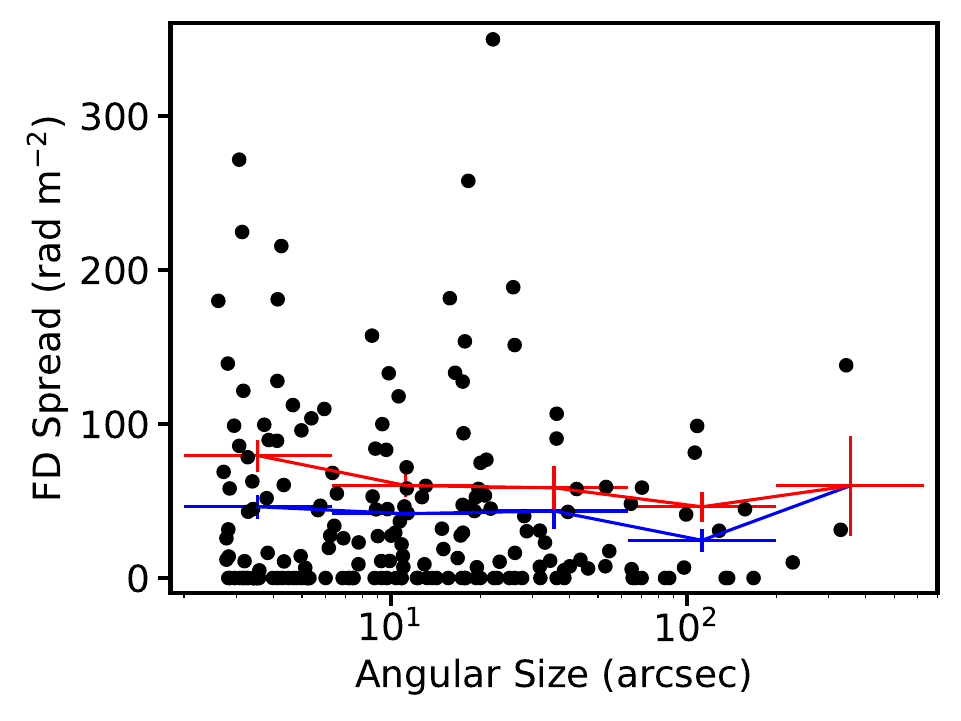}
\caption{The FD spread against the angular size ($\theta$) of the 191 polarised EGSs, plotted individually as the black data points with error bars included. The data are then averaged in independent angular size bins with constant bin widths in logarithmic space of $\log_{10}(\theta/{\rm arcsec}) = 0.5$, and shown as the blue and red data points for the cases of binning all data and binning non-zero data only, respectively. Note that the $x$-axis is in logarithmic scale.}
\label{fig:srcsize}
\end{figure}

\begin{figure}
\includegraphics[width=0.47\textwidth]{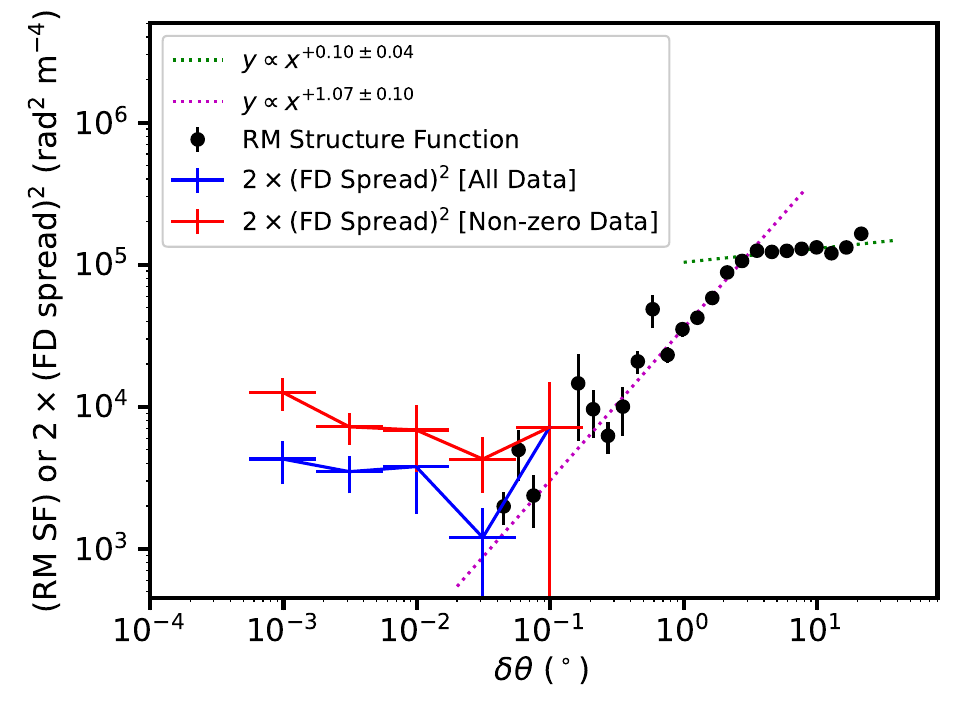}
\caption{The RM SF of the 191 target sources using the peak RM from the Faraday spectra shown in black data points. The RM SF can be represented by a broken power-law function, shown as the green and magenta dotted lines. We also include the trend of $2 \times $(FD Spread)$^2$ as a function of angular scale for the 191 EGSs that we consider, as shown by the blue and red data points (see Figure~\ref{fig:srcsize}).}
\label{fig:rmsf}
\end{figure}

\begin{table*}
\caption{Resulting $p$-values from the two-sample KS-test from comparing between $\ell$ bins}
\label{table:longitude_ks}
\centering
\begin{tabular}{lcccccccc}
\hline
\multicolumn{9}{c}{All Data (Including EGSs with Zero FD Spread)}\\
\hline
& \multicolumn{8}{c}{$\ell$ Bins}\\
$\ell$ Bins & $20^\circ$--$24^\circ$ & $24^\circ$--$28^\circ$ & $28^\circ$--$32^\circ$ & $32^\circ$--$36^\circ$ & $36^\circ$--$40^\circ$ & $40^\circ$--$44^\circ$ & $44^\circ$--$48^\circ$ & $48^\circ$--$52^\circ$\\
\hline
$20^\circ$--$24^\circ$ & -- & 0.386 & \textbf{0.004} & 0.726 & 0.771 & 0.974 & 0.572 & 0.352 \\
$24^\circ$--$28^\circ$ & 0.386 & -- & 0.164 & 0.167 & 0.876 & 0.594 & 0.932 & 0.940 \\
$28^\circ$--$32^\circ$ & \textbf{0.004} & 0.164 & -- & \textbf{0.002} & \textbf{0.033} & \textbf{0.036} & 0.135 & 0.067 \\
$32^\circ$--$36^\circ$ & 0.726 & 0.167 & \textbf{0.002} & -- & 0.344 & 0.818 & 0.087 & 0.247 \\
$36^\circ$--$40^\circ$ & 0.771 & 0.876 & \textbf{0.033} & 0.344 & -- & 0.955 & 0.978 & 0.490 \\
$40^\circ$--$44^\circ$ & 0.974 & 0.594 & \textbf{0.036} & 0.818 & 0.955 & -- & 0.517 & 0.660 \\
$44^\circ$--$48^\circ$ & 0.572 & 0.932 & 0.135 & 0.087 & 0.978 & 0.517 & -- & 0.946 \\
$48^\circ$--$52^\circ$ & 0.352 & 0.940 & 0.067 & 0.247 & 0.490 & 0.660 & 0.946 & -- \\
\hline
\end{tabular}
\\
\begin{tabular}{lcccccccc}
\hline
\multicolumn{9}{c}{Only EGSs with Non-zero FD Spread}\\
\hline
& \multicolumn{8}{c}{$\ell$ Bins}\\
$\ell$ Bins & $20^\circ$--$24^\circ$ & $24^\circ$--$28^\circ$ & $28^\circ$--$32^\circ$ & $32^\circ$--$36^\circ$ & $36^\circ$--$40^\circ$ & $40^\circ$--$44^\circ$ & $44^\circ$--$48^\circ$ & $48^\circ$--$52^\circ$\\
\hline
$20^\circ$--$24^\circ$ & -- & 0.979 & 0.151 & 0.196 & 0.997 & 0.866 & 0.866 & 0.242 \\
$24^\circ$--$28^\circ$ & 0.979 & -- & 0.181 & 0.300 & 0.998 & 0.909 & 0.812 & 0.532 \\
$28^\circ$--$32^\circ$ & 0.151 & 0.181 & -- & 0.579 & 0.148 & 0.360 & 0.270 & 0.746 \\
$32^\circ$--$36^\circ$ & 0.196 & 0.300 & 0.579 & -- & 0.184 & 0.512 & 0.373 & 0.855 \\
$36^\circ$--$40^\circ$ & 0.997 & 0.998 & 0.148 & 0.184 & -- & 0.553 & 0.986 & 0.235 \\
$40^\circ$--$44^\circ$ & 0.866 & 0.909 & 0.360 & 0.512 & 0.553 & -- & 0.898 & 0.543 \\
$44^\circ$--$48^\circ$ & 0.866 & 0.812 & 0.270 & 0.373 & 0.986 & 0.898 & -- & 0.519 \\
$48^\circ$--$52^\circ$ & 0.242 & 0.532 & 0.746 & 0.855 & 0.235 & 0.543 & 0.519 & -- \\
\hline
\multicolumn{9}{l}{\textsc{NOTE} -- $p$-values below 0.05 are bold-faced.} \\
\end{tabular}
\end{table*}

\subsubsection{As a function of source angular size} \label{sec:angularsize}

We show the FD spread as a function of the source angular size in Figure~\ref{fig:srcsize}, from which we find that the measured FD spread is essentially constant across the two decades in angular size ($2\farcs5$--$300^{\prime\prime}$). This is further supported by the near-zero Pearson correlation coefficient of $-0.019$ between these two parameters, as well as the corresponding $p$-value of $0.803$, both suggesting that the two are not correlated. This appears to be in contrast with the findings from previous RM structure function (SF) studies, which quantify the spatial fluctuations of RM due to the turbulent magneto-ionic medium by considering the RM differences between EGS pairs:
\begin{equation}
{\rm RM~SF} (\delta \theta) = \langle [{\rm RM}(\boldsymbol{\theta}) - {\rm RM}(\boldsymbol{\theta} + \boldsymbol{\delta \theta}) ]^2 \rangle{\rm ,}
\end{equation}
where $\boldsymbol{\theta}$ is the 2D sky position vector, $\delta \theta$ is the angular distance between two EGSs (and in bold-face the corresponding 2D vector), and the angled bracket $\langle...\rangle$ denotes averaging all pairs with a distance of $\delta \theta$ between them. For the case of the Milky Way \citep[e.g.][]{haverkorn08,stil11}, the RM SF is typically flat at angular scales larger than a few degrees and follows a power law with a slope of about $0.3$--$1.1$ at smaller angular scales. The former regime is typically interpreted as probing scales larger than the outer scale of MHD turbulence, while the latter corresponds to the inertial range of turbulence cascade of the supernova-driven turbulence. Studies of the RM SF slopes in the inertial range of the Small and Large Magellanic Clouds report similar slopes of $1.4$--$1.5$ \citep{seta23}. Since our FD spread values probe the angular scales of about $2\farcs5$--$300^{\prime\prime}$ corresponding to the inertial range, we would naively expect to see an equivalent decrease of FD spread towards smaller angular sizes, which we do not.

\begin{figure}
\includegraphics[width=0.47\textwidth]{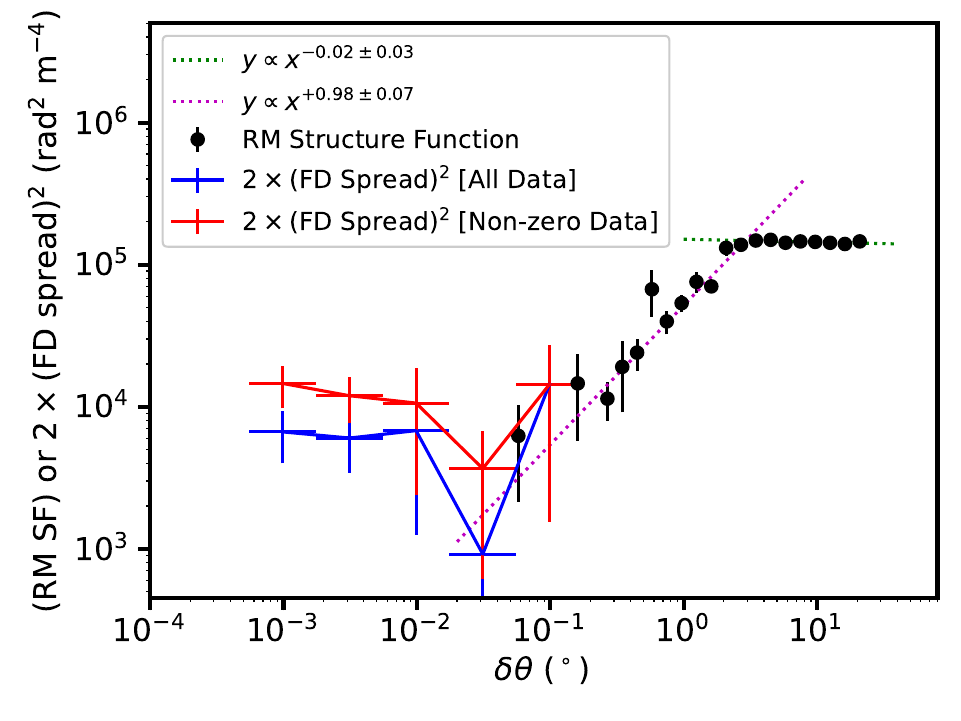}
\caption{Similar to Figure~\ref{fig:rmsf}, but both the RM SF and $2 \times$(FD Spread)$^2$ are constructed from the 94 polarised EGSs situated at $|b| < 3^\circ$ only.}
\label{fig:rmsf2}
\end{figure}

To aid interpretation, we construct the RM SF from the 191 polarised EGSs, using the RM for each source obtained from peak-finding from the Faraday spectra derived from the foreground-subtracted Stokes \textit{IQU} values (Section~\ref{sec:fg_sub}). Specifically, we divided the source pairs into 30 independent bins in $\delta\theta$ space with constant width in logarithmic scale, and discarded bins with less than five pairs within them. As shown in Figure~\ref{fig:rmsf}, the RM SF can be well represented by a broken power-law function, with a break scale of $\delta \theta = 3\fdg0 \pm 0\fdg3$, a spectral slope of $+0.10 \pm 0.04$ at large scales, and $+1.07 \pm 0.10$ at small scales. At scales $\gtrsim 10^\circ$, we see signs of steepening of the RM SF, which can be tracing the Galactic-scale regular magnetic fields \citep[e.g.][]{mao15,seta24}. In addition, we plot $2 \times ({\rm FD~spread})^2$ as a function of the angular size of the EGSs in the same Figure. The reason is that, if one samples pairs of FD measurements from a Gaussian distribution with a standard deviation of $\sigma_{\rm FD}$ and proceeds to calculate the RM SF, the resulting amplitude will be $2 \sigma_{\rm FD}^2$ \citep[see e.g.][]{seta23}.

From this direct comparison, it is clear that the inertial range of the supernova-driven MHD turbulence cascade (shown by the purple dotted line) cannot explain our observed FD spread, since the former is up to two orders of magnitude lower than the latter in the angular scale range of $2\farcs5$--$300^{\prime\prime}$. Finally, given that there appears to be an enhancement of the FD spread amplitude for EGSs at $|b| < 3^\circ$ with respect to their high-latitude counterparts (Section~\ref{sec:latitude}), we repeat the above exercise by using the 94 EGSs in that sky area only. The results are shown in Figure~\ref{fig:rmsf2}.

\begin{figure}
\includegraphics[width=0.47\textwidth]{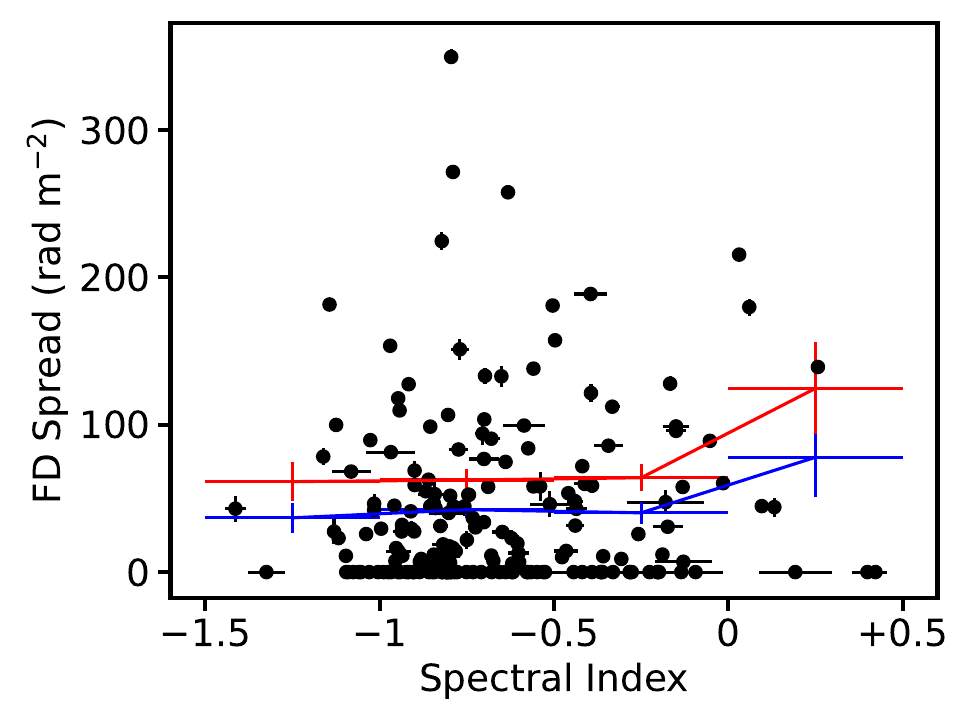}
\caption{The FD spread against the spectral index ($S(\nu) \propto \nu^\alpha$) of the 191 polarised EGSs, each plotted as a black data point with the error bars included. The data points are further averaged within spectral index bins with a constant bin width of $0.5$, and shown as the blue and red data points for the cases of binning all data and binning non-zero data only, respectively.}
\label{fig:spectral_index}
\end{figure}

\subsubsection{As a function of source spectral index} \label{sec:spectralindex}

The spectral index of EGSs can be used as an indicator of the astrophysical properties of these emitting sources \citep[e.g.][]{degasperin18}. We therefore explore if the FD spread shows any relationships with the reported spectral index from \cite{ma20}. Any dependencies between the two would imply that the observed FD spread is an intrinsic property of the EGSs themselves. The results are shown in Figure~\ref{fig:spectral_index}, from which we cannot see any statistically significant trends (the Pearson correlation coefficient between the two is $0.053$ with $p$-value of 0.465, showing that the two are unrelated). This means we do not find evidence that the FD spread is an intrinsic property of the EGSs (see Section~\ref{sec:exgal}).

\subsubsection{Contributions of Galactic ionised structures} \label{sec:haintensity}

\begin{figure}
\includegraphics[width=0.47\textwidth]{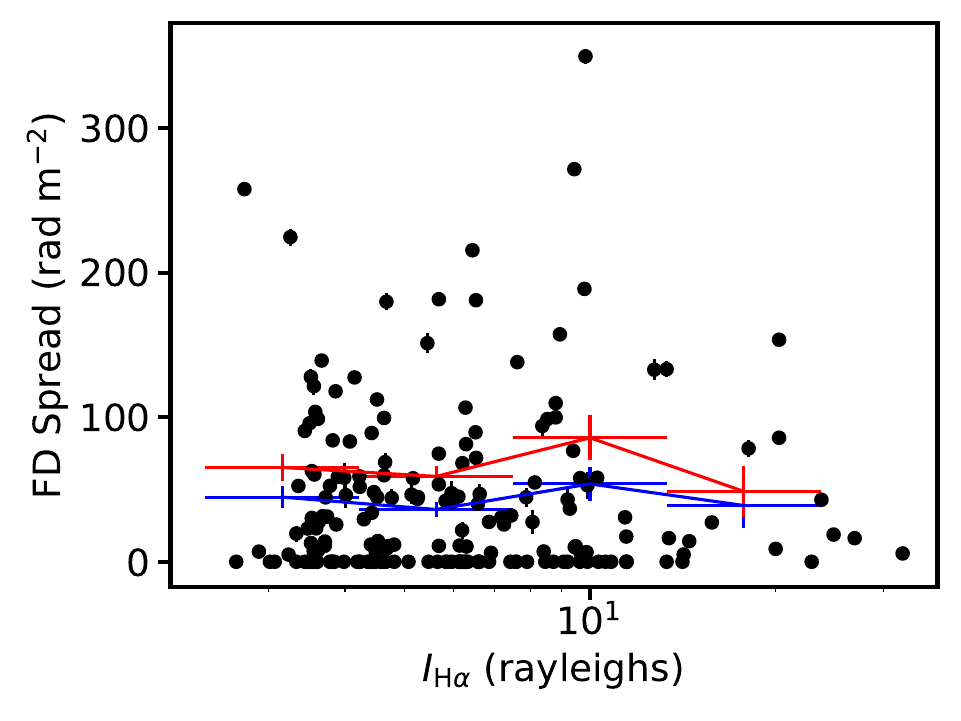}
\caption{The FD spread of the EGSs against the H$\alpha$ intensity of the Galactic ISM from WHAMSS \citep{haffner03,haffner10}. The data points shown in black are further averaged within bins with a constant width in logarithmic space of $\log_{10} (I_{{\rm H}\alpha} / {\rm rayleighs}) = 0.25$, for both cases of using all data points (blue) and non-zero data points only (red).}
\label{fig:fds_ha}
\end{figure}

Finally, we explore if the enhanced FD spread can be attributed to some isolated ionised structures in the Galactic ISM. In Figure~\ref{fig:fds_ha}, we show the FD spread against H$\alpha$ intensity from the WHAMSS survey \citep{haffner03,haffner10}. From the binned data across $I_{{\rm H}\alpha}$ as shown by the blue (all data) and red (non-zero data) lines, we conclude that there are no monotonically increasing trends of FD spread across $I_{{\rm H}\alpha}$. This is further supported by the near-zero Pearson correlation coefficient between the two parameters ($-0.025$).

Furthermore, we investigate if there are obvious positional overlaps between our high-FD-spread EGSs and ionised Galactic objects, namely supernova remnants (SNRs) and H\textsc{ii} regions. We first compare with the \cite{green19} catalogue of SNRs\footnote{Obtained from: \href{https://cdsarc.cds.unistra.fr/viz-bin/cat/VII/284}{https://cdsarc.cds.unistra.fr/viz-bin/cat/VII/284}; 2019 June version is used here.}, and find that only six of our EGSs can be affected -- SNR G25.1$-$2.3 is intercepting NVSS J184249$-$075604, NVSS J184629$-$081333, and NVSS J184617$-$081126; SNR G28.8$+$1.5 is situated in front of NVSS J183652$-$024606 and NVSS J183939$-$030047; and SNR G42.8$+$0.6 impedes NVSS J190741$+$090717. We therefore conclude that Galactic SNRs cannot have been the main cause of our observed FD spread. Meanwhile, we compare with the \textit{WISE} catalogue of Galactic H\textsc{ii} regions \citep{anderson14}\footnote{Obtained from: \href{http://astro.phys.wvu.edu/wise/}{http://astro.phys.wvu.edu/wise/}; v2.2 is used here.}, and find that five sources are intercepted by H\textsc{ii} regions -- NVSS J183717$-$015034, NVSS J185513$+$052158, NVSS J190343$+$055256, NVSS J191158$+$161147, and NVSS J192203$+$162243. We again deem this as insufficient to have caused our observed significant FD spread values.

\begin{table*}
\caption{Source morphology against the best-fit Stokes \textit{QU}-fitting model}
\label{table:morph_model}
\centering
\begin{tabular}{lccccc}
\hline
& \multicolumn{4}{c}{Source Morphology} & \\
Best-fit Model & 1 & 2 & 3 & c & Total\\
\hline
1T & 27 (41\,\%) & 21 (28\,\%) & 5 (31\,\%) & 13 (39\,\%) & 66 (35\,\%) \\
2T & 16 (24\,\%) & 14 (18\,\%) & 3 (19\,\%) & 6 (18\,\%) & 39 (20\,\%) \\
1Ed & 8 (12\,\%) & 14 (18\,\%) & 3 (19\,\%) & 5 (15\,\%)& 30 (16\,\%) \\
2Ed-c & 4 (6\,\%) & 8 (11\,\%) & 2 (13\,\%) & 3 (9\,\%) & 17 (9\,\%) \\
1S & 9 (14\,\%) & 18 (24\,\%) & 3 (19\,\%) & 5 (15\,\%) & 35 (18\,\%) \\
2S & 1 (2\,\%) & 1 (1\,\%) & 0 (0\,\%) & 0 (0\,\%) & 2 (1\,\%) \\
1Id & 1 (2\,\%) & 0 (0\,\%) & 0 (0\,\%) & 1 (3\,\%) & 2 (1\,\%) \\
\hline
Total & 66 (100\,\%) & 76 (100\,\%) & 16 (100\,\%) & 33 (100\,\%) & 191 (100\,\%) \\
\hline
\multicolumn{6}{l}{\textsc{NOTE} -- No sources are best-fit by the 2Ed-s model, and therefore it is omitted here.} \\
\multicolumn{6}{l}{\phantom{\textsc{NOTE} -- }The numbers in the parentheses are the percentages within the respective column.} \\
\end{tabular}
\end{table*}

\subsection{The best-fit Stokes \textit{QU}-fitting model}

Next, we turn our attention to the best-fit Stokes \textit{QU}-fitting model. We investigate whether it varies with sky position, EGS morphology, and EGS angular size.

\subsubsection{Spatial distribution} \label{sec:qumodel_spatial}

From the top panel of Figure~\ref{fig:map_fdspread}, we find that the spatial distribution of the Stokes \textit{QU}-fitting best-fit models is not entirely uniform. The most notable anomaly can be found within the Galactic longitude range of $24^\circ$--$33^\circ$ (roughly coinciding with the tangent point of the Scutum arm at $31\fdg0$; Section~\ref{sec:longitude}), where only seven out of the 47 EGSs (15\,\%) are Faraday simple (1T; shown in red). In comparison, for the 144 sources outside of this longitude range, 59 of them (41\,\%) are Faraday simple. We further investigate the Galactic longitude range of $43^\circ$--$52^\circ$ that covers the Sagittarius arm tangent (at $\ell = 50\fdg6$), and find that 20 out of the 53 EGSs (38\,\%) are Faraday simple. This appears to suggest that the Scutum arm tangent is introducing significant Faraday complexity to the background EGSs, which we will further discuss in Section~\ref{sec:origin}.

\subsubsection{Relation with source morphology} \label{sec:morph_npol}

In Table~\ref{table:morph_model}, we summarise the fraction of EGSs within each morphological class (Section~\ref{sec:highres_images}) that are best represented by each of the Stokes \textit{QU}-fitting models. Surprisingly, we find that there are no correlations between the two, while one could have expected that, for example, spatial doubles would often be represented by polarisation models with two components (i.e.\ 2T, 2Ed-c, 2Ed-s, or 2S). Previous works studying EGSs away from the Galactic plane suggested that there can be relations between the source morphology and the number of polarisation components identified from Stokes \textit{QU}-fitting \citep[e.g.][]{osullivan17,ma19a}, with the source morphologies being determined from high angular resolution radio images (at a few arcseconds, or even sub-arcsec). This will be further discussed in Section~\ref{sec:gal}.

\subsubsection{Relation with source angular size} \label{sec:qumodel_size}

We divide our EGSs into seven groups by their best-fit Stokes \textit{QU}-fitting model, and explore whether they have different statistical distributions in the source angular size. In particular, we perform two-sample KS-tests on the angular size for each of the 21 permutations of the groups. The resulting $p$-values are all high, with the lowest being $0.07$ from comparing between 2T and 1Ed, followed by $0.09$ between 1T and 1Ed and $0.10$ between 2T and 1S. We conclude that there are no clear differences in the angular size distributions between EGSs best-fit by different Stokes \textit{QU}-fitting models.

From the above, combined with the fact that the 1T sources span almost the entire source angular size range of $2\farcs5$--$300^{\prime\prime}$ (see the $\text{FD spread} = 0$ points in Figure~\ref{fig:srcsize}), we suggest that the Faraday simple sources that we see are indeed probing through locally homogeneous lines-of-sight through the Milky Way ISM, instead of them being spatially compact and therefore unable to reveal the fluctuating ISM. We therefore deem the use of the full data set above (instead of using non-zero data only) to be more appropriate.

\section{Discussion} \label{sec:discussion}

As shown in Section~\ref{sec:fdspread}, we find an enhanced FD spread near the Galactic mid-plane, which is especially notable within $|b| < 3^\circ$. Furthermore, the FD spread appears to modulate across Galactic longitude, but remains constant with respect to angular scale of the EGS within $2.5^{\prime\prime}$--$300^{\prime\prime}$. Finally, we do not find any clear relations between FD spread and the EGS spectral index, or the H$\alpha$ intensity in the Galactic ISM.

These observations require some considerations of the origin of these variations. Here, we discuss some possible sources of these variations, as well as the astrophysics implied by the scenarios, and implications for future works.

\subsection{Origin of the observed Faraday depth spread} \label{sec:origin}

There are three obvious possible causes for the observed FD spread: Galactic ISM structures, extragalactic contributions, or instrumental effects. We investigate each of these possibilities below.

\subsubsection{Galactic origin} \label{sec:gal}

We consider small-scale magneto-ionic structures in the Milky Way to be the most probable primary explanation for the observed Faraday complexity of our target EGSs, especially those situated at $|b| < 3^\circ$. This view is supported by the exponential trend of the FD spread with the amplitude of the Galactic latitude (see Section~\ref{sec:latitude}). We further point out the higher fraction of Faraday complex sources near the Scutum arm tangent (Section~\ref{sec:qumodel_spatial}) that is in agreement with the findings from the southern spiral arm tangents \citep{ranchod24}. All these combined are challenging to explain if the FD spread is dominated by extragalactic effects. In addition, as we will explore in more detail in Section~\ref{sec:exgal} below, the amplitude of FD spread that we observe exceeds the expectation from extragalactic contributions.

However, we point out that it is somewhat surprising that the FD spread does not show dependence on the angular scale of the sources (Section~\ref{sec:angularsize}). This means that the FD spread that we observe cannot be attributed to the cascade of the isotropic turbulent magnetic field injected at $10$--$100\,{\rm pc}$ scales \citep[e.g.][]{haverkorn08,stil11}, but would instead be caused by magneto-ionic structures at $< 2\farcs5$ scale (elaborated and explored further in Section~\ref{sec:nature}).

Finally, the lack of correlation between the source morphology and the number of polarisation components (Section~\ref{sec:morph_npol}) can be explained if the observed Faraday complexity is dominated by the Galactic contribution. As mentioned, previous works studying EGSs with Faraday complexity likely dominated by extragalactic contributions \citep[e.g.][]{osullivan17,ma19a} suggested a possible relationship between the two. However, if the Faraday complexity for our work here can be attributed to magneto-ionic structures in the Milky Way at $< 2\farcs5$ scale, the two do not necessarily have to be related.

\begin{figure}
\includegraphics[width=0.47\textwidth]{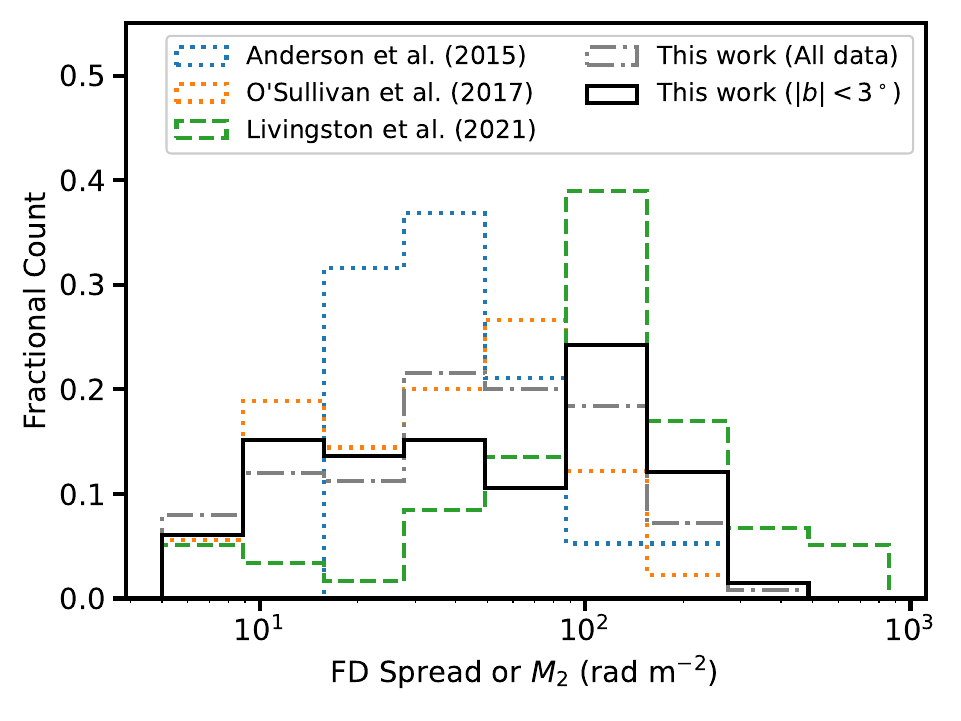}
\caption{The statistical distribution of FD spread \citep[for][as well as this work]{osullivan17} and the second moment of Faraday spectrum \citep[$M_2$; for][]{anderson15,livingston21}.}
\label{fig:fds_m2}
\end{figure}

\subsubsection{Extragalactic origin} \label{sec:exgal}

We deem it less likely that the Faraday complexities observed in our EGS sample are dominated by extragalactic contributions (from, e.g.\ the environments within or surrounding the EGSs). This can be seen when we compare our observed FD spread amplitude with those from previous studies, as summarised below.
\begin{itemize}
\item \cite{anderson15} observed 563 EGSs at high Galactic latitude ($b \approx -55^\circ$) using the ATCA at $1.3$--$2.0\,{\rm GHz}$, and found 19 sources exhibiting Faraday complexity. The Faraday complexity is quantified as the second moment of the RM-Clean component ($\sigma_\phi$ in their work; $M_2$ from hereon). The mean reported $M_2$ value of the Faraday complex sources is $48 \pm 9\,{\rm rad\,m}^{-2}$, with a minimum (maximum) of $17\,{\rm rad\,m}^{-2}$ ($188\,{\rm rad\,m}^{-2}$). Note that they have pointed out that their uncovered Faraday complexity could have been induced by the turbulent Galactic ISM interface undergoing phase transitions, the intervening Fornax galaxy cluster, or intrinsic to the EGSs. This means that their reported $M_2$ can be regarded as upper limits to the extragalactic contributions.
\item \cite{osullivan17} used the ATCA at $1$--$3\,{\rm GHz}$ to observe 100 EGSs at $|b| > 20^\circ$, and characterised their Faraday complexity using Stokes \textit{QU}-fitting. They have found that 90 of the target sources are Faraday complex. We use their reported parameters to calculate the FD spread of each of those 90 sources and find a mean of $50 \pm 4\,{\rm rad\,m}^{-2}$, as well as minimum (maximum) of $6\,{\rm rad\,m}^{-2}$ ($169\,{\rm rad\,m}^{-2}$). We postulate that the observed Faraday complexity of their work can be dominated by extragalactic contributions.
\item \cite{livingston21} studied the Faraday complexity of 62 EGSs within $10^\circ$ from the Galactic centre in projection using the ATCA at $1.1$--$3.1\,{\rm GHz}$, and found using $M_2$ that 59 of their targets exhibit Faraday complexity. The mean $M_2$ of those 59 sources from their work is $154 \pm 20\,{\rm rad\,m}^{-2}$. with minimum (maximum) values of $6\,{\rm rad\,m}^{-2}$ ($857\,{\rm rad\,m}^{-2}$). The enhanced Faraday complexity has been attributed to the small-scale turbulent magneto-ionic structures near the Galactic centre.
\item For our work here, we use the \cite{ma20} VLA data at $1$--$2\,{\rm GHz}$ and find that 125 out of the 191 polarised EGSs are Faraday complex in the Galactic plane region of $20^\circ \leq \ell \leq 52^\circ$ and $|b| \leq 5^\circ$. The mean FD spread of our Faraday complex sources is $65 \pm 5\,{\rm rad\,m}^{-2}$, with minimum (maximum) values of $5\,{\rm rad\,m}^{-2}$ ($350\,{\rm rad\,m}^{-2}$). Meanwhile, if we restrict to $|b| < 3^\circ$ only (see Section~\ref{sec:latitude}), we find that 66 out of the 94 polarised EGSs are Faraday complex, with the mean FD spread being $78 \pm 9\,{\rm rad\,m}^{-2}$, and minimum (maximum) values are also $5\,{\rm rad\,m}^{-2}$ ($350\,{\rm rad\,m}^{-2}$).
\end{itemize}
For the above, the uncertainties of the mean values are calculated using the SEM (Equation~\ref{eq:sem}). We find that both the mean and maximum FD spread from our data are considerably higher than the equivalents from \cite{anderson15} and \cite{osullivan17}, while lower than the $M_2$ statistics from \cite{livingston21}. The same conclusion can be drawn from inspecting the statistical distributions of FD spread or $M_2$ of the Faraday complex sources between these studies, as shown in Figure~\ref{fig:fds_m2}. First, upon applying the two-sample KS-test between our data across the full $|b| \leq 5^\circ$ range and the three works above, we find resulting $p$-values of $0.13$, $0.33$, and $7.5 \times 10^{-8}$ when comparing against \cite{anderson15}, \cite{osullivan17}, and \cite{livingston21}, respectively. By replacing our data with the $|b| < 3^\circ$ subset instead, we get KS-test $p$-values of $0.058$, $8.6 \times 10^{-3}$, and $7.1 \times 10^{-4}$ when compared against the respective works above. Finally, applying the same KS-test between \cite{anderson15} and \cite{osullivan17} yields a $p$-value of 0.14, with the low value possibly due to the limited sample size (19) of \cite{anderson15}. Summarising all these above, we argue that our observed FD spread, especially within $|b| < 3^\circ$, is unlikely to be dominated by extragalactic contributions. Instead, we believe that the FD spread of our EGSs is the aggregate of the extragalactic and the (dominating) Galactic components.

We further point out that we do not find any apparent relations between the FD spread and the spectral index of our target sources (Section~\ref{sec:spectralindex}). The latter can reflect the intrinsic properties of these EGSs, and therefore any relations between the two (which we do not see) would suggest that the FD spread is also an intrinsic property of the EGSs themselves. However, we note that the converse of this statement is not necessarily true (i.e.\ a lack of correlation between the two does not rule out that the FD spread originates from the magneto-ionic medium within / surrounding the EGSs). Regarding the near-constant FD spread as a function of the EGS angular size, while it can be naturally explained if the FD spread is an intrinsic property of the sources \citep[and hence, is independent of the apparent source sizes; e.g.][]{seta23}, it can equally be explained in the Galactic scenario by the presence of magneto-ionic structures at $< 2\farcs5$ scales (see Section~\ref{sec:nature}).

\begin{figure*}
\includegraphics[width=0.47\textwidth]{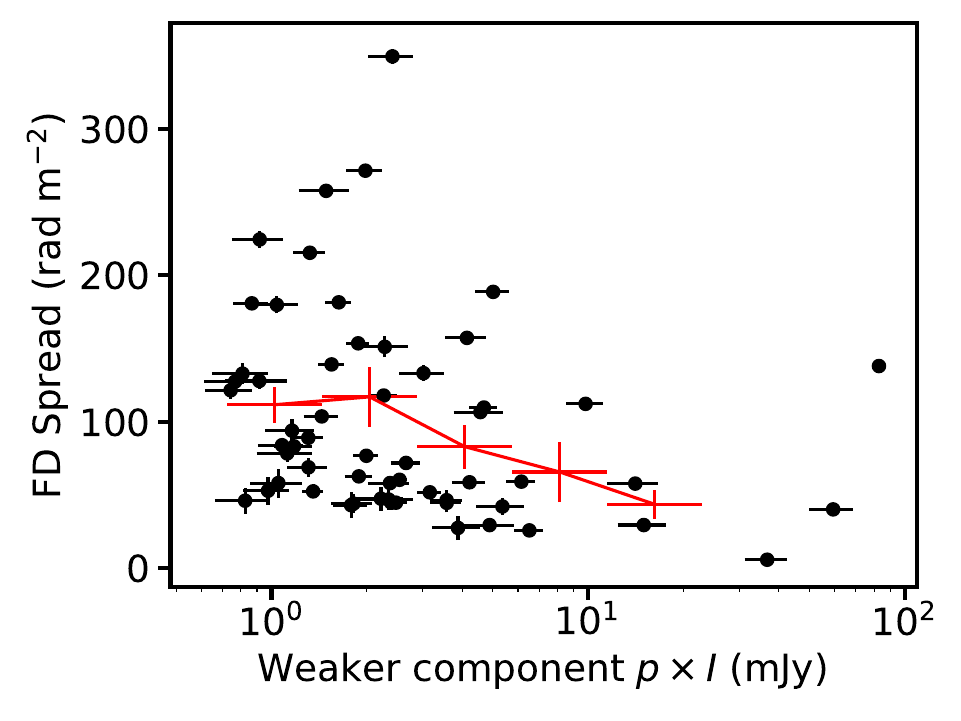}
\includegraphics[width=0.47\textwidth]{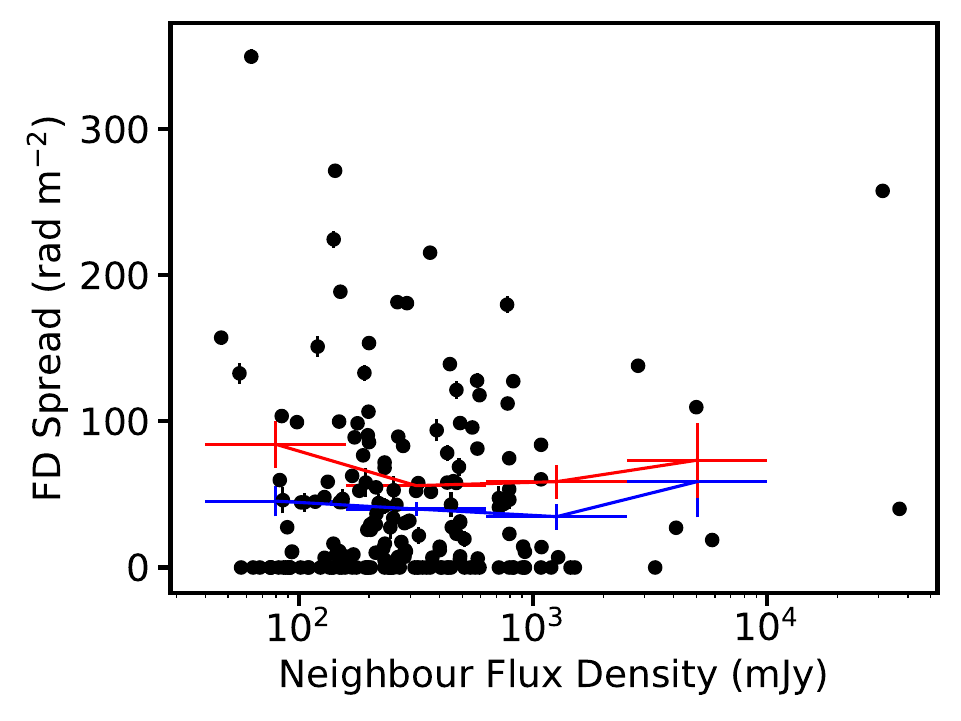}
\includegraphics[width=0.47\textwidth]{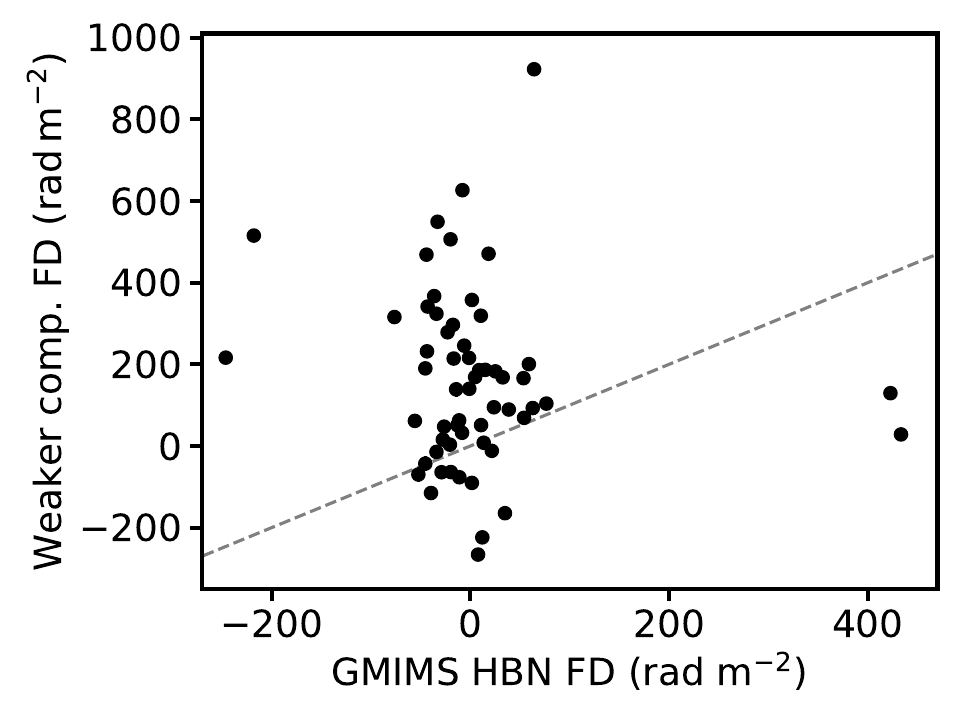}
\includegraphics[width=0.47\textwidth]{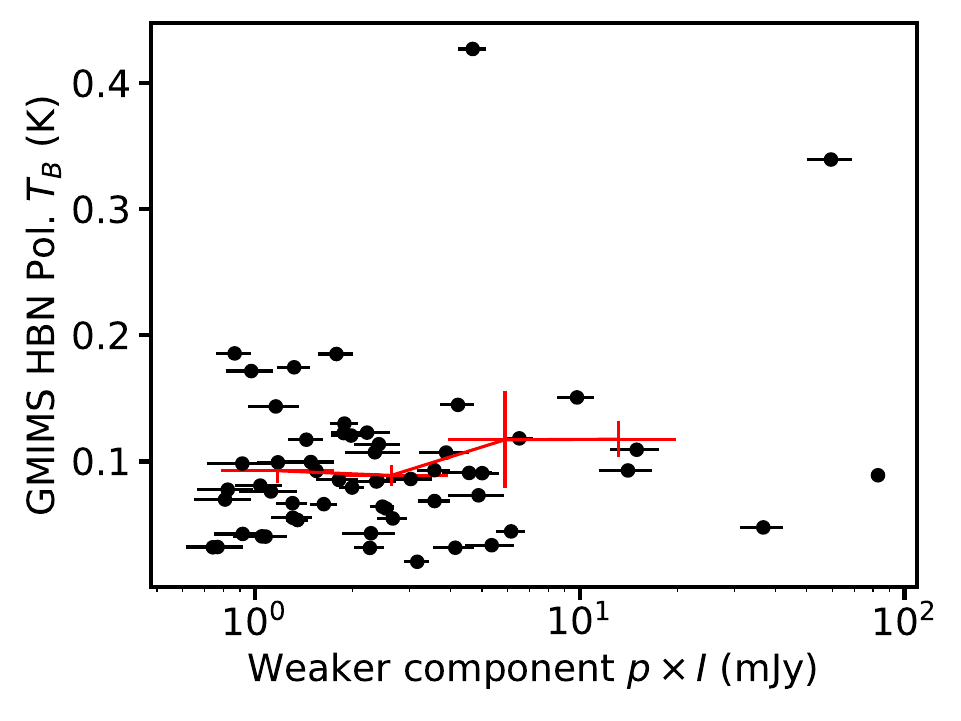}
\caption{Tests for various potential sources of instrumental effects. \textbf{(Upper left)} The FD spread against the PI (obtained from multiplying the $p$ from Stokes \textit{QU}-fitting to the Stokes \textit{I} flux density at 1400\,MHz) of the weaker component. \textbf{(Upper right)} The FD spread against the Stokes \textit{I} flux density of the closest neighbour within $40^\prime$ from the NVSS catalogue \citep{condon98}, with the blue and red data points showing the mean FD spread values within the independent bins from considering all data points and only non-zero data points, respectively. \textbf{(Bottom left)} The FD value of the weaker component against the FD of the Galactic diffuse polarised emission from the GMIMS HBN survey \citep{wolleben21}, with the diagonal grey line showing where the $x$- and $y$-values agree. \textbf{(Bottom right)} The polarised flux density of the Galactic diffuse polarised emission from the GMIMS HBN survey against the PI of the weaker component. Since we do not see any convincing trends, we conclude that instrumental effects do not play a major role in the FD spread values of our sample.}
\label{fig:weak_comp}
\end{figure*}

Finally, we acknowledge that high FD contributions of $\sim 10^3\,{\rm rad\,m}^{-2}$ \citep[e.g.][]{takamura23}, or even $\sim 10^5\,{\rm rad\,m}^{-2}$ \citep[e.g.][]{hovatta19}, from the EGSs themselves have been reported. One can postulate that, in such an intense magneto-ionic medium causing the extreme FD, any inhomogeneities in the medium can easily lead to FD spread of $\gtrsim 100\,{\rm rad\,m}^{-2}$. However, we point out that these observations are often performed at much higher frequency ($\sim 10$--$100\,{\rm GHz}$) than the \cite{ma20} observations ($1$--$2\,{\rm GHz}$), and therefore should not be compared directly. At high radio frequencies, the detected synchrotron emission likely originates from near the core region of active galactic nuclei (AGNs). Meanwhile, at low radio frequencies, most of the emission corresponds to volumes much further away from the core (e.g.\ kpc-scale jets and radio lobes). For our concerned frequency range, intrinsic FD contributions of $\gtrsim 100\,{\rm rad\,m}^{-2}$ are considered rare, but have been found in some cases \citep[e.g.][]{pasetto18,baidoo23}.

\subsubsection{Instrumental origin}

Here, we carefully assess if the enhanced FD spread towards the Galactic mid-plane can be erroneously perceived due to various instrumental effects. As one observes closer to the Galactic mid-plane, the increasing Galactic emission in both total and polarised intensities can affect linear polarisation measurements of the EGSs. One specific example is that if there is a bright Stokes \textit{I} source situated away from the pointing centre, the uncorrected off-axis instrumental polarisation \citep[e.g.][]{ma19b} can lead to non-zero measured Stokes \textit{Q} and \textit{U} values, even if the source itself is unpolarised. If this source is not properly deconvolved in the image plane in Stokes \textit{QU}, the resulting sidelobes across the Stokes \textit{QU} images, presumably at a low level of polarised intensity (PI), can be picked up and mistaken as genuine polarised emission from our target EGSs. While such cases can have been mitigated by our foreground subtraction routine (Section~\ref{sec:fg_sub}), we investigate below if there are residual effects that can have misled our interpretations.

Our first test starts by the hypothesis that, there can be low-level ripples of instrumental polarised emission in the \cite{ma20} data that we can have mistakenly extracted alongside the true emission of the EGSs. To test if this can indeed be the case, we use the 58 sources best-fit by two polarisation components in Stokes \textit{QU}-fitting, and assume that the weaker component (i.e.\ one with lower $p$) can be attributed to the above instrumental effect, and the lower the values of PI, the more likely that the signal can be of instrumental origin. If the statistical distributions of FD spread with Galactic versus instrumental origins are different, it can be possible for the FD spread to have different distributions across PI values. We plot the FD spread of these 58 sources against PI (obtained from multiplying $p$ by the $1400\,{\rm MHz}$ Stokes \textit{I} values reported by \citealt{ma20} in their appendix), as shown in Figure~\ref{fig:weak_comp} top left panel. While the Pearson correlation coefficient between the two is considerable ($-0.12$), the corresponding $p$-value of $0.39$ suggests that the two are likely uncorrelated. We therefore conclude that we do not find, from this test, obvious signs of our results being affected by this scenario.

Next, we explore directly if nearby bright sources in total intensity can be the true culprits of our results. This is tested by drawing a circle of radius of $40^\prime$ (roughly the primary beam of the VLA in L-band) around each of our 191 sources, within which we get the brightest neighbouring sources from the NVSS catalogue \citep{condon98}. We then plot the FD spread values against this neighbour flux density in Stokes \textit{I}, as shown in Figure~\ref{fig:weak_comp} top right panel. From this, we note that sources with modest neighbour flux densities ($\sim 100\,{\rm mJy}$; believed to be unlikely to corrupt our target sources) can still have significant FD spread values of $\gtrsim 200\,{\rm rad\,m}^{-2}$. This, in addition to the flat FD spread across neighbour flux density, shows that the enhanced FD spread observed towards our 191 target sources is unlikely to be due to instrumental effects related to bright nearby sources.

We further briefly discuss the effects of the off-axis sources from the \cite{ma20} observations. In total, they have identified and included 26 polarised off-axis sources within $5^\prime$ from the pointing centres. In their work, they have pointed out that the off-axis instrumental polarisation effect \citep{jagannathan17} has not been corrected for, and therefore it is plausible that up to $0.5\,\%$ of the total intensity has been leaked into the linear polarisation and mistaken as true astrophysical signal at ${\rm FD} \approx 0$. We investigate if this can have happened by looking into the best-fit parameters from Stokes \textit{QU}-fitting for these 26 EGSs, and identify sources with any polarisation components of $p \leq 1\,\%$. From this, only one source has been identified: NVSS J183414$-$030119 with $p_{0,2} = 0.52 \pm 0.06\,\%$ and ${\rm FD}_2 = +51 \pm 7\,{\rm rad\,m}^{-2}$. While this one source might have been affected by the instrumental polarisation effect, we conclude that the off-axis instrumental polarisation effect cannot have misled our overall results.

Finally, we compare our results from the 191 background EGSs with the Global Magneto-Ionic Medium Survey (GMIMS) High Band North (HBN) data \citep{wolleben21} that trace the Galactic polarised emission to verify that there are no residual effects remaining following the foreground subtraction routine (Section~\ref{sec:fg_sub}). We first consider again the case of EGSs best-fit by two polarisation components and investigate if the weaker component (in $p$) can be attributed to the residual Galactic diffuse polarised emission. The FD of the weaker component is compared against the FD from the GMIMS HBN survey, as shown in the lower left panel of Figure~\ref{fig:weak_comp}. We do not see a one-to-one trend between the two, suggesting that the weaker polarisation components of the EGSs do not correspond to the Galactic diffuse emission\footnote{We point out a caveat that interferometric observations of extended polarised emission beyond the interferometer's largest angular scale can subtly alter the obtained FD values \citep{gaensler01}.}. Next, we compare GMIMS HBN PI against the weaker component PI, which should show a linear correlation between the two if the weaker component is actually the diffuse Galactic emission, with $T_B/{\rm K} \approx 0.25 ({\rm PI}/{\rm mJy})$ for $50^{\prime\prime} \times 42^{\prime\prime}$ beam at $1.5\,{\rm GHz}$. This is shown in the lower right panel of Figure~\ref{fig:weak_comp}, which again do not show obvious trends. Though, the obvious caveat here is that the possible effects of beam depolarisation has not been taken into account. To summarise the above, we do not see any indications that there are residual instrumental effects that can have misguided our interpretations in this work.

\begin{figure*}
\includegraphics[width=0.80\textwidth]{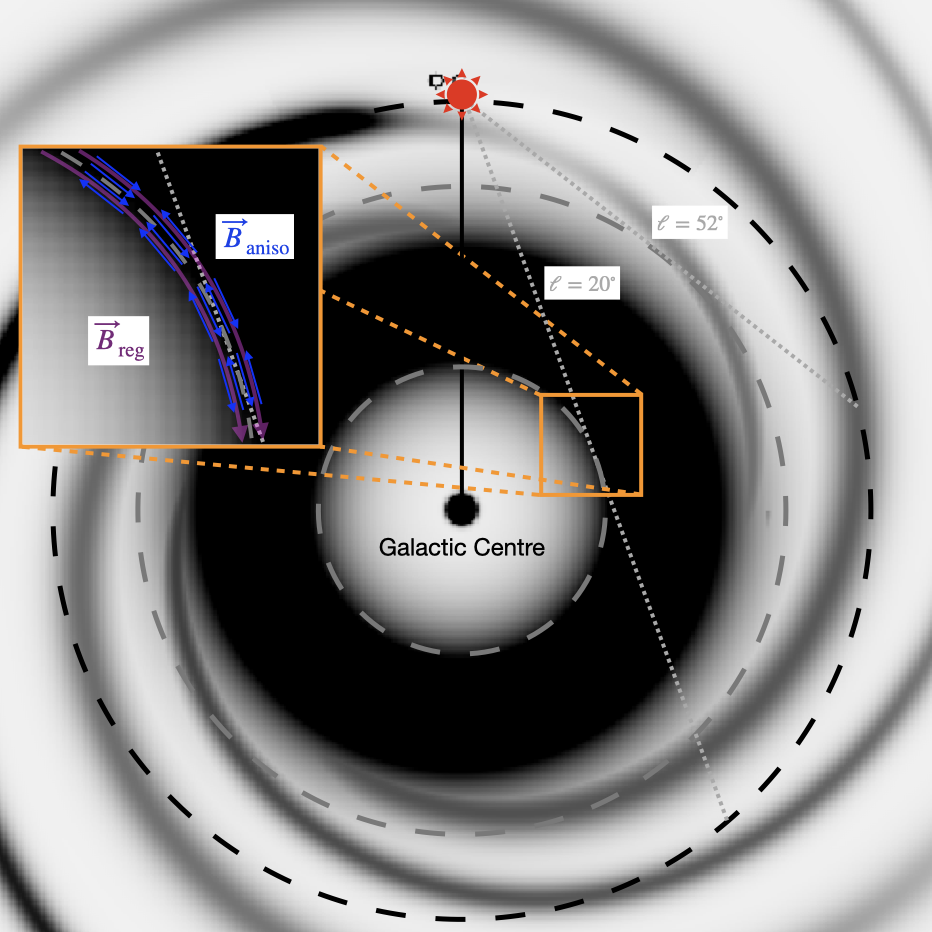}
\caption{A schematic of the anisotropic turbulent magnetic fields of the Milky Way ($\vec{B}_{\rm aniso}$) that can explain our observed FD spread towards background EGSs, showing a top-down view of the Milky Way disk as seen from the North Galactic Pole. The background greyscale image shows the thermal electron number density on the Galactic mid-plane from \citet{ymw16}. The black dashed circle represents the Solar circle with a radius of $8.18\,{\rm kpc}$ \citep{gravity19}, with the Sun marked by the red icon. Meanwhile, the two grey dashed circles centred at the Galactic centre have their tangent points intercepted by our sight lines at $\ell = 20^\circ$ and $52^\circ$ (with these sight lines shown by the dotted grey lines). Note that the coherence lengths and the filling factor of the $\vec{B}_{\rm amiso}$ are not in scale, and the illustrated pitch angles of $0^\circ$ for both $\vec{B}_{\rm aniso}$ and the regular magnetic field ($\vec{B}_{\rm reg}$) do not reflect the actual values.}
\label{fig:illustration}
\end{figure*}

\subsection{The astrophysical nature of the Faraday depth spread} \label{sec:nature}

Through the analysis in Section~\ref{sec:origin}, we believe that the observed FD spread near the Galactic mid-plane is likely dominated by Galactic contributions. In this Section, we explore the magneto-ionic structures that may be causing the enhanced FD spread towards our EGSs.

\subsubsection{Compact ionised structures} \label{sec:ese}

We first assess the plausibility that the enhanced FD spread towards our target EGSs are tracing compact ionised structures in the Milky Way. First, as pointed out in Section~\ref{sec:haintensity}, we do not see an increased H$\alpha$ intensity towards EGSs with higher FD spread. While this means that we do not find evidence from the H$\alpha$ map that there can be contributions by compact ionised structures, we point out that this can be attributed to the rather limited angular resolution of $1^\circ$ for the WHAMSS data \citep{haffner03,haffner10}, as well as the possibility of severe extinction of H$\alpha$ emission towards the distant parts of the Milky Way. In other words, we may not be able to identify such structures from the H$\alpha$ maps even if they are present. However, we also find that SNRs and H\textsc{ii} regions identified from various observations cannot explain our observed FD spread (Section~\ref{sec:haintensity}).

We further argue that ultra compact ionised structures causing Extreme Scattering Events \citep[ESEs; e.g.][]{bannister16} are unlikely to be the culprit here. These elusive Galactic structures can focus/defocus the emission from compact background sources such as pulsars and AGNs, with the observational signature of modulations in the detected flux densities of the background sources over $\approx 10$--$100\,{\rm days}$. The responsible ionised structures have thermal electron number densities of $\approx 1000\,{\rm cm}^{-3}$ and physical scales of $\approx 3 \times 10^{-4}\,{\rm pc}$ \citep{bannister16}. By modifying Equation~\ref{eq:fd}, we obtain:
\begin{equation}
B_\parallel = \frac{\text{FD spread}}{0.81 \cdot n_e \cdot \text{Path-length}}{\rm .} \label{eq:pl_fdspread}
\end{equation}
Assuming that the FD spread of $\approx 50\,{\rm rad\,m}^{-2}$ that we observe from our EGSs are due to the partial obscuration of our EGSs by these ionised structures, we estimate using the above parameters and Equation~\ref{eq:pl_fdspread} that the required coherent magnetic field strength would be $\approx 200\,\mu{\rm G}$. This is significantly higher than the current best upper limit of the magnetic field strength within these ionised structures \citep[$\lesssim 40\,\mu{\rm G}$;][]{clegg96}, and therefore we deem these compact ionised structures as an unlikely explanation to our observed FD spread. However, we point out that if this might be established to be the case, this will prove broadband spectro-polarimetric observations of EGSs to be an exceptionally powerful approach to identifying these elusive over-densities of ionised gas.

\subsubsection{The regular magnetic field}

We consider it impossible for the Galactic regular magnetic field to be the cause for our observed enhancement of FD spread. This conclusion can be arrived at by considering the spatial scales involved. The regular magnetic fields of galaxies are coherent on $\gtrsim\,{\rm kpc}$ scales \citep[e.g.][]{beck16}, and that of the Milky Way as seen from our perspective leads to FD structures at angular scales of larger than several degrees \citep[e.g.][]{vaneck11,ma20,hutschenreuter22}. Given that the enhanced FD spread can be seen towards EGSs with angular scales of $\lesssim 10^{\prime\prime}$, the regular magnetic field cannot be the source of our observed phenomenon.

\subsubsection{The isotropic turbulent magnetic field} \label{sec:dis_iso}

The case for the supernova-driven isotropic turbulent magnetic field being the root of the enhanced FD spread can be similarly ruled out. The coherence length of such magnetic structures is estimated to be about 10\,pc within spiral arms, and about 100\,pc in inter-arm regions \citep[e.g.][]{haverkorn08}. The spiral arm value translates to about $3^\prime$--$30^\prime$ at assumed distances of $1$--$10\,{\rm kpc}$, and even larger if we adopt the inter-arm value. These compare unfavourably to our EGS sample that can have angular scales of $\lesssim 10^{\prime\prime}$. Furthermore, we showed in Figure~\ref{fig:srcsize} that the FD spread exhibits a constant value across angular scale, which is distinct from the expected decaying power-law towards smaller scales as the result of turbulence cascade \citep[e.g.][]{haverkorn08,stil11,seta23}. Both these show that our observed enhanced FD spread cannot be explained by this component of the galactic magnetic field.

We note that the turbulence driven by stellar feedback processes such as stellar wind can be a possible explanation to our observational results. Early-type stars can drive a wind with a mechanical power of $\sim 10^{35}\,{\rm erg\,s}^{-1}$ \citep{draine11}, leading to turbulence in the surrounding ISM at $\sim$\,pc scale \citep{norman96,offner15,mackey20}. However, given that OB-type stars are rare \citep[$\sim 10^4$ known in the Milky Way;][]{reed03,liu19} and are concentrated within $\lesssim 100\,{\rm pc}$ from the Galactic mid-plane \citep{reed00}, it is unlikely to be the primary explanation to our observed widespread enhanced FD-spread for EGSs within $|b| < 3^\circ$. Meanwhile, late-type stars can similarly deposit turbulent energy into the ISM via their winds \citep{elmegreen04}. For instance, those from Solar-type stars can have a mechanical power of $\sim 10^{27}\,{\rm erg\,s}^{-1}$ \citep{liu21}, which can develop turbulence at $\sim 10^{-3}\,{\rm pc}$ scale \citep{mackey20}. Since late-type stars are innumerable and are situated up to about $300$--$400\,{\rm pc}$ from the Galactic plane \citep{sparke06}, the wind bubbles that they create can possibly explain the observed FD spread of our background target sources. However, the confirmation of this scenario will require detailed assessment of the spatial distribution of the magnitude as well as the occurrence of the Galactic FD spread (Section~\ref{sec:future}).

\subsubsection{The anisotropic turbulent magnetic field}

We finally consider the contributions of the anisotropic turbulent magnetic field. As mentioned in Section~\ref{sec:intro}, this galactic magnetic field component can be produced by exerting a shock compression (by, e.g.\ spiral arms) and/or a shearing motion (by the differential rotation of the galaxy) on an initially isotropic turbulent magnetic field \citep[which can have a coherence length of about $10\,{\rm pc}$ in spiral arms;][]{haverkorn08}. These actions can diminish the magnetic field component as well as the turbulence coherence length along the shock direction (or perpendicular to the shearing motion), while simultaneously increasing the magnetic field strength. The resulting anisotropic turbulent magnetic field can therefore be highly elongated, and approximately directed along the tangents of galacto-centric rings (i.e.\ aligned approximately parallel to the lines-of-sight near the Galactic tangent points from our perspective; Figure~\ref{fig:illustration}).

It is therefore possible to attribute our observed enhanced FD spread of EGSs near the Galactic mid-plane to the anisotropic turbulent magnetic field near the Galactic tangent points. This is because, at those locations, this galactic magnetic field component can appear to have a short coherence length ($< 2\farcs5$) in the plane-of-sky from our perspective. The amplified magnetic field strength at small physical scales can therefore lead to FD spread that we detect from our data. However, given our currently limited understanding in this galactic magnetic field component \citep[e.g.][]{jansson12b,beck13,haverkorn15,beck19}, as well as our scarce EGS sample, we cannot further validate if this is indeed the case.

We end our discussions on the anisotropic turbulent magnetic fields here by pointing out a few predictions that can be tested against with future polarimetric observations. First, we anticipate that high-angular-resolution ($\approx 1^{\prime\prime}$) spectro-polarimetric observations towards our EGS sample will be able to directly spatially resolve the magneto-ionic structures leading to the observed FD spread. In other words, the high-resolution FD maps should see patches with different FD values that reflect the plane-of-sky spatial scale as well as the areal filling factor of the responsible magneto-ionic structures. Second, further to the above point, if the underlying magneto-ionic structure is indeed the anisotropic turbulent magnetic field caused by Galactic shocks and shears, we expect that the coherence lengths directly measurable from the above FD maps will be much shorter along the Galactic longitude axis than along the Galactic latitude axis. This is because, for both of the astrophysical processes mentioned above, the interfaces can be 2D sheets perpendicular to the Galactic plane in the simplest cases \citep[however, see also e.g.][]{gomez02,ordog17}. Third and finally, as we expect the anisotropic turbulent magnetic fields to be directed approximately along the tangents of Galacto-centric rings, we should not see any enhanced FD spread from EGSs towards the Galactic anti-centre ($\ell = 180^\circ$), since within this Galactic region, the anisotropic turbulent magnetic fields are expected to be perpendicular to the line-of-sight, and therefore will not contribute to FD or FD spread. In particular, we point out that the second and third predictions will allow one to distinguish between the anisotropic turbulent magnetic field case from the small-scale turbulent magnetic structures from stellar feedback processes (Section~\ref{sec:dis_iso}).

\subsection{Implications for Galactic magnetism studies}

Our study here that fully exploits the capabilities of broadband spectro-polarimetric observations finds that small-scale ($< 2\farcs5$) magneto-ionic structures with significant contributions to the FD spread of EGSs ($\approx 50\,{\rm rad\,m}^{-2}$) appear to be prevalent on the Galactic plane. Regardless of the exact astrophysical nature of the responsible magneto-ionic structures (Section~\ref{sec:nature}), we point out that their impacts on future Galactic magnetism studies can be profound, as elaborated below.

The resulting significant Faraday complexity of the background EGSs means that each source can be emitting at multiple FD values. The challenge for Galactic magnetism studies is then how exactly can one identify the single FD value for each EGS that can best represent the Milky Way magnetic fields, be it for the study of the Galactic-scale magnetic field \citep[e.g.][]{vaneck11,jansson12,ordog17,ma20} or the small-scale turbulent magnetic field \citep[e.g.][]{haverkorn08,stil11}. We suggest that the optimal approach will depend on the volume filling factor of the responsible magneto-ionic structures:
\begin{itemize}
\item For the case where the filling factor is high, we believe that spatially extended EGSs ($\gtrsim 10^{\prime\prime}$) will be less affected, as the FD contributions of the small-scale magneto-ionic structures will statistically average out across the plane-of-sky. This means that despite the significant Faraday complexity, the average FD can be the most representative of the other components of the Galactic magnetic fields. Meanwhile, spatially compact EGSs ($\ll 10^{\prime\prime}$) can be less useful for this case, as their narrow beam volumes make them more susceptible to the FD contributions of this small-scale magneto-ionic structure (i.e.\ no statistical cancellation across the plane-of-sky). This will be true even if they appear to be Faraday simple. All these mean that the optimal approach to obtaining the most ``pristine'' FD values that best reflect the coherent and isotropic turbulent magnetic fields can be to select the spatially extended sources and to discard the spatially compact sources. Or, in other words, the best method can be to counter-intuitively select the Faraday complex sources and discard the Faraday simple sources.
\item On the other hand, if the filling factor is low, then most EGSs may remain unaffected, though the accuracy of Galactic magnetism studies can still benefit from filtering out the occasional sources that can be affected by such small-scale magneto-ionic structures. This is especially the case for spatially compact sources that can again be more susceptible.
\end{itemize}

Our work here suggests that, at least for our Galactic region of interest ($20^\circ \leq \ell \leq 52^\circ$; $|b| \leq 5^\circ$), the filling factor of the small-scale magneto-ionic structures is high. This means that the former approach may be optimal for this sky area.

In addition, as illustrated by \cite{jansson12b}, the inclusion of the anisotropic turbulent magnetic field component is crucial for accurate modelling of the Galactic magnetic field. While instrument limitations have long been the obstacle precluding us from characterising this Galactic magnetic field component in detail, our work here suggests that broadband spectro-polarimetric observations might now be capable of identifying such elusive magnetic structures. We ought to further exploit this observational window to significantly advance our knowledge in the anisotropic turbulent magnetic fields in the Milky Way, to match the recent rapid progress through the multitude of observational tracers to chart the magnetic field structure of our Galaxy \citep[see][]{boulanger18}.
 
\subsection{Future prospects} \label{sec:future}

Our work here finds that some small-scale magneto-ionic structures, which can be best explained by the Galactic anisotropic turbulent magnetic fields or the isotropic turbulent magnetic fields driven by stellar feedback processes in the Milky Way, can be contributing a significant amount of FD spread to EGSs situated near the Galactic mid-plane in projection. Future detailed systematic studies of such magnetic structures are warranted to improve our understanding of their demographics and true astrophysical nature. Eventually, we can gain a deeper understanding of the origins and roles of these galactic magnetic field components, as well as improve the modelling accuracy of the other magnetic components (namely, the regular and the supernova-driven isotropic turbulent magnetic fields) in the Milky Way and, by extrapolation, in other galaxies.

The recent study of the Faraday complexity of EGSs in the southern Galactic plane using the ATCA \citep{ranchod24} complements the northern sample presented in this paper. Meanwhile, The H\textsc{i}/OH/Recombination line survey of the inner Milky Way \citep[THOR;][]{beuther16,shanahan19} carried out also using the VLA in L-band but instead in C-array configuration (with an angular resolution of $\approx 15^{\prime\prime}$) can be an interesting immediate extension of our project presented here. Their survey coverage of $14\fdg5$--$67\fdg25$ in Galactic longitude and within $\pm 1\fdg25$ in latitude presents a good spatial overlap for comparison. Their focus right on the Galactic mid-plane means that we expect many of their detected polarised sources to be significantly Faraday complex (with $\text{FD spread} \approx 50\,{\rm rad\,m}^{-2}$, in addition to the intrinsic FD spread from the EGSs themselves), while their wide longitude coverage can be the key asset to clearly identifying any Galactic longitude dependence of FD spread.

In the near future, the various on-going broadband spectro-polarimetric surveys in both hemispheres will certainly help further clarify the details of the concerned small-scale magneto-ionic structures on the Galactic plane. Both the SPICE-RACS \citep[][Thomson et al.\ in prep.]{thomson23} and the Polarization Sky Survey of the Universe's Magnetism \citep[POSSUM;][]{gaensler10,gaensler25} surveys using ASKAP \citep{hotan21} will cover a large area across the southern sky. Their broad $\lambda^2$ coverage can easily reveal the spatial FD spread within the angular resolution of $\approx 20^{\prime\prime}$, while the wide coverage of both the southern Galactic plane and high Galactic latitude regions at similar sensitivity levels will allow us to more accurately separate the Galactic and extragalactic contributions using the same, homogeneous dataset. Meanwhile, the MeerKAT radio telescope survey led by the Max-Planck-Institut f\"ur Radioastronomie (MPIfR), known as the MPIfR-MeerKAT Galactic Plane Survey \citep[MMGPS;][]{padmanabh23}, will also cover the southern Galactic plane ($298^\circ$--$15^\circ$\footnote{\label{gl}Through Galactic longitude of $0^\circ$.} / $260^\circ$--$350^\circ$ / $280^\circ$--$15^\circ$\footnoteref{gl} in longitude; within $11^\circ$ / $5\fdg2$ / $1\fdg5$ in latitude in their UHF- / L- / S-band components) across a wide frequency range ($544$--$2844\,{\rm MHz}$) that can even further improve the accuracy of Faraday complexity characterisation, but will not have the equivalent off-plane control region as the ASKAP surveys above. Finally, the polarisation component of VLASS \citep{mao14,lacy20} covering the northern sky in S-band ($2$--$4\,{\rm GHz}$) will have the prospect of spatially resolving the small-scale magneto-ionic structures identified here with their superior $2\farcs5$ angular resolution. We particularly stress the significance of such high-spatial-resolution observations in directly revealing the physical scales of the small-scale magneto-ionic structures, and in informing us whether sinc-component sources can indeed be caused by spatial FD variations in the Galactic foreground (Section~\ref{sec:fds_definition}), which can have profound implications on future Faraday complexity studies. Overall, the key to fully unlock this relatively unknown aspect of the Galactic magnetic fields will certainly require the combination of the strengths of all these surveys.

\section{Conclusions} \label{sec:conclusions}

Using the \cite{ma20} spectro-polarimetric data of 191 polarised EGSs on the Galactic plane ($|b| \leq 5^\circ$; $20^\circ \leq \ell \leq 52^\circ$) obtained using the VLA at $1$--$2\,{\rm GHz}$, we measure the spatial fluctuations of FD across the EGSs by applying the Stokes \textit{QU}-fitting algorithm, with the results converted to a newly devised parameter called the FD spread. Out of the full sample, we find significant (mean of $42 \pm 4\,{\rm rad\,m}^{-2}$; up to $350 \pm 5\,{\rm rad\,m}^{-2}$) FD spread that approximately follow an exponential decay as a function of the angular distance from the Galactic mid-plane. The FD spread is particularly enhanced for EGSs within Galactic latitude of $\pm 3^\circ$ (mean of $55 \pm 7\,{\rm rad\,m}^{-2}$). In addition, we find hints of modulation of FD spread as a function of Galactic longitude, in agreement with the recent finding of a higher concentration of Faraday complex EGSs along spiral arm tangents \citep{ranchod24}. From these, we believe that the observed FD spread is dominated by Galactic plane contributions.

By using the high angular resolution total intensity images from the VLASS \citep{lacy20} and RACS-low1 \citep{mcconnell20} surveys, we accurately measure the angular size of each of the EGSs that serve as an upper limit to the magneto-ionic structures responsible for the FD spread. The FD spread appears to be constant across the sampled angular size of $\approx 2\farcs5$--$300^{\prime\prime}$, suggesting that the angular scale of the underlying magneto-ionic structure is smaller than $2\farcs5$. We further construct the RM structure function amongst the 191 EGSs, and compare it against $2 \times (\text{FD spread})^2$ as a function of angular scale. From this, we find that the amplitude of the latter is far higher (by up to two orders of magnitude) than that of the RM structure function extrapolated to the same angular scale range. We, therefore, conclude that the observed FD spread towards the 191 EGSs cannot be explained by the conventional supernova-driven isotropic turbulent magnetic fields.

We consider the anisotropic turbulent magnetic field to be a likely explanation for the enhanced FD spread towards our target EGSs. This galactic magnetic field component can be created by imposing shocks (by, e.g.\ spiral arms) and shears (by, e.g.\ the differential rotation of the Galactic disk) onto an originally isotropic turbulent magnetic field. Our current limited knowledge in this galactic magnetic field component, compounded with our scarce EGS sample, precludes us from an extensive validation of this case. We further point out the possibility of a stellar feedback-driven isotropic turbulent magnetic field as the explanation to the enhanced FD spread, while noting that stellar wind from early-type stars alone cannot fully explain our observations, meaning that late-type stars may be responsible to our observed phenomenon. We propose several observational tests that can be pursued using broadband spectro-polarimetric data from the many on-going surveys in both hemispheres. Overall, we demonstrate that the Faraday complexity induced by the Milky Way presents as both a challenge and a new opportunity for Galactic magnetism studies. Along with other recent studies such as \cite{livingston21} and \cite{ranchod24}, we highlight the importance of fully utilising the capability of broadband spectro-polarimetric observations for accurate characterisation of the Galactic magnetic structures at all scales.

\section*{Acknowledgements}

We thank the anonymous referee for their insightful comments that have led to significant improvements in the presentation of the paper. We thank Rainer Beck, John Dickey, Katia Ferri\'ere, Martin Houde, Shinsuke Ideguchi, and Sarwar Khan for fruitful discussions on this work. The authors acknowledge Interstellar Institute's program ``Anything Interstellar'' and the Paris-Saclay University's Institut Pascal for hosting discussions that nourished the development of the ideas behind this work. The authors acknowledge that this work has benefited from discussions during the programme ``Towards a Comprehensive Model of the Galactic Magnetic Field'' at Nordita in April 2023, partly supported by the NordForsk and Royal Astronomical Society. YKM, AS, NMc-G, CLVE, LO, and CSA acknowledge the Ngunnawal and Ngambri peoples who are the traditional custodians of the land on which the Research School of Astronomy \& Astrophysics is sited at Mount Stromlo. AS acknowledges support from the Australian Research Council's Discovery Early Career Researcher Award (DECRA, project~DE250100003). This work was partially funded by the Australian Government through an Australian Research Council Australian Laureate Fellowship (project number FL210100039) to NMc-G and through the Discovery Projects funding scheme (project number DP220101558). AO is partly supported by the Dunlap Institute at the University of Toronto. This research was supported by the International Research Exchange Support Program of the National Institutes of Natural Sciences and by JSPS KAKENHI Grant Number, JP21H01135 (TA, KK). The National Radio Astronomy Observatory is a facility of the National Science Foundation operated under cooperative agreement by Associated Universities, Inc. This scientific work uses data obtained from Inyarrimanha Ilgari Bundara / the Murchison Radio-astronomy Observatory. We acknowledge the Wajarri Yamaji People as the Traditional Owners and native title holders of the Observatory site. CSIRO’s ASKAP radio telescope is part of the Australia Telescope National Facility (\href{https://ror.org/05qajvd42}{https://ror.org/05qajvd42}). Operation of ASKAP is funded by the Australian Government with support from the National Collaborative Research Infrastructure Strategy. ASKAP uses the resources of the Pawsey Supercomputing Research Centre. Establishment of ASKAP, Inyarrimanha Ilgari Bundara, the CSIRO Murchison Radio-astronomy Observatory and the Pawsey Supercomputing Research Centre are initiatives of the Australian Government, with support from the Government of Western Australia and the Science and Industry Endowment Fund. This research has made use of the CIRADA cutout service at \href{http://cutouts.cirada.ca/}{cutouts.cirada.ca}, operated by the Canadian Initiative for Radio Astronomy Data Analysis (CIRADA). CIRADA is funded by a grant from the Canada Foundation for Innovation 2017 Innovation Fund (Project 35999), as well as by the Provinces of Ontario, British Columbia, Alberta, Manitoba and Quebec, in collaboration with the National Research Council of Canada, the US National Radio Astronomy Observatory and Australia’s Commonwealth Scientific and Industrial Research Organisation. This research has been made possible by the \textsc{Python 3} \citep{python3} software packages of \textsc{NumPy} \citep{numpy}, \textsc{Matplotlib} \citep{matplotlib}, \textsc{SciPy} \citep{scipy}, \textsc{Astropy} \citep{astropy1,astropy2,astropy3}, \textsc{Bilby} \citep{ashton19}, and \textsc{RM-Tools} \citep[][Van Eck et al.\ in prep.]{purcell20}.

\section*{Data Availability}

The main results presented in Table~\ref{table:qufit} in machine-readable format, as well as the Stokes \textit{IQU} spectra in PolSpectra format \citep{vaneck23}, are available in the Online Supplementary Materials as \texttt{best\_model\_table.fits} and \texttt{polspectra.fits}, respectively. The description of each column for \texttt{best\_model\_table.fits} is embedded within the table file. Furthermore, the Stokes \textit{QU}-fitting results in RMTable format \citep{vaneck23} will be included in the RMTable consolidated catalogue\footnote{\href{https://github.com/CIRADA-Tools/RMTable}{https://github.com/CIRADA-Tools/RMTable}} upon publication of this manuscript.

\bibliographystyle{mnras}
\bibliography{ms}

\begin{thebibliography}{}
\makeatletter
\relax
\def\mn@urlcharsother{\let\do\@makeother \do\$\do\&\do\#\do\^\do\_\do\%\do\~}
\def\mn@doi{\begingroup\mn@urlcharsother \@ifnextchar [ {\mn@doi@}
  {\mn@doi@[]}}
\def\mn@doi@[#1]#2{\def\@tempa{#1}\ifx\@tempa\@empty \href
  {http://dx.doi.org/#2} {doi:#2}\else \href {http://dx.doi.org/#2} {#1}\fi
  \endgroup}
\def\mn@eprint#1#2{\mn@eprint@#1:#2::\@nil}
\def\mn@eprint@arXiv#1{\href {http://arxiv.org/abs/#1} {{\tt arXiv:#1}}}
\def\mn@eprint@dblp#1{\href {http://dblp.uni-trier.de/rec/bibtex/#1.xml}
  {dblp:#1}}
\def\mn@eprint@#1:#2:#3:#4\@nil{\def\@tempa {#1}\def\@tempb {#2}\def\@tempc
  {#3}\ifx \@tempc \@empty \let \@tempc \@tempb \let \@tempb \@tempa \fi \ifx
  \@tempb \@empty \def\@tempb {arXiv}\fi \@ifundefined
  {mn@eprint@\@tempb}{\@tempb:\@tempc}{\expandafter \expandafter \csname
  mn@eprint@\@tempb\endcsname \expandafter{\@tempc}}}

\bibitem[\protect\citeauthoryear{{Anderson}, {Bania}, {Balser}, {Cunningham},
  {Wenger}, {Johnstone}  \& {Armentrout}}{{Anderson} et~al.}{2014}]{anderson14}
{Anderson} L.~D.,  {Bania} T.~M.,  {Balser} D.~S.,  {Cunningham} V.,  {Wenger}
  T.~V.,  {Johnstone} B.~M.,   {Armentrout} W.~P.,  2014, \mn@doi [\apjs]
  {10.1088/0067-0049/212/1/1}, \href
  {https://ui.adsabs.harvard.edu/abs/2014ApJS..212....1A} {212, 1}

\bibitem[\protect\citeauthoryear{{Anderson}, {Gaensler}, {Feain}  \&
  {Franzen}}{{Anderson} et~al.}{2015}]{anderson15}
{Anderson} C.~S.,  {Gaensler} B.~M.,  {Feain} I.~J.,   {Franzen} T.~M.~O.,
  2015, \mn@doi [\apj] {10.1088/0004-637X/815/1/49}, \href
  {https://ui.adsabs.harvard.edu/abs/2015ApJ...815...49A} {815, 49}

\bibitem[\protect\citeauthoryear{{Arshakian}, {Beck}, {Krause}  \&
  {Sokoloff}}{{Arshakian} et~al.}{2009}]{arshakian09}
{Arshakian} T.~G.,  {Beck} R.,  {Krause} M.,   {Sokoloff} D.,  2009, \mn@doi
  [\aap] {10.1051/0004-6361:200810964}, \href
  {https://ui.adsabs.harvard.edu/abs/2009A&A...494...21A} {494, 21}

\bibitem[\protect\citeauthoryear{{Ashton} et~al.,}{{Ashton}
  et~al.}{2019}]{ashton19}
{Ashton} G.,  et~al., 2019, \mn@doi [\apjs] {10.3847/1538-4365/ab06fc}, \href
  {https://ui.adsabs.harvard.edu/abs/2019ApJS..241...27A} {241, 27}

\bibitem[\protect\citeauthoryear{{Astropy Collaboration} et~al.,}{{Astropy
  Collaboration} et~al.}{2013}]{astropy1}
{Astropy Collaboration} et~al., 2013, \mn@doi [\aap]
  {10.1051/0004-6361/201322068}, \href
  {https://ui.adsabs.harvard.edu/abs/2013A&A...558A..33A} {558, A33}

\bibitem[\protect\citeauthoryear{{Astropy Collaboration} et~al.,}{{Astropy
  Collaboration} et~al.}{2018}]{astropy2}
{Astropy Collaboration} et~al., 2018, \mn@doi [\aj] {10.3847/1538-3881/aabc4f},
  \href {https://ui.adsabs.harvard.edu/abs/2018AJ....156..123A} {156, 123}

\bibitem[\protect\citeauthoryear{{Astropy Collaboration} et~al.,}{{Astropy
  Collaboration} et~al.}{2022}]{astropy3}
{Astropy Collaboration} et~al., 2022, \mn@doi [\apj]
  {10.3847/1538-4357/ac7c74}, \href
  {https://ui.adsabs.harvard.edu/abs/2022ApJ...935..167A} {935, 167}

\bibitem[\protect\citeauthoryear{{Baidoo}, {Perley}, {Eilek}, {Smirnov},
  {Vacca}  \& {En{\ss}lin}}{{Baidoo} et~al.}{2023}]{baidoo23}
{Baidoo} L.,  {Perley} R.~A.,  {Eilek} J.,  {Smirnov} O.,  {Vacca} V.,
  {En{\ss}lin} T.,  2023, \mn@doi [\apj] {10.3847/1538-4357/acebc5}, \href
  {https://ui.adsabs.harvard.edu/abs/2023ApJ...955...16B} {955, 16}

\bibitem[\protect\citeauthoryear{{Bannister}, {Stevens}, {Tuntsov}, {Walker},
  {Johnston}, {Reynolds}  \& {Bignall}}{{Bannister} et~al.}{2016}]{bannister16}
{Bannister} K.~W.,  {Stevens} J.,  {Tuntsov} A.~V.,  {Walker} M.~A.,
  {Johnston} S.,  {Reynolds} C.,   {Bignall} H.,  2016, \mn@doi [Science]
  {10.1126/science.aac7673}, \href
  {https://ui.adsabs.harvard.edu/abs/2016Sci...351..354B} {351, 354}

\bibitem[\protect\citeauthoryear{{Beck}}{{Beck}}{2007}]{beck07}
{Beck} R.,  2007, \mn@doi [\aap] {10.1051/0004-6361:20066988}, \href
  {https://ui.adsabs.harvard.edu/abs/2007A&A...470..539B} {470, 539}

\bibitem[\protect\citeauthoryear{{Beck}}{{Beck}}{2015}]{beck15}
{Beck} R.,  2015, \mn@doi [\aap] {10.1051/0004-6361/201425572}, \href
  {https://ui.adsabs.harvard.edu/abs/2015A&A...578A..93B} {578, A93}

\bibitem[\protect\citeauthoryear{{Beck}}{{Beck}}{2016}]{beck16}
{Beck} R.,  2016, \mn@doi [\aapr] {10.1007/s00159-015-0084-4}, \href
  {http://adsabs.harvard.edu/abs/2015A\%26ARv..24....4B} {24, 4}

\bibitem[\protect\citeauthoryear{{Beck} \& {Wielebinski}}{{Beck} \&
  {Wielebinski}}{2013}]{beck13}
{Beck} R.,  {Wielebinski} R.,  2013, in {Oswalt} T.~D.,  {Gilmore} G.,  eds,
  Planets, Stars and Stellar Systems.~Vol.\ 5: Galactic Structure and Stellar
  Populations. Springer, Berlin, p.~641

\bibitem[\protect\citeauthoryear{{Beck}, {Brandenburg}, {Moss}, {Shukurov}  \&
  {Sokoloff}}{{Beck} et~al.}{1996}]{beck96}
{Beck} R.,  {Brandenburg} A.,  {Moss} D.,  {Shukurov} A.,   {Sokoloff} D.,
  1996, \mn@doi [\araa] {10.1146/annurev.astro.34.1.155}, \href
  {http://adsabs.harvard.edu/abs/1996ARA\%26A..34..155B} {34, 155}

\bibitem[\protect\citeauthoryear{{Beck}, {Fletcher}, {Shukurov}, {Snodin},
  {Sokoloff}, {Ehle}, {Moss}  \& {Shoutenkov}}{{Beck} et~al.}{2005}]{beck05}
{Beck} R.,  {Fletcher} A.,  {Shukurov} A.,  {Snodin} A.,  {Sokoloff} D.~D.,
  {Ehle} M.,  {Moss} D.,   {Shoutenkov} V.,  2005, \mn@doi [\aap]
  {10.1051/0004-6361:20053556}, \href
  {https://ui.adsabs.harvard.edu/abs/2005A&A...444..739B} {444, 739}

\bibitem[\protect\citeauthoryear{{Beck}, {Chamandy}, {Elson}  \&
  {Blackman}}{{Beck} et~al.}{2019}]{beck19}
{Beck} R.,  {Chamandy} L.,  {Elson} E.,   {Blackman} E.~G.,  2019, \mn@doi
  [Galaxies] {10.3390/galaxies8010004}, \href
  {https://ui.adsabs.harvard.edu/abs/2019Galax...8....4B} {8, 4}

\bibitem[\protect\citeauthoryear{{Beck}, {Berkhuijsen}, {Gie{\ss}{\"u}bel}  \&
  {Mulcahy}}{{Beck} et~al.}{2020}]{beck20}
{Beck} R.,  {Berkhuijsen} E.~M.,  {Gie{\ss}{\"u}bel} R.,   {Mulcahy} D.~D.,
  2020, \mn@doi [\aap] {10.1051/0004-6361/201936481}, \href
  {https://ui.adsabs.harvard.edu/abs/2020A&A...633A...5B} {633, A5}

\bibitem[\protect\citeauthoryear{{Beuther} et~al.,}{{Beuther}
  et~al.}{2016}]{beuther16}
{Beuther} H.,  et~al., 2016, \mn@doi [\aap] {10.1051/0004-6361/201629143},
  \href {https://ui.adsabs.harvard.edu/abs/2016A&A...595A..32B} {595, A32}

\bibitem[\protect\citeauthoryear{{Boulanger} et~al.,}{{Boulanger}
  et~al.}{2018}]{boulanger18}
{Boulanger} F.,  et~al., 2018, \mn@doi [\jcap] {10.1088/1475-7516/2018/08/049},
  \href {https://ui.adsabs.harvard.edu/abs/2018JCAP...08..049B} {2018, 049}

\bibitem[\protect\citeauthoryear{{Brandenburg} \& {Ntormousi}}{{Brandenburg} \&
  {Ntormousi}}{2023}]{brandenburg23}
{Brandenburg} A.,  {Ntormousi} E.,  2023, \mn@doi [\araa]
  {10.1146/annurev-astro-071221-052807}, \href
  {https://ui.adsabs.harvard.edu/abs/2023ARA&A..61..561B} {61, 561}

\bibitem[\protect\citeauthoryear{{Brentjens} \& {de Bruyn}}{{Brentjens} \& {de
  Bruyn}}{2005}]{brentjens05}
{Brentjens} M.~A.,  {de Bruyn} A.~G.,  2005, \mn@doi [\aap]
  {10.1051/0004-6361:20052990}, \href
  {http://adsabs.harvard.edu/abs/2005A\%26A...441.1217B} {441, 1217}

\bibitem[\protect\citeauthoryear{{Brown} \& {Taylor}}{{Brown} \&
  {Taylor}}{2001}]{brown01}
{Brown} J.~C.,  {Taylor} A.~R.,  2001, \mn@doi [\apjl] {10.1086/338358}, \href
  {https://ui.adsabs.harvard.edu/abs/2001ApJ...563L..31B} {563, L31}

\bibitem[\protect\citeauthoryear{{Burn}}{{Burn}}{1966}]{burn66}
{Burn} B.~J.,  1966, \mn@doi [\mnras] {10.1093/mnras/133.1.67}, \href
  {https://ui.adsabs.harvard.edu/abs/1966MNRAS.133...67B} {133, 67}

\bibitem[\protect\citeauthoryear{{CASA Team} et~al.,}{{CASA Team}
  et~al.}{2022}]{casa22}
{CASA Team} et~al., 2022, \mn@doi [\pasp] {10.1088/1538-3873/ac9642}, \href
  {https://ui.adsabs.harvard.edu/abs/2022PASP..134k4501C} {134, 114501}

\bibitem[\protect\citeauthoryear{{Chan} \& {Del Popolo}}{{Chan} \& {Del
  Popolo}}{2022}]{chan22}
{Chan} M.~H.,  {Del Popolo} A.,  2022, \mn@doi [\mnras]
  {10.1093/mnrasl/slac091}, \href
  {https://ui.adsabs.harvard.edu/abs/2022MNRAS.516L..72C} {516, L72}

\bibitem[\protect\citeauthoryear{{Chen}, {Lopez-Rodriguez}, {Ivison}, {Geach},
  {Dye}, {Liu}  \& {Bendo}}{{Chen} et~al.}{2024}]{chen24}
{Chen} J.,  {Lopez-Rodriguez} E.,  {Ivison} R.~J.,  {Geach} J.~E.,  {Dye} S.,
  {Liu} X.,   {Bendo} G.,  2024, \mn@doi [A&A] {10.1051/0004-6361/202450969},
  692, A34

\bibitem[\protect\citeauthoryear{{Cho} \& {Lazarian}}{{Cho} \&
  {Lazarian}}{2003}]{cho03}
{Cho} J.,  {Lazarian} A.,  2003, \mn@doi [\mnras]
  {10.1046/j.1365-8711.2003.06941.x}, \href
  {https://ui.adsabs.harvard.edu/abs/2003MNRAS.345..325C} {345, 325}

\bibitem[\protect\citeauthoryear{{Cho}, {Lazarian}  \& {Vishniac}}{{Cho}
  et~al.}{2002}]{cho02}
{Cho} J.,  {Lazarian} A.,   {Vishniac} E.~T.,  2002, \mn@doi [\apj]
  {10.1086/324186}, \href
  {https://ui.adsabs.harvard.edu/abs/2002ApJ...564..291C} {564, 291}

\bibitem[\protect\citeauthoryear{{Clegg}, {Fey}  \& {Fiedler}}{{Clegg}
  et~al.}{1996}]{clegg96}
{Clegg} A.~W.,  {Fey} A.~L.,   {Fiedler} R.~L.,  1996, \mn@doi [\apjl]
  {10.1086/309884}, \href
  {https://ui.adsabs.harvard.edu/abs/1996ApJ...457L..23C} {457, L23}

\bibitem[\protect\citeauthoryear{{Condon}, {Cotton}, {Greisen}, {Yin},
  {Perley}, {Taylor}  \& {Broderick}}{{Condon} et~al.}{1998}]{condon98}
{Condon} J.~J.,  {Cotton} W.~D.,  {Greisen} E.~W.,  {Yin} Q.~F.,  {Perley}
  R.~A.,  {Taylor} G.~B.,   {Broderick} J.~J.,  1998, \mn@doi [\aj]
  {10.1086/300337}, \href {http://adsabs.harvard.edu/abs/1998AJ....115.1693C}
  {115, 1693}

\bibitem[\protect\citeauthoryear{{Dickey} et~al.,}{{Dickey}
  et~al.}{2019}]{dickey19}
{Dickey} J.~M.,  et~al., 2019, \mn@doi [\apj] {10.3847/1538-4357/aaf85f}, \href
  {https://ui.adsabs.harvard.edu/abs/2019ApJ...871..106D} {871, 106}

\bibitem[\protect\citeauthoryear{{Dickey} et~al.,}{{Dickey}
  et~al.}{2022}]{dickey22}
{Dickey} J.~M.,  et~al., 2022, \mn@doi [\apj] {10.3847/1538-4357/ac94ce}, \href
  {https://ui.adsabs.harvard.edu/abs/2022ApJ...940...75D} {940, 75}

\bibitem[\protect\citeauthoryear{{Draine}}{{Draine}}{2011}]{draine11}
{Draine} B.~T.,  2011, {Physics of the Interstellar and Intergalactic Medium}.
Princeton University Press

\bibitem[\protect\citeauthoryear{{Elmegreen} \& {Scalo}}{{Elmegreen} \&
  {Scalo}}{2004}]{elmegreen04}
{Elmegreen} B.~G.,  {Scalo} J.,  2004, \mn@doi [\araa]
  {10.1146/annurev.astro.41.011802.094859}, \href
  {https://ui.adsabs.harvard.edu/abs/2004ARA&A..42..211E} {42, 211}

\bibitem[\protect\citeauthoryear{{Englmaier} \& {Gerhard}}{{Englmaier} \&
  {Gerhard}}{1999}]{englmaier99}
{Englmaier} P.,  {Gerhard} O.,  1999, \mn@doi [\mnras]
  {10.1046/j.1365-8711.1999.02280.x}, \href
  {https://ui.adsabs.harvard.edu/abs/1999MNRAS.304..512E} {304, 512}

\bibitem[\protect\citeauthoryear{{Farnsworth}, {Rudnick}  \&
  {Brown}}{{Farnsworth} et~al.}{2011}]{farnsworth11}
{Farnsworth} D.,  {Rudnick} L.,   {Brown} S.,  2011, \mn@doi [\aj]
  {10.1088/0004-6256/141/6/191}, \href
  {http://adsabs.harvard.edu/abs/2011AJ....141..191F} {141, 191}

\bibitem[\protect\citeauthoryear{{Federrath} \& {Klessen}}{{Federrath} \&
  {Klessen}}{2012}]{federrath12}
{Federrath} C.,  {Klessen} R.~S.,  2012, \mn@doi [\apj]
  {10.1088/0004-637X/761/2/156}, \href
  {https://ui.adsabs.harvard.edu/abs/2012ApJ...761..156F} {761, 156}

\bibitem[\protect\citeauthoryear{{Ferri{\`e}re}}{{Ferri{\`e}re}}{2020}]{ferriere20}
{Ferri{\`e}re} K.,  2020, \mn@doi [Plasma Physics and Controlled Fusion]
  {10.1088/1361-6587/ab49eb}, \href
  {https://ui.adsabs.harvard.edu/abs/2020PPCF...62a4014F} {62, 014014}

\bibitem[\protect\citeauthoryear{{Ferriere}, {Mac Low}  \&
  {Zweibel}}{{Ferriere} et~al.}{1991}]{ferriere91}
{Ferriere} K.~M.,  {Mac Low} M.-M.,   {Zweibel} E.~G.,  1991, \mn@doi [\apj]
  {10.1086/170185}, \href
  {https://ui.adsabs.harvard.edu/abs/1991ApJ...375..239F} {375, 239}

\bibitem[\protect\citeauthoryear{{Ferri{\`e}re}, {West}  \&
  {Jaffe}}{{Ferri{\`e}re} et~al.}{2021}]{ferriere21}
{Ferri{\`e}re} K.,  {West} J.~L.,   {Jaffe} T.~R.,  2021, \mn@doi [\mnras]
  {10.1093/mnras/stab1641}, \href
  {https://ui.adsabs.harvard.edu/abs/2021MNRAS.507.4968F} {507, 4968}

\bibitem[\protect\citeauthoryear{{Fletcher}, {Beck}, {Shukurov}, {Berkhuijsen}
  \& {Horellou}}{{Fletcher} et~al.}{2011}]{fletcher11}
{Fletcher} A.,  {Beck} R.,  {Shukurov} A.,  {Berkhuijsen} E.~M.,   {Horellou}
  C.,  2011, \mn@doi [\mnras] {10.1111/j.1365-2966.2010.18065.x}, \href
  {https://ui.adsabs.harvard.edu/abs/2011MNRAS.412.2396F} {412, 2396}

\bibitem[\protect\citeauthoryear{{GRAVITY Collaboration} et~al.,}{{GRAVITY
  Collaboration} et~al.}{2019}]{gravity19}
{GRAVITY Collaboration} et~al., 2019, \mn@doi [\aap]
  {10.1051/0004-6361/201935656}, \href
  {https://ui.adsabs.harvard.edu/abs/2019A&A...625L..10G} {625, L10}

\bibitem[\protect\citeauthoryear{{Gaensler}, {Dickey}, {McClure-Griffiths},
  {Green}, {Wieringa}  \& {Haynes}}{{Gaensler} et~al.}{2001}]{gaensler01}
{Gaensler} B.~M.,  {Dickey} J.~M.,  {McClure-Griffiths} N.~M.,  {Green} A.~J.,
  {Wieringa} M.~H.,   {Haynes} R.~F.,  2001, \mn@doi [\apj] {10.1086/319468},
  \href {http://adsabs.harvard.edu/abs/2001ApJ...549..959G} {549, 959}

\bibitem[\protect\citeauthoryear{{Gaensler}, {Landecker}, {Taylor}  \& {POSSUM
  Collaboration}}{{Gaensler} et~al.}{2010}]{gaensler10}
{Gaensler} B.~M.,  {Landecker} T.~L.,  {Taylor} A.~R.,   {POSSUM Collaboration}
  2010, BAAS, \href {http://adsabs.harvard.edu/abs/2010AAS...21547013G} {42,
  515}

\bibitem[\protect\citeauthoryear{{Gaensler} et~al.,}{{Gaensler}
  et~al.}{2025}]{gaensler25}
{Gaensler} B.~M.,  et~al., 2025, \mn@doi [\pasa] {10.48550/arXiv.2505.08272},
  \href {https://ui.adsabs.harvard.edu/abs/2025arXiv250508272G} {in press,
  arXiv:2505.08272}

\bibitem[\protect\citeauthoryear{{Geach}, {Lopez-Rodriguez}, {Doherty}, {Chen},
  {Ivison}, {Bendo}, {Dye}  \& {Coppin}}{{Geach} et~al.}{2023}]{geach23}
{Geach} J.~E.,  {Lopez-Rodriguez} E.,  {Doherty} M.~J.,  {Chen} J.,  {Ivison}
  R.~J.,  {Bendo} G.~J.,  {Dye} S.,   {Coppin} K.~E.~K.,  2023, \mn@doi [\nat]
  {10.1038/s41586-023-06346-4}, \href
  {https://ui.adsabs.harvard.edu/abs/2023Natur.621..483G} {621, 483}

\bibitem[\protect\citeauthoryear{{Goldreich} \& {Sridhar}}{{Goldreich} \&
  {Sridhar}}{1995}]{goldreich95}
{Goldreich} P.,  {Sridhar} S.,  1995, \mn@doi [\apj] {10.1086/175121}, \href
  {https://ui.adsabs.harvard.edu/abs/1995ApJ...438..763G} {438, 763}

\bibitem[\protect\citeauthoryear{{G{\'o}mez} \& {Cox}}{{G{\'o}mez} \&
  {Cox}}{2002}]{gomez02}
{G{\'o}mez} G.~C.,  {Cox} D.~P.,  2002, \mn@doi [\apj] {10.1086/343129}, \href
  {https://ui.adsabs.harvard.edu/abs/2002ApJ...580..235G} {580, 235}

\bibitem[\protect\citeauthoryear{{Gordon} et~al.,}{{Gordon}
  et~al.}{2021}]{gordon21}
{Gordon} Y.~A.,  et~al., 2021, \mn@doi [\apjs] {10.3847/1538-4365/ac05c0},
  \href {https://ui.adsabs.harvard.edu/abs/2021ApJS..255...30G} {255, 30}

\bibitem[\protect\citeauthoryear{{Green}}{{Green}}{2011}]{green11}
{Green} D.~A.,  2011, \mn@doi [Bulletin of the Astronomical Society of India]
  {10.48550/arXiv.1108.5083}, \href
  {https://ui.adsabs.harvard.edu/abs/2011BASI...39..289G} {39, 289}

\bibitem[\protect\citeauthoryear{{Green}}{{Green}}{2019}]{green19}
{Green} D.~A.,  2019, \mn@doi [Journal of Astrophysics and Astronomy]
  {10.1007/s12036-019-9601-6}, \href
  {https://ui.adsabs.harvard.edu/abs/2019JApA...40...36G} {40, 36}

\bibitem[\protect\citeauthoryear{{Haffner}, {Reynolds}, {Tufte}, {Madsen},
  {Jaehnig}  \& {Percival}}{{Haffner} et~al.}{2003}]{haffner03}
{Haffner} L.~M.,  {Reynolds} R.~J.,  {Tufte} S.~L.,  {Madsen} G.~J.,  {Jaehnig}
  K.~P.,   {Percival} J.~W.,  2003, \mn@doi [\apjs] {10.1086/378850}, \href
  {http://adsabs.harvard.edu/abs/2003ApJS..149..405H} {149, 405}

\bibitem[\protect\citeauthoryear{{Haffner}, {Reynolds}, {Madsen}, {Hill},
  {Barger}, {Jaehnig}, {Mierkiewicz}  \& {Percival}}{{Haffner}
  et~al.}{2010}]{haffner10}
{Haffner} L.~M.,  {Reynolds} R.~J.,  {Madsen} G.~J.,  {Hill} A.~S.,  {Barger}
  K.~A.,  {Jaehnig} K.~P.,  {Mierkiewicz} E.~J.,   {Percival} J.~W.,  2010,
  BAAS, \href {http://adsabs.harvard.edu/abs/2010AAS...21541528H} {42, 265}

\bibitem[\protect\citeauthoryear{{Hale} et~al.,}{{Hale} et~al.}{2021}]{hale21}
{Hale} C.~L.,  et~al., 2021, \mn@doi [\pasa] {10.1017/pasa.2021.47}, \href
  {https://ui.adsabs.harvard.edu/abs/2021PASA...38...58H} {38, e058}

\bibitem[\protect\citeauthoryear{{Han}, {Beck}, {Ehle}, {Haynes}  \&
  {Wielebinski}}{{Han} et~al.}{1999}]{han99}
{Han} J.~L.,  {Beck} R.,  {Ehle} M.,  {Haynes} R.~F.,   {Wielebinski} R.,
  1999, \aap, \href {https://ui.adsabs.harvard.edu/abs/1999A&A...348..405H}
  {348, 405}

\bibitem[\protect\citeauthoryear{Harris et~al.,}{Harris et~al.}{2020}]{numpy}
Harris C.~R.,  et~al., 2020, \mn@doi [Nature] {10.1038/s41586-020-2649-2}, 585,
  357

\bibitem[\protect\citeauthoryear{{Hassani}, {Tabatabaei}, {Hughes},
  {Chastenet}, {McLeod}, {Schinnerer}  \& {Nasiri}}{{Hassani}
  et~al.}{2022}]{hassani22}
{Hassani} H.,  {Tabatabaei} F.,  {Hughes} A.,  {Chastenet} J.,  {McLeod} A.~F.,
   {Schinnerer} E.,   {Nasiri} S.,  2022, \mn@doi [\mnras]
  {10.1093/mnras/stab3202}, \href
  {https://ui.adsabs.harvard.edu/abs/2022MNRAS.510...11H} {510, 11}

\bibitem[\protect\citeauthoryear{{Haverkorn}}{{Haverkorn}}{2015}]{haverkorn15}
{Haverkorn} M.,  2015, in {Lazarian} A.,  {de Gouveia Dal Pino} E.~M.,
  {Melioli} C.,  eds,  Astrophysics and Space Science Library Vol. 407,
  Magnetic Fields in Diffuse Media. Springer-Verlag, Berlin, p.~483

\bibitem[\protect\citeauthoryear{{Haverkorn} \& {Spangler}}{{Haverkorn} \&
  {Spangler}}{2013}]{haverkorn13}
{Haverkorn} M.,  {Spangler} S.~R.,  2013, \mn@doi [\ssr]
  {10.1007/s11214-013-0014-6}, \href
  {https://ui.adsabs.harvard.edu/abs/2013SSRv..178..483H} {178, 483}

\bibitem[\protect\citeauthoryear{{Haverkorn}, {Brown}, {Gaensler}  \&
  {McClure-Griffiths}}{{Haverkorn} et~al.}{2008}]{haverkorn08}
{Haverkorn} M.,  {Brown} J.~C.,  {Gaensler} B.~M.,   {McClure-Griffiths} N.~M.,
   2008, \mn@doi [\apj] {10.1086/587165}, \href
  {http://adsabs.harvard.edu/abs/2008ApJ...680..362H} {680, 362}

\bibitem[\protect\citeauthoryear{{Hotan} et~al.,}{{Hotan}
  et~al.}{2021}]{hotan21}
{Hotan} A.~W.,  et~al., 2021, \mn@doi [\pasa] {10.1017/pasa.2021.1}, \href
  {https://ui.adsabs.harvard.edu/abs/2021PASA...38....9H} {38, e009}

\bibitem[\protect\citeauthoryear{{Hou} \& {Han}}{{Hou} \& {Han}}{2015}]{hou15}
{Hou} L.~G.,  {Han} J.~L.,  2015, \mn@doi [\mnras] {10.1093/mnras/stv1904},
  \href {https://ui.adsabs.harvard.edu/abs/2015MNRAS.454..626H} {454, 626}

\bibitem[\protect\citeauthoryear{{Houde}, {Fletcher}, {Beck}, {Hildebrand},
  {Vaillancourt}  \& {Stil}}{{Houde} et~al.}{2013}]{houde13}
{Houde} M.,  {Fletcher} A.,  {Beck} R.,  {Hildebrand} R.~H.,  {Vaillancourt}
  J.~E.,   {Stil} J.~M.,  2013, \mn@doi [\apj] {10.1088/0004-637X/766/1/49},
  \href {https://ui.adsabs.harvard.edu/abs/2013ApJ...766...49H} {766, 49}

\bibitem[\protect\citeauthoryear{{Hovatta}, {O'Sullivan}, {Mart{\'\i}-Vidal},
  {Savolainen}  \& {Tchekhovskoy}}{{Hovatta} et~al.}{2019}]{hovatta19}
{Hovatta} T.,  {O'Sullivan} S.,  {Mart{\'\i}-Vidal} I.,  {Savolainen} T.,
  {Tchekhovskoy} A.,  2019, \mn@doi [\aap] {10.1051/0004-6361/201832587}, \href
  {https://ui.adsabs.harvard.edu/abs/2019A&A...623A.111H} {623, A111}

\bibitem[\protect\citeauthoryear{Hunter}{Hunter}{2007}]{matplotlib}
Hunter J.~D.,  2007, \mn@doi [Computing in Science \& Engineering]
  {10.1109/MCSE.2007.55}, 9, 90

\bibitem[\protect\citeauthoryear{{Hutschenreuter} et~al.,}{{Hutschenreuter}
  et~al.}{2022}]{hutschenreuter22}
{Hutschenreuter} S.,  et~al., 2022, \mn@doi [\aap]
  {10.1051/0004-6361/202140486}, \href
  {https://ui.adsabs.harvard.edu/abs/2022A&A...657A..43H} {657, A43}

\bibitem[\protect\citeauthoryear{{Jaffe}, {Leahy}, {Banday}, {Leach}, {Lowe}
  \& {Wilkinson}}{{Jaffe} et~al.}{2010}]{jaffe10}
{Jaffe} T.~R.,  {Leahy} J.~P.,  {Banday} A.~J.,  {Leach} S.~M.,  {Lowe} S.~R.,
   {Wilkinson} A.,  2010, \mn@doi [\mnras] {10.1111/j.1365-2966.2009.15745.x},
  \href {https://ui.adsabs.harvard.edu/abs/2010MNRAS.401.1013J} {401, 1013}

\bibitem[\protect\citeauthoryear{{Jagannathan}, {Bhatnagar}, {Rau}  \&
  {Taylor}}{{Jagannathan} et~al.}{2017}]{jagannathan17}
{Jagannathan} P.,  {Bhatnagar} S.,  {Rau} U.,   {Taylor} A.~R.,  2017, \mn@doi
  [\aj] {10.3847/1538-3881/aa77f8}, \href
  {http://adsabs.harvard.edu/abs/2017AJ....154...56J} {154, 56}

\bibitem[\protect\citeauthoryear{{Jansson} \& {Farrar}}{{Jansson} \&
  {Farrar}}{2012a}]{jansson12}
{Jansson} R.,  {Farrar} G.~R.,  2012a, \mn@doi [\apj]
  {10.1088/0004-637X/757/1/14}, \href
  {http://adsabs.harvard.edu/abs/2012ApJ...757...14J} {757, 14}

\bibitem[\protect\citeauthoryear{{Jansson} \& {Farrar}}{{Jansson} \&
  {Farrar}}{2012b}]{jansson12b}
{Jansson} R.,  {Farrar} G.~R.,  2012b, \mn@doi [\apjl]
  {10.1088/2041-8205/761/1/L11}, \href
  {https://ui.adsabs.harvard.edu/abs/2012ApJ...761L..11J} {761, L11}

\bibitem[\protect\citeauthoryear{{Khademi}, {Nasiri}  \&
  {Tabatabaei}}{{Khademi} et~al.}{2023}]{khademi23}
{Khademi} M.,  {Nasiri} S.,   {Tabatabaei} F.~S.,  2023, \mn@doi [\apj]
  {10.3847/1538-4357/acb99b}, \href
  {https://ui.adsabs.harvard.edu/abs/2023ApJ...945...36K} {945, 36}

\bibitem[\protect\citeauthoryear{{Kierdorf} et~al.,}{{Kierdorf}
  et~al.}{2020}]{kierdorf20}
{Kierdorf} M.,  et~al., 2020, \mn@doi [\aap] {10.1051/0004-6361/202037847},
  \href {https://ui.adsabs.harvard.edu/abs/2020A&A...642A.118K} {642, A118}

\bibitem[\protect\citeauthoryear{{Krumholz} \& {Federrath}}{{Krumholz} \&
  {Federrath}}{2019}]{krumholz19}
{Krumholz} M.~R.,  {Federrath} C.,  2019, \mn@doi [Frontiers in Astronomy and
  Space Sciences] {10.3389/fspas.2019.00007}, \href
  {https://ui.adsabs.harvard.edu/abs/2019FrASS...6....7K} {6, 7}

\bibitem[\protect\citeauthoryear{{Lacy} et~al.,}{{Lacy} et~al.}{2020}]{lacy20}
{Lacy} M.,  et~al., 2020, \mn@doi [\pasp] {10.1088/1538-3873/ab63eb}, \href
  {https://ui.adsabs.harvard.edu/abs/2020PASP..132c5001L} {132, 035001}

\bibitem[\protect\citeauthoryear{{Laing}}{{Laing}}{1980}]{laing80}
{Laing} R.~A.,  1980, \mn@doi [\mnras] {10.1093/mnras/193.3.439}, \href
  {https://ui.adsabs.harvard.edu/abs/1980MNRAS.193..439L} {193, 439}

\bibitem[\protect\citeauthoryear{{Liu}, {Cui}, {Liu}, {Huang}, {Zhao}  \&
  {Zhang}}{{Liu} et~al.}{2019}]{liu19}
{Liu} Z.,  {Cui} W.,  {Liu} C.,  {Huang} Y.,  {Zhao} G.,   {Zhang} B.,  2019,
  \mn@doi [\apjs] {10.3847/1538-4365/ab0a0d}, \href
  {https://ui.adsabs.harvard.edu/abs/2019ApJS..241...32L} {241, 32}

\bibitem[\protect\citeauthoryear{{Liu} et~al.,}{{Liu} et~al.}{2021}]{liu21}
{Liu} M.,  et~al., 2021, \mn@doi [\aap] {10.1051/0004-6361/202039615}, \href
  {https://ui.adsabs.harvard.edu/abs/2021A&A...650A..14L} {650, A14}

\bibitem[\protect\citeauthoryear{{Livingston}, {McClure-Griffiths}, {Gaensler},
  {Seta}  \& {Alger}}{{Livingston} et~al.}{2021}]{livingston21}
{Livingston} J.~D.,  {McClure-Griffiths} N.~M.,  {Gaensler} B.~M.,  {Seta} A.,
   {Alger} M.~J.,  2021, \mn@doi [\mnras] {10.1093/mnras/stab253}, \href
  {https://ui.adsabs.harvard.edu/abs/2021MNRAS.502.3814L} {502, 3814}

\bibitem[\protect\citeauthoryear{{Livingston}, {McClure-Griffiths}, {Mao},
  {Ma}, {Gaensler}, {Heald}  \& {Seta}}{{Livingston}
  et~al.}{2022}]{livingston22}
{Livingston} J.~D.,  {McClure-Griffiths} N.~M.,  {Mao} S.~A.,  {Ma} Y.~K.,
  {Gaensler} B.~M.,  {Heald} G.,   {Seta} A.,  2022, \mn@doi [\mnras]
  {10.1093/mnras/stab3375}, \href
  {https://ui.adsabs.harvard.edu/abs/2022MNRAS.510..260L} {510, 260}

\bibitem[\protect\citeauthoryear{{Ma}, {Mao}, {Stil}, {Basu}, {West}, {Heiles},
  {Hill}  \& {Betti}}{{Ma} et~al.}{2019a}]{ma19a}
{Ma} Y.~K.,  {Mao} S.~A.,  {Stil} J.,  {Basu} A.,  {West} J.,  {Heiles} C.,
  {Hill} A.~S.,   {Betti} S.~K.,  2019a, \mn@doi [\mnras]
  {10.1093/mnras/stz1325}, \href
  {https://ui.adsabs.harvard.edu/abs/2019MNRAS.487.3432M} {487, 3432}

\bibitem[\protect\citeauthoryear{{Ma}, {Mao}, {Stil}, {Basu}, {West}, {Heiles},
  {Hill}  \& {Betti}}{{Ma} et~al.}{2019b}]{ma19b}
{Ma} Y.~K.,  {Mao} S.~A.,  {Stil} J.,  {Basu} A.,  {West} J.,  {Heiles} C.,
  {Hill} A.~S.,   {Betti} S.~K.,  2019b, \mn@doi [\mnras]
  {10.1093/mnras/stz1328}, \href
  {https://ui.adsabs.harvard.edu/abs/2019MNRAS.487.3454M} {487, 3454}

\bibitem[\protect\citeauthoryear{{Ma}, {Mao}, {Ordog}  \& {Brown}}{{Ma}
  et~al.}{2020}]{ma20}
{Ma} Y.~K.,  {Mao} S.~A.,  {Ordog} A.,   {Brown} J.~C.,  2020, \mn@doi [\mnras]
  {10.1093/mnras/staa2105}, \href
  {https://ui.adsabs.harvard.edu/abs/2020MNRAS.497.3097M} {497, 3097}

\bibitem[\protect\citeauthoryear{{Mac Low} \& {Klessen}}{{Mac Low} \&
  {Klessen}}{2004}]{maclow04}
{Mac Low} M.-M.,  {Klessen} R.~S.,  2004, \mn@doi [Reviews of Modern Physics]
  {10.1103/RevModPhys.76.125}, \href
  {http://adsabs.harvard.edu/abs/2004RvMP...76..125M} {76, 125}

\bibitem[\protect\citeauthoryear{{Mackey}, {Green}  \& {Moutzouri}}{{Mackey}
  et~al.}{2020}]{mackey20}
{Mackey} J.,  {Green} S.,   {Moutzouri} M.,  2020, in Journal of Physics
  Conference Series. IOP, p. 012012, \mn@doi{10.1088/1742-6596/1620/1/012012}

\bibitem[\protect\citeauthoryear{{Mao} et~al.,}{{Mao} et~al.}{2014}]{mao14}
{Mao} S.~A.,  et~al., 2014, \mn@doi [arXiv e-prints]
  {10.48550/arXiv.1401.1875}, \href
  {https://ui.adsabs.harvard.edu/abs/2014arXiv1401.1875M} {p. arXiv:1401.1875}

\bibitem[\protect\citeauthoryear{{Mao}, {Zweibel}, {Fletcher}, {Ott}  \&
  {Tabatabaei}}{{Mao} et~al.}{2015}]{mao15}
{Mao} S.~A.,  {Zweibel} E.,  {Fletcher} A.,  {Ott} J.,   {Tabatabaei} F.,
  2015, \mn@doi [\apj] {10.1088/0004-637X/800/2/92}, \href
  {https://ui.adsabs.harvard.edu/abs/2015ApJ...800...92M} {800, 92}

\bibitem[\protect\citeauthoryear{{Mao} et~al.,}{{Mao} et~al.}{2017}]{mao17}
{Mao} S.~A.,  et~al., 2017, \mn@doi [Nature Astronomy]
  {10.1038/s41550-017-0218-x}, \href
  {http://adsabs.harvard.edu/abs/2017NatAs...1..621M} {1, 621}

\bibitem[\protect\citeauthoryear{{McConnell} et~al.,}{{McConnell}
  et~al.}{2020}]{mcconnell20}
{McConnell} D.,  et~al., 2020, \mn@doi [\pasa] {10.1017/pasa.2020.41}, \href
  {https://ui.adsabs.harvard.edu/abs/2020PASA...37...48M} {37, e048}

\bibitem[\protect\citeauthoryear{{McMullin}, {Waters}, {Schiebel}, {Young}  \&
  {Golap}}{{McMullin} et~al.}{2007}]{mcmullin07}
{McMullin} J.~P.,  {Waters} B.,  {Schiebel} D.,  {Young} W.,   {Golap} K.,
  2007, in {Shaw} R.~A.,  {Hill} F.,   {Bell} D.~J.,  eds,  ASP Conf. Ser. Vol.
  376, Astronomical Data Analysis Software and Systems XVI. ASP, San Francisco,
  CA, p.~127

\bibitem[\protect\citeauthoryear{{Mulcahy}, {Beck}  \& {Heald}}{{Mulcahy}
  et~al.}{2017}]{mulcahy17}
{Mulcahy} D.~D.,  {Beck} R.,   {Heald} G.~H.,  2017, \mn@doi [\aap]
  {10.1051/0004-6361/201629907}, \href
  {https://ui.adsabs.harvard.edu/abs/2017A&A...600A...6M} {600, A6}

\bibitem[\protect\citeauthoryear{{Nakanishi} \& {Sofue}}{{Nakanishi} \&
  {Sofue}}{2016}]{nakanishi16}
{Nakanishi} H.,  {Sofue} Y.,  2016, \mn@doi [\pasj] {10.1093/pasj/psv108},
  \href {https://ui.adsabs.harvard.edu/abs/2016PASJ...68....5N} {68, 5}

\bibitem[\protect\citeauthoryear{{Norman} \& {Ferrara}}{{Norman} \&
  {Ferrara}}{1996}]{norman96}
{Norman} C.~A.,  {Ferrara} A.,  1996, \mn@doi [\apj] {10.1086/177603}, \href
  {http://adsabs.harvard.edu/abs/1996ApJ...467..280N} {467, 280}

\bibitem[\protect\citeauthoryear{{Ntormousi}, {Tassis}, {Del Sordo},
  {Fragkoudi}  \& {Pakmor}}{{Ntormousi} et~al.}{2020}]{ntormousi20}
{Ntormousi} E.,  {Tassis} K.,  {Del Sordo} F.,  {Fragkoudi} F.,   {Pakmor} R.,
  2020, \mn@doi [\aap] {10.1051/0004-6361/202037835}, \href
  {https://ui.adsabs.harvard.edu/abs/2020A&A...641A.165N} {641, A165}

\bibitem[\protect\citeauthoryear{{O'Sullivan} et~al.,}{{O'Sullivan}
  et~al.}{2012}]{osullivan12}
{O'Sullivan} S.~P.,  et~al., 2012, \mn@doi [\mnras]
  {10.1111/j.1365-2966.2012.20554.x}, \href
  {http://adsabs.harvard.edu/abs/2012MNRAS.421.3300O} {421, 3300}

\bibitem[\protect\citeauthoryear{{O'Sullivan}, {Purcell}, {Anderson}, {Farnes},
  {Sun}  \& {Gaensler}}{{O'Sullivan} et~al.}{2017}]{osullivan17}
{O'Sullivan} S.~P.,  {Purcell} C.~R.,  {Anderson} C.~S.,  {Farnes} J.~S.,
  {Sun} X.~H.,   {Gaensler} B.~M.,  2017, \mn@doi [\mnras]
  {10.1093/mnras/stx1133}, \href
  {https://ui.adsabs.harvard.edu/abs/2017MNRAS.469.4034O} {469, 4034}

\bibitem[\protect\citeauthoryear{{Oberhelman}, {Van Eck}, {McClure-Griffiths}
  \& {Vanderwoude}}{{Oberhelman} et~al.}{2024}]{oberhelman24}
{Oberhelman} L.,  {Van Eck} C.,  {McClure-Griffiths} N.,   {Vanderwoude} S.,
  2024, Technical Report~71, Diffuse Emission Subtraction for POSSUM Survey.
POSSUM

\bibitem[\protect\citeauthoryear{{Offner} \& {Arce}}{{Offner} \&
  {Arce}}{2015}]{offner15}
{Offner} S. S.~R.,  {Arce} H.~G.,  2015, \mn@doi [\apj]
  {10.1088/0004-637X/811/2/146}, \href
  {https://ui.adsabs.harvard.edu/abs/2015ApJ...811..146O} {811, 146}

\bibitem[\protect\citeauthoryear{{Ordog}, {Brown}, {Kothes}  \&
  {Landecker}}{{Ordog} et~al.}{2017}]{ordog17}
{Ordog} A.,  {Brown} J.~C.,  {Kothes} R.,   {Landecker} T.~L.,  2017, \mn@doi
  [\aap] {10.1051/0004-6361/201730740}, \href
  {http://adsabs.harvard.edu/abs/2017A\%26A...603A..15O} {603, A15}

\bibitem[\protect\citeauthoryear{{Padmanabh} et~al.,}{{Padmanabh}
  et~al.}{2023}]{padmanabh23}
{Padmanabh} P.~V.,  et~al., 2023, \mn@doi [\mnras] {10.1093/mnras/stad1900},
  \href {https://ui.adsabs.harvard.edu/abs/2023MNRAS.524.1291P} {524, 1291}

\bibitem[\protect\citeauthoryear{{Pakmor}, {Marinacci}  \& {Springel}}{{Pakmor}
  et~al.}{2014}]{pakmor14}
{Pakmor} R.,  {Marinacci} F.,   {Springel} V.,  2014, \mn@doi [\apjl]
  {10.1088/2041-8205/783/1/L20}, \href
  {https://ui.adsabs.harvard.edu/abs/2014ApJ...783L..20P} {783, L20}

\bibitem[\protect\citeauthoryear{{Pasetto}, {Carrasco-Gonz{\'a}lez},
  {O'Sullivan}, {Basu}, {Bruni}, {Kraus}, {Curiel}  \& {Mack}}{{Pasetto}
  et~al.}{2018}]{pasetto18}
{Pasetto} A.,  {Carrasco-Gonz{\'a}lez} C.,  {O'Sullivan} S.,  {Basu} A.,
  {Bruni} G.,  {Kraus} A.,  {Curiel} S.,   {Mack} K.-H.,  2018, \mn@doi [\aap]
  {10.1051/0004-6361/201731804}, \href
  {https://ui.adsabs.harvard.edu/abs/2018A&A...613A..74P} {613, A74}

\bibitem[\protect\citeauthoryear{{Purcell}, {Van Eck}, {West}, {Sun}  \&
  {Gaensler}}{{Purcell} et~al.}{2020}]{purcell20}
{Purcell} C.~R.,  {Van Eck} C.~L.,  {West} J.,  {Sun} X.~H.,   {Gaensler}
  B.~M.,  2020, {RM-Tools: Rotation measure (RM) synthesis and Stokes
  QU-fitting}, Astrophysics Source Code Library, record ascl:2005.003
  (\mn@eprint {ascl} {2005.003})

\bibitem[\protect\citeauthoryear{{Ranchod}, {Mao}, {Deane}, {Sridhar},
  {Damas-Segovia}, {Livingston}  \& {Ma}}{{Ranchod} et~al.}{2024}]{ranchod24}
{Ranchod} S.,  {Mao} S.~A.,  {Deane} R.,  {Sridhar} S.~S.,  {Damas-Segovia} A.,
   {Livingston} J.~D.,   {Ma} Y.~K.,  2024, \mn@doi [\aap]
  {10.1051/0004-6361/202348993}, \href
  {https://ui.adsabs.harvard.edu/abs/2024A&A...686A.104R} {686, A104}

\bibitem[\protect\citeauthoryear{{Reed}}{{Reed}}{2000}]{reed00}
{Reed} B.~C.,  2000, \mn@doi [\aj] {10.1086/301421}, \href
  {https://ui.adsabs.harvard.edu/abs/2000AJ....120..314R} {120, 314}

\bibitem[\protect\citeauthoryear{{Reed}}{{Reed}}{2003}]{reed03}
{Reed} B.~C.,  2003, \mn@doi [\aj] {10.1086/374771}, \href
  {https://ui.adsabs.harvard.edu/abs/2003AJ....125.2531R} {125, 2531}

\bibitem[\protect\citeauthoryear{{Ruzmaikin}, {Sokolov}  \&
  {Shukurov}}{{Ruzmaikin} et~al.}{1988}]{ruzmaikin88}
{Ruzmaikin} A.,  {Sokolov} D.,   {Shukurov} A.,  1988, {Magnetic Fields of
  Galaxies}.
(Springer-Verlag Berlin Heidelberg)

\bibitem[\protect\citeauthoryear{{Schnitzeler}}{{Schnitzeler}}{2018}]{schnitzeler18}
{Schnitzeler} D.~H.~F.~M.,  2018, \mn@doi [\mnras] {10.1093/mnras/stx2754},
  \href {https://ui.adsabs.harvard.edu/abs/2018MNRAS.474..300S} {474, 300}

\bibitem[\protect\citeauthoryear{{Seta} \& {Federrath}}{{Seta} \&
  {Federrath}}{2022}]{seta22}
{Seta} A.,  {Federrath} C.,  2022, \mn@doi [\mnras] {10.1093/mnras/stac1400},
  \href {https://ui.adsabs.harvard.edu/abs/2022MNRAS.514..957S} {514, 957}

\bibitem[\protect\citeauthoryear{{Seta} \& {Federrath}}{{Seta} \&
  {Federrath}}{2024}]{seta24}
{Seta} A.,  {Federrath} C.,  2024, \mn@doi [\mnras] {10.1093/mnras/stae1935},
  \href {https://ui.adsabs.harvard.edu/abs/2024MNRAS.tmp.1898S} {}

\bibitem[\protect\citeauthoryear{{Seta}, {Shukurov}, {Wood}, {Bushby}  \&
  {Snodin}}{{Seta} et~al.}{2018}]{seta18}
{Seta} A.,  {Shukurov} A.,  {Wood} T.~S.,  {Bushby} P.~J.,   {Snodin} A.~P.,
  2018, \mn@doi [\mnras] {10.1093/mnras/stx2606}, \href
  {https://ui.adsabs.harvard.edu/abs/2018MNRAS.473.4544S} {473, 4544}

\bibitem[\protect\citeauthoryear{{Seta}, {Federrath}, {Livingston}  \&
  {McClure-Griffiths}}{{Seta} et~al.}{2023}]{seta23}
{Seta} A.,  {Federrath} C.,  {Livingston} J.~D.,   {McClure-Griffiths} N.~M.,
  2023, \mn@doi [\mnras] {10.1093/mnras/stac2972}, \href
  {https://ui.adsabs.harvard.edu/abs/2023MNRAS.518..919S} {518, 919}

\bibitem[\protect\citeauthoryear{{Shanahan} et~al.,}{{Shanahan}
  et~al.}{2019}]{shanahan19}
{Shanahan} R.,  et~al., 2019, \mn@doi [\apjl] {10.3847/2041-8213/ab58d4}, \href
  {https://ui.adsabs.harvard.edu/abs/2019ApJ...887L...7S} {887, L7}

\bibitem[\protect\citeauthoryear{{Simonetti}, {Cordes}  \&
  {Spangler}}{{Simonetti} et~al.}{1984}]{simonetti84}
{Simonetti} J.~H.,  {Cordes} J.~M.,   {Spangler} S.~R.,  1984, \mn@doi [\apj]
  {10.1086/162391}, \href
  {https://ui.adsabs.harvard.edu/abs/1984ApJ...284..126S} {284, 126}

\bibitem[\protect\citeauthoryear{{Skilling}}{{Skilling}}{2004}]{skilling04}
{Skilling} J.,  2004, in {Fischer} R.,  {Preuss} R.,   {Toussaint} U.~V.,  eds,
   American Institute of Physics Conference Series Vol. 735, Bayesian Inference
  and Maximum Entropy Methods in Science and Engineering: 24th International
  Workshop on Bayesian Inference and Maximum Entropy Methods in Science and
  Engineering. pp 395--405, \mn@doi{10.1063/1.1835238}

\bibitem[\protect\citeauthoryear{{Sokoloff}, {Bykov}, {Shukurov},
  {Berkhuijsen}, {Beck}  \& {Poezd}}{{Sokoloff} et~al.}{1998}]{sokoloff98}
{Sokoloff} D.~D.,  {Bykov} A.~A.,  {Shukurov} A.,  {Berkhuijsen} E.~M.,  {Beck}
  R.,   {Poezd} A.~D.,  1998, \mn@doi [\mnras]
  {10.1046/j.1365-8711.1998.01782.x}, \href
  {https://ui.adsabs.harvard.edu/abs/1998MNRAS.299..189S} {299, 189}

\bibitem[\protect\citeauthoryear{{Sparke} \& {Gallagher}}{{Sparke} \&
  {Gallagher}}{2006}]{sparke06}
{Sparke} L.~S.,  {Gallagher} III J.~S.,  2006, {Galaxies in the Universe}.
(Cambridge Univ.\ Press, Cambridge), \mn@doi{10.2277/0521855934}

\bibitem[\protect\citeauthoryear{{Stil} \& {Taylor}}{{Stil} \&
  {Taylor}}{2007}]{stil07}
{Stil} J.~M.,  {Taylor} A.~R.,  2007, \mn@doi [\apj] {10.1086/519791}, \href
  {https://ui.adsabs.harvard.edu/abs/2007ApJ...663L..21S} {663, L21}

\bibitem[\protect\citeauthoryear{{Stil}, {Taylor}  \& {Sunstrum}}{{Stil}
  et~al.}{2011}]{stil11}
{Stil} J.~M.,  {Taylor} A.~R.,   {Sunstrum} C.,  2011, \mn@doi [\apj]
  {10.1088/0004-637X/726/1/4}, \href
  {http://adsabs.harvard.edu/abs/2011ApJ...726....4S} {726, 4}

\bibitem[\protect\citeauthoryear{{Sun} et~al.,}{{Sun} et~al.}{2015}]{sun15}
{Sun} X.~H.,  et~al., 2015, \mn@doi [\aj] {10.1088/0004-6256/149/2/60}, \href
  {https://ui.adsabs.harvard.edu/abs/2015AJ....149...60S} {149, 60}

\bibitem[\protect\citeauthoryear{{Takamura} et~al.,}{{Takamura}
  et~al.}{2023}]{takamura23}
{Takamura} M.,  et~al., 2023, \mn@doi [\apj] {10.3847/1538-4357/acd9a8}, \href
  {https://ui.adsabs.harvard.edu/abs/2023ApJ...952...47T} {952, 47}

\bibitem[\protect\citeauthoryear{{Thomson} et~al.,}{{Thomson}
  et~al.}{2019}]{thomson19}
{Thomson} A. J.~M.,  et~al., 2019, \mn@doi [\mnras] {10.1093/mnras/stz1438},
  \href {https://ui.adsabs.harvard.edu/abs/2019MNRAS.487.4751T} {487, 4751}

\bibitem[\protect\citeauthoryear{{Thomson} et~al.,}{{Thomson}
  et~al.}{2023}]{thomson23}
{Thomson} A. J.~M.,  et~al., 2023, \mn@doi [\pasa] {10.1017/pasa.2023.38},
  \href {https://ui.adsabs.harvard.edu/abs/2023PASA...40...40T} {40, e040}

\bibitem[\protect\citeauthoryear{{Vall{\'e}e}}{{Vall{\'e}e}}{2022}]{vallee22}
{Vall{\'e}e} J.~P.,  2022, \mn@doi [\na] {10.1016/j.newast.2022.101896}, \href
  {https://ui.adsabs.harvard.edu/abs/2022NewA...9701896V} {97, 101896}

\bibitem[\protect\citeauthoryear{{Van Eck} et~al.,}{{Van Eck}
  et~al.}{2011}]{vaneck11}
{Van Eck} C.~L.,  et~al., 2011, \mn@doi [\apj] {10.1088/0004-637X/728/2/97},
  \href {http://adsabs.harvard.edu/abs/2011ApJ...728...97V} {728, 97}

\bibitem[\protect\citeauthoryear{{Van Eck} et~al.,}{{Van Eck}
  et~al.}{2023}]{vaneck23}
{Van Eck} C.~L.,  et~al., 2023, \mn@doi [\apjs] {10.3847/1538-4365/acda24},
  \href {https://ui.adsabs.harvard.edu/abs/2023ApJS..267...28V} {267, 28}

\bibitem[\protect\citeauthoryear{Van~Rossum \& Drake}{Van~Rossum \&
  Drake}{2009}]{python3}
Van~Rossum G.,  Drake F.~L.,  2009, Python 3 Reference Manual.
CreateSpace, Scotts Valley, CA

\bibitem[\protect\citeauthoryear{Virtanen et~al.,}{Virtanen
  et~al.}{2020}]{scipy}
Virtanen P.,  et~al., 2020, \mn@doi [Nature Methods]
  {10.1038/s41592-019-0686-2}, \href {https://rdcu.be/b08Wh} {17, 261}

\bibitem[\protect\citeauthoryear{{Wolleben} et~al.,}{{Wolleben}
  et~al.}{2021}]{wolleben21}
{Wolleben} M.,  et~al., 2021, \mn@doi [\aj] {10.3847/1538-3881/abf7c1}, \href
  {https://ui.adsabs.harvard.edu/abs/2021AJ....162...35W} {162, 35}

\bibitem[\protect\citeauthoryear{{Yao}, {Manchester}  \& {Wang}}{{Yao}
  et~al.}{2017}]{ymw16}
{Yao} J.~M.,  {Manchester} R.~N.,   {Wang} N.,  2017, \mn@doi [\apj]
  {10.3847/1538-4357/835/1/29}, \href
  {https://ui.adsabs.harvard.edu/\#abs/2017ApJ...835...29Y} {835, 29}

\bibitem[\protect\citeauthoryear{{de Gasperin}, {Intema}  \& {Frail}}{{de
  Gasperin} et~al.}{2018}]{degasperin18}
{de Gasperin} F.,  {Intema} H.~T.,   {Frail} D.~A.,  2018, \mn@doi [\mnras]
  {10.1093/mnras/stx3125}, \href
  {https://ui.adsabs.harvard.edu/abs/2018MNRAS.474.5008D} {474, 5008}

\makeatother
\end{thebibliography}

\appendix

\section{Stokes \textit{QU}-fitting results and spatial structure information} \label{sec:qufit_results}

In Table~\ref{table:qufit}, we list the results obtained from Stokes \textit{QU}-fitting (Section~\ref{sec:qufit}) alongside the subsequently derived FD spread values (Section~\ref{sec:fdspread}). In addition, we provide information on the angular extent and morphology of the EGSs obtained from the RACS \citep{mcconnell20} and VLASS \citep{lacy20} images (Section~\ref{sec:highres_images}). All information can be found in machine-readable format as \texttt{best\_model\_table.fits} in the Online Supplementary Materials.

\begin{landscape}
\begin{table}
\caption{Stokes \textit{QU}-fitting results and spatial structure information}
\label{table:qufit}
\begin{tabular}{lccccccccccccc}
\hline
Target source & $\ell$ & $b$ & Angular & Morphology & Best-fit & $\phi$ & $\sigma_\phi$ & $\Delta\phi$ & $p_0$ & ${\rm PA}_0$ & FD spread & BIC & $\Delta {\rm BIC}$ \\
(NVSS) & ($^\circ$) & ($^\circ$) & size ($^{\prime\prime}$) && model & (${\rm rad\,m}^{-2}$) & (${\rm rad\,m}^{-2}$) & (${\rm rad\,m}^{-2}$) & (per cent) & ($^\circ$) & (${\rm rad\,m}^{-2}$) && \\
\hline
J184415$-$131243 & $20.34$ & $-4.42$ & 10.8 & 2 & 1T & $-98.5 \pm 2.1$ & -- & -- & $3.44 \pm 0.13$ & $153.6 \pm 4.8$ & $0.0$ & $-921.1$ & --\\
J181343$-$090743 & $20.49$ & $+4.11$ & 21.0 & 3 & 2T & $+166.0 \pm 5.7$ & -- & -- & $5.63 \pm 0.50$ & $157.2 \pm 12.6$ & $76.8 \pm 4.3$ & $-426.0$ & $107.6$\\
&&&&&& $+12.4 \pm 6.4$ & -- & -- & $5.80 \pm 0.52$ & $130.4 \pm 12.1$ && \\
J183519$-$111559 & $21.08$ & $-1.59$ & 9.3 & c & 1T & $-62.4 \pm 0.9$ & -- & -- & $2.32 \pm 0.04$ & $78.0 \pm 2.0$ & $0.0$ & $-1140.5$ & --\\
J181851$-$090659 & $21.10$ & $+3.00$ & 54.5 & 2 & 1Ed & $+230.5 \pm 1.9$ & $17.5 \pm 0.8$ & -- & $8.60 \pm 0.53$ & $41.3 \pm 3.3$ & $17.4 \pm 0.8$ & $-935.0$ & $0.0$\\
J181931$-$091059 & $21.12$ & $+2.82$ & 137.4 & 2 & 1T & $+196.1 \pm 2.6$ & -- & -- & $3.59 \pm 0.15$ & $11.3 \pm 5.9$ & $0.0$ & $-819.5$ & --\\
J183759$-$112627 & $21.23$ & $-2.25$ & 7.8 & 2 & 1Ed & $-91.7 \pm 0.8$ & $8.9 \pm 0.7$ & -- & $15.23 \pm 0.57$ & $164.6 \pm 1.6$ & $8.9 \pm 0.7$ & $-737.0$ & $0.8$\\
J182443$-$092933 & $21.44$ & $+1.54$ & 15.0 & 2 & 1Ed & $+48.5 \pm 2.7$ & $18.8 \pm 1.0$ & -- & $14.44 \pm 1.37$ & $41.3 \pm 4.4$ & $18.8 \pm 1.0$ & $-698.3$ & $3.6$\\
J184555$-$115813 & $21.64$ & $-4.22$ & 3.4 & 1 & 1T & $-117.0 \pm 4.2$ & -- & -- & $11.73 \pm 0.85$ & $43.7 \pm 8.8$ & $0.0$ & $-384.7$ & --\\
J184606$-$115808 & $21.66$ & $-4.26$ & 6.6 & 1 & 1Id & $-178.2 \pm 2.9$ & $20.5 \pm 2.5$ & $50.9 \pm 2.7$ & $7.24 \pm 0.57$ & $85.3 \pm 4.4$ & $54.9 \pm 2.6$ & $-1458.0$ & $32.2$\\
J181419$-$073733 & $21.88$ & $+4.69$ & 3.4 & 1 & 1T & $+67.8 \pm 2.2$ & -- & -- & $13.14 \pm 0.54$ & $171.6 \pm 3.9$ & $0.0$ & $-411.3$ & --\\
J184059$-$110139 & $21.93$ & $-2.72$ & 4.5 & 1 & 1T & $-26.7 \pm 1.6$ & -- & -- & $3.67 \pm 0.11$ & $171.9 \pm 3.4$ & $0.0$ & $-1078.0$ & --\\
J182503$-$085445 & $22.00$ & $+1.74$ & 3.3 & 2 & 2Ed-c & $+163.9 \pm 0.9$ & $5.8 \pm 0.9$ & -- & $11.10 \pm 0.34$ & $92.0 \pm 1.7$ & $78.4 \pm 5.8$ & $-788.8$ & $19.4$\\
&&&&&& $+7.9 \pm 11.6$ & $5.8 \pm 0.9$ & -- & $0.95 \pm 0.19$ & $50.7 \pm 21.2$ && \\
J182535$-$083948 & $22.28$ & $+1.74$ & 4.0 & 1 & 1T & $+15.5 \pm 5.5$ & -- & -- & $9.65 \pm 0.99$ & $113.3 \pm 10.4$ & $0.0$ & $-209.6$ & --\\
J182542$-$083723 & $22.33$ & $+1.73$ & 16.4 & 2 & 2Ed-c & $+41.3 \pm 1.5$ & $9.4 \pm 1.1$ & -- & $4.10 \pm 0.26$ & $147.4 \pm 2.9$ & $133.2 \pm 5.4$ & $-1048.3$ & $11.7$\\
&&&&&& $-223.8 \pm 10.7$ & $9.4 \pm 1.1$ & -- & $0.68 \pm 0.11$ & $70.8 \pm 19.2$ && \\
J183931$-$101336 & $22.48$ & $-2.03$ & 3.3 & 1 & -- & -- & -- & -- & -- & -- & -- & -- & --\\
J183942$-$101038 & $22.54$ & $-2.05$ & 26.2 & 2 & 1Ed & $+132.1 \pm 1.7$ & $16.3 \pm 0.7$ & -- & $22.51 \pm 1.37$ & $14.4 \pm 2.7$ & $16.3 \pm 0.7$ & $-743.0$ & $4.3$\\
J184750$-$110658 & $22.61$ & $-4.25$ & 10.7 & 2 & 1S & $-67.9 \pm 1.4$ & -- & $36.8 \pm 1.6$ & $10.91 \pm 0.43$ & $101.8 \pm 2.1$ & $36.8 \pm 1.6$ & $-1071.0$ & $0.5$\\
J184906$-$111430 & $22.64$ & $-4.59$ & 3.3 & 1 & 2T & $+9.4 \pm 11.8$ & -- & -- & $4.28 \pm 0.45$ & $127.7 \pm 24.0$ & $43.0 \pm 8.8$ & $-596.9$ & $67.8$\\
&&&&&& $-76.6 \pm 13.0$ & -- & -- & $3.67 \pm 0.45$ & $164.3 \pm 28.3$ && \\
J184911$-$111241 & $22.68$ & $-4.59$ & 7.5 & 2 & 1T & $-6.4 \pm 1.6$ & -- & -- & $2.62 \pm 0.07$ & $3.9 \pm 3.0$ & $0.0$ & $-1204.9$ & --\\
J182530$-$080945 & $22.71$ & $+1.99$ & 11.3 & 2 & 2T & $-149.7 \pm 3.5$ & -- & -- & $3.41 \pm 0.19$ & $91.5 \pm 7.7$ & $58.1 \pm 4.6$ & $-790.3$ & $28.7$\\
&&&&&& $-265.9 \pm 8.4$ & -- & -- & $1.42 \pm 0.21$ & $112.6 \pm 16.9$ && \\
J184808$-$105535 & $22.82$ & $-4.23$ & 7.2 & 2 & 1T & $+15.0 \pm 3.9$ & -- & -- & $8.61 \pm 0.61$ & $109.3 \pm 8.3$ & $0.0$ & $-395.6$ & --\\
J184812$-$105133 & $22.88$ & $-4.22$ & 8.7 & 2 & 2T & $+138.7 \pm 18.7$ & -- & -- & $4.03 \pm 0.67$ & $159.2 \pm 26.9$ & $52.9 \pm 9.5$ & $-431.7$ & $18.3$\\
&&&&&& $+32.9 \pm 3.0$ & -- & -- & $16.41 \pm 0.61$ & $151.3 \pm 6.1$ && \\
J184552$-$103126 & $22.93$ & $-3.56$ & 3.2 & 1 & 1T & $+1.2 \pm 2.1$ & -- & -- & $8.49 \pm 0.31$ & $174.8 \pm 3.9$ & $0.0$ & $-682.9$ & $66.5$\\
J181949$-$065524 & $23.15$ & $+3.81$ & 21.7 & 3 & 1S & $+106.8 \pm 3.5$ & -- & $45.0 \pm 3.6$ & $4.97 \pm 0.40$ & $10.8 \pm 5.5$ & $45.0 \pm 3.6$ & $-756.3$ & $1.3$\\
J182537$-$073729 & $23.20$ & $+2.21$ & 4.3 & 1 & 1T & $-57.4 \pm 0.9$ & -- & -- & $2.18 \pm 0.03$ & $84.8 \pm 1.5$ & $0.0$ & $-1255.7$ & --\\
J182431$-$072714 & $23.23$ & $+2.53$ & 10.8 & c & 1T & $+24.4 \pm 1.0$ & -- & -- & $6.48 \pm 0.11$ & $30.8 \pm 1.9$ & $0.0$ & $-996.3$ & --\\
J182920$-$073400 & $23.68$ & $+1.42$ & 25.9 & c & 2Ed-c & $+361.3 \pm 1.9$ & $12.6 \pm 1.0$ & -- & $12.70 \pm 0.88$ & $163.5 \pm 3.8$ & $188.7 \pm 3.6$ & $-608.3$ & $7.8$\\
&&&&&& $-14.5 \pm 6.9$ & $12.6 \pm 1.0$ & -- & $4.47 \pm 0.55$ & $162.2 \pm 11.2$ && \\
J184644$-$094654 & $23.68$ & $-3.41$ & 8.8 & 2 & 1T & $+90.7 \pm 2.3$ & -- & -- & $5.00 \pm 0.15$ & $173.3 \pm 4.7$ & $0.0$ & $-835.4$ & --\\
J184547$-$093821 & $23.70$ & $-3.14$ & 5.0 & 1 & -- & -- & -- & -- & -- & -- & -- & -- & --\\
J183052$-$074402 & $23.71$ & $+1.01$ & 31.8 & 3 & 1S & $+503.8 \pm 2.6$ & -- & $30.8 \pm 3.6$ & $14.30 \pm 0.92$ & $60.2 \pm 3.2$ & $30.8 \pm 3.6$ & $-671.4$ & $0.3$\\
J182043$-$062415 & $23.71$ & $+3.86$ & 10.8 & 2 & 1S & $-18.5 \pm 3.4$ & -- & $21.8 \pm 5.9$ & $7.14 \pm 0.42$ & $147.4 \pm 2.7$ & $21.8 \pm 5.9$ & $-646.0$ & $0.0$\\
J184541$-$093643 & $23.72$ & $-3.10$ & 4.1 & 1 & 1T & $+189.0 \pm 1.4$ & -- & -- & $3.62 \pm 0.10$ & $39.6 \pm 2.9$ & $0.0$ & $-1125.9$ & --\\
J185239$-$101324 & $23.95$ & $-4.91$ & 227.2 & c & 1Ed & $+76.6 \pm 1.4$ & $10.1 \pm 1.2$ & -- & $5.85 \pm 0.41$ & $150.9 \pm 3.0$ & $10.1 \pm 1.2$ & $-1114.5$ & $1.1$\\
J183902$-$083023 & $23.95$ & $-1.14$ & 3.8 & 1 & 1S & $+518.8 \pm 0.7$ & -- & $16.3 \pm 1.3$ & $13.51 \pm 0.17$ & $173.1 \pm 0.6$ & $16.3 \pm 1.3$ & $-1140.0$ & $0.1$\\
J182104$-$060915 & $23.98$ & $+3.90$ & 5.2 & 1 & 1T & $+9.9 \pm 3.9$ & -- & -- & $4.56 \pm 0.31$ & $110.7 \pm 8.6$ & $0.0$ & $-551.3$ & --\\
J183321$-$073121 & $24.18$ & $+0.56$ & 9.0 & c & 1Ed & $+756.3 \pm 4.5$ & $27.2 \pm 1.2$ & -- & $6.89 \pm 0.69$ & $30.8 \pm 6.2$ & $27.2 \pm 1.2$ & $-1314.8$ & $0.6$\\
J185027$-$091037 & $24.64$ & $-3.96$ & 19.7 & c & 1S & $+107.5 \pm 5.1$ & -- & $57.9 \pm 2.6$ & $4.44 \pm 0.38$ & $107.4 \pm 8.0$ & $57.9 \pm 2.6$ & $-1143.8$ & $1.4$\\
\hline
\end{tabular}
\end{table}
\end{landscape}

\begin{landscape}
\begin{table}
\ContinuedFloat
\caption{\textit{continued}}
\begin{tabular}{lccccccccccccc}
\hline
Target source & $\ell$ & $b$ & Angular & Morphology & Best-fit & $\phi$ & $\sigma_\phi$ & $\Delta\phi$ & $p_0$ & ${\rm PA}_0$ & FD spread & BIC & $\Delta {\rm BIC}$ \\
(NVSS) & ($^\circ$) & ($^\circ$) & size ($^{\prime\prime}$) && model & (${\rm rad\,m}^{-2}$) & (${\rm rad\,m}^{-2}$) & (${\rm rad\,m}^{-2}$) & (per cent) & ($^\circ$) & (${\rm rad\,m}^{-2}$) && \\
\hline
J185030$-$090659 & $24.70$ & $-3.94$ & 97.6 & 3 & 1Ed & $+165.4 \pm 0.9$ & $6.8 \pm 1.2$ & -- & $6.77 \pm 0.31$ & $174.3 \pm 1.8$ & $6.8 \pm 1.2$ & $-1183.0$ & $0.2$\\
J184249$-$075604 & $24.89$ & $-1.71$ & 64.9 & 2 & 2T & $+934.0 \pm 0.8$ & -- & -- & $6.70 \pm 0.36$ & $32.4 \pm 2.1$ & $5.7 \pm 1.4$ & $-1312.1$ & $272.8$\\
&&&&&& $+922.5 \pm 2.7$ & -- & -- & $2.38 \pm 0.36$ & $112.6 \pm 5.7$ && \\
J182058$-$050223 & $24.95$ & $+4.44$ & 20.1 & c & 1Id & $+88.7 \pm 3.7$ & $10.9 \pm 1.4$ & $74.0 \pm 1.3$ & $22.12 \pm 0.78$ & $22.0 \pm 6.0$ & $74.8 \pm 1.3$ & $-778.9$ & $19.2$\\
J182351$-$052429 & $24.96$ & $+3.63$ & 36.2 & c & 1T & $+28.9 \pm 4.1$ & -- & -- & $7.11 \pm 0.54$ & $168.4 \pm 8.2$ & $0.0$ & $-486.5$ & --\\
J182111$-$050219 & $24.98$ & $+4.39$ & 20.7 & c & 1S & $+119.0 \pm 2.9$ & -- & $53.6 \pm 1.9$ & $4.99 \pm 0.30$ & $4.9 \pm 4.5$ & $53.6 \pm 1.9$ & $-1061.9$ & $1.6$\\
J184629$-$081333 & $25.05$ & $-2.65$ & 17.8 & 3 & 2T & $+476.0 \pm 0.9$ & -- & -- & $5.94 \pm 0.09$ & $109.2 \pm 1.9$ & $153.6 \pm 2.3$ & $-1136.8$ & $155.5$\\
&&&&&& $+168.8 \pm 4.6$ & -- & -- & $1.19 \pm 0.10$ & $90.7 \pm 9.7$ && \\
J184617$-$081126 & $25.05$ & $-2.59$ & 3.1 & 1 & 1S & $+205.0 \pm 4.9$ & -- & $85.8 \pm 2.3$ & $19.54 \pm 1.77$ & $42.4 \pm 8.2$ & $85.8 \pm 2.3$ & $-425.2$ & $4.0$\\
J182013$-$042541 & $25.41$ & $+4.89$ & 38.5 & 3 & 1T & $+68.1 \pm 0.6$ & -- & -- & $5.21 \pm 0.06$ & $127.7 \pm 1.3$ & $0.0$ & $-1165.2$ & --\\
J184511$-$060146 & $26.85$ & $-1.36$ & 39.5 & 2 & 1S & $+99.6 \pm 2.6$ & -- & $42.9 \pm 2.9$ & $11.96 \pm 0.93$ & $175.2 \pm 4.2$ & $42.9 \pm 2.9$ & $-747.7$ & $1.0$\\
J183253$-$042628 & $26.86$ & $+2.09$ & 17.2 & 2 & 1S & $+173.2 \pm 2.3$ & -- & $27.6 \pm 3.7$ & $4.51 \pm 0.23$ & $125.7 \pm 2.6$ & $27.6 \pm 3.8$ & $-1099.3$ & $0.2$\\
J182634$-$030927 & $27.27$ & $+4.08$ & 6.4 & c & 1S & $+205.5 \pm 1.8$ & -- & $33.9 \pm 2.5$ & $5.61 \pm 0.26$ & $139.8 \pm 2.4$ & $33.9 \pm 2.5$ & $-1233.8$ & $0.5$\\
J182644$-$030952 & $27.29$ & $+4.04$ & 3.5 & 2 & 1T & $+211.6 \pm 7.9$ & -- & -- & $5.24 \pm 0.68$ & $159.2 \pm 12.3$ & $0.0$ & $-457.6$ & --\\
J183847$-$040042 & $27.92$ & $+0.98$ & 5.0 & 1 & 1Ed & $+312.8 \pm 0.3$ & $14.2 \pm 0.1$ & -- & $7.22 \pm 0.07$ & $148.7 \pm 0.5$ & $14.2 \pm 0.1$ & $-1626.7$ & $13.6$\\
J183400$-$030340 & $28.22$ & $+2.48$ & 3.9 & 1 & 1S & $+50.3 \pm 9.0$ & -- & $89.6 \pm 4.5$ & $1.12 \pm 0.21$ & $95.4 \pm 15.3$ & $89.6 \pm 4.5$ & $-1445.6$ & --\\
J185054$-$050942 & $28.27$ & $-2.23$ & 9.8 & c & 2T & $+584.7 \pm 0.9$ & -- & -- & $4.44 \pm 0.06$ & $161.6 \pm 1.8$ & $132.9 \pm 7.3$ & $-1256.6$ & $12.2$\\
&&&&&& $+318.9 \pm 14.5$ & -- & -- & $0.35 \pm 0.07$ & $96.0 \pm 27.7$ && \\
J183414$-$030119 & $28.28$ & $+2.44$ & 4.3 & 1 & 2T & $+482.0 \pm 0.7$ & -- & -- & $4.43 \pm 0.06$ & $168.8 \pm 1.2$ & $215.4 \pm 3.6$ & $-1248.8$ & $69.9$\\
&&&&&& $+51.1 \pm 7.2$ & -- & -- & $0.52 \pm 0.06$ & $42.0 \pm 14.4$ && \\
J184415$-$041757 & $28.29$ & $-0.36$ & 109.8 & 2 & -- & -- & -- & -- & -- & -- & -- & -- & --\\
J185523$-$053804 & $28.36$ & $-3.44$ & 5.1 & 1 & 1Ed & $+169.0 \pm 1.0$ & $6.6 \pm 1.1$ & -- & $13.66 \pm 0.67$ & $133.0 \pm 1.9$ & $6.6 \pm 1.1$ & $-798.0$ & $0.1$\\
J183652$-$024606 & $28.81$ & $+1.97$ & 9.3 & 2 & 1S & $+478.6 \pm 3.4$ & -- & $99.9 \pm 0.2$ & $6.69 \pm 0.32$ & $38.4 \pm 7.8$ & $99.9 \pm 0.2$ & $-1099.2$ & $119.7$\\
J185744$-$052527 & $28.81$ & $-3.87$ & 156.7 & 3 & 2T & $+323.8 \pm 12.8$ & -- & -- & $2.65 \pm 0.30$ & $12.7 \pm 28.0$ & $44.6 \pm 6.5$ & $-739.8$ & $13.9$\\
&&&&&& $+234.6 \pm 2.3$ & -- & -- & $12.13 \pm 0.32$ & $154.4 \pm 4.9$ && \\
J183939$-$030047 & $28.91$ & $+1.24$ & 38.4 & 2 & 1Ed & $+674.7 \pm 0.8$ & $5.0 \pm 1.5$ & -- & $10.35 \pm 0.41$ & $102.6 \pm 1.6$ & $5.0 \pm 1.5$ & $-991.3$ & $0.1$\\
J183701$-$015140 & $29.63$ & $+2.36$ & 2.9 & 1 & 1T & $+294.9 \pm 6.7$ & -- & -- & $1.37 \pm 0.15$ & $106.7 \pm 12.4$ & $0.0$ & $-1034.0$ & --\\
J183717$-$015034 & $29.68$ & $+2.30$ & 36.2 & 2 & 2Ed-c & $+302.1 \pm 0.7$ & $14.6 \pm 0.4$ & -- & $6.90 \pm 0.22$ & $88.5 \pm 1.5$ & $106.6 \pm 3.6$ & $-1473.1$ & $24.2$\\
&&&&&& $+92.9 \pm 7.4$ & $14.6 \pm 0.4$ & -- & $0.62 \pm 0.11$ & $161.9 \pm 14.6$ && \\
J182900$-$002018 & $30.07$ & $+4.84$ & 12.7 & 2 & 1S & $-95.5 \pm 6.0$ & -- & $52.5 \pm 3.7$ & $3.68 \pm 0.43$ & $21.2 \pm 9.4$ & $52.5 \pm 3.6$ & $-1058.7$ & $1.1$\\
J183827$-$013111 & $30.10$ & $+2.19$ & 17.4 & 2 & 2T & $+277.6 \pm 10.2$ & -- & -- & $10.12 \pm 0.98$ & $160.9 \pm 17.5$ & $47.5 \pm 8.1$ & $-342.3$ & $8.2$\\
&&&&&& $+182.7 \pm 12.6$ & -- & -- & $6.35 \pm 1.09$ & $124.0 \pm 22.5$ && \\
J184124$-$015255 & $30.11$ & $+1.37$ & 344.3 & 2 & 2Ed-c & $+296.5 \pm 1.4$ & $12.4 \pm 0.4$ & -- & $1.24 \pm 0.04$ & $56.7 \pm 3.2$ & $138.1 \pm 0.9$ & $-1479.2$ & $293.1$\\
&&&&&& $+22.5 \pm 1.0$ & $12.4 \pm 0.4$ & -- & $1.63 \pm 0.05$ & $106.4 \pm 2.2$ && \\
J183840$-$012957 & $30.14$ & $+2.16$ & 99.2 & 2 & 1S & $+327.4 \pm 1.3$ & -- & $41.2 \pm 1.5$ & $11.36 \pm 0.45$ & $61.1 \pm 2.2$ & $41.2 \pm 1.5$ & $-1020.5$ & $1.0$\\
J183551$-$005941 & $30.27$ & $+3.01$ & 17.5 & 2 & 2T & $+200.4 \pm 9.5$ & -- & -- & $2.24 \pm 0.44$ & $122.8 \pm 12.7$ & $29.4 \pm 5.0$ & $-1223.5$ & $20.9$\\
&&&&&& $+141.6 \pm 2.9$ & -- & -- & $6.00 \pm 0.46$ & $48.0 \pm 4.8$ && \\
J183603$-$005747 & $30.32$ & $+2.98$ & 5.7 & 1 & 2T & $+157.0 \pm 8.9$ & -- & -- & $1.84 \pm 0.23$ & $96.8 \pm 15.2$ & $44.1 \pm 6.4$ & $-1018.1$ & $24.2$\\
&&&&&& $+68.9 \pm 9.2$ & -- & -- & $1.69 \pm 0.25$ & $161.2 \pm 15.7$ && \\
J190014$-$033504 & $30.74$ & $-3.59$ & 5.8 & 2 & 2T & $+626.3 \pm 12.7$ & -- & -- & $1.24 \pm 0.24$ & $28.4 \pm 21.2$ & $46.9 \pm 7.1$ & $-1206.7$ & $66.1$\\
&&&&&& $+532.6 \pm 6.4$ & -- & -- & $2.78 \pm 0.25$ & $52.9 \pm 9.9$ && \\
J184959$-$013256 & $31.39$ & $-0.38$ & 6.0 & 2 & 2Ed-c & $+206.2 \pm 1.6$ & $8.6 \pm 1.2$ & -- & $0.58 \pm 0.04$ & $74.3 \pm 3.1$ & $109.7 \pm 1.7$ & $-1742.8$ & $6.1$\\
&&&&&& $-11.9 \pm 3.0$ & $8.6 \pm 1.2$ & -- & $0.30 \pm 0.03$ & $172.5 \pm 6.2$ && \\
\hline
\end{tabular}
\end{table}
\end{landscape}

\begin{landscape}
\begin{table}
\ContinuedFloat
\caption{\textit{continued}}
\begin{tabular}{lccccccccccccc}
\hline
Target source & $\ell$ & $b$ & Angular & Morphology & Best-fit & $\phi$ & $\sigma_\phi$ & $\Delta\phi$ & $p_0$ & ${\rm PA}_0$ & FD spread & BIC & $\Delta {\rm BIC}$ \\
(NVSS) & ($^\circ$) & ($^\circ$) & size ($^{\prime\prime}$) && model & (${\rm rad\,m}^{-2}$) & (${\rm rad\,m}^{-2}$) & (${\rm rad\,m}^{-2}$) & (per cent) & ($^\circ$) & (${\rm rad\,m}^{-2}$) && \\
\hline
J183838$+$000858 & $31.60$ & $+2.92$ & 5.4 & 1 & 2T & $+116.9 \pm 1.2$ & -- & -- & $3.61 \pm 0.08$ & $3.0 \pm 2.2$ & $103.7 \pm 3.3$ & $-1262.3$ & $27.5$\\
&&&&&& $-90.4 \pm 6.4$ & -- & -- & $0.73 \pm 0.09$ & $133.6 \pm 10.9$ && \\
J183415$+$004451 & $31.64$ & $+4.16$ & 3.0 & 1 & 1S & $-42.0 \pm 4.6$ & -- & $98.8 \pm 1.3$ & $23.46 \pm 2.31$ & $120.0 \pm 9.5$ & $98.8 \pm 1.3$ & $-645.3$ & $2.1$\\
J183418$+$004852 & $31.70$ & $+4.18$ & 52.9 & 2 & 1Ed & $+76.7 \pm 0.7$ & $7.6 \pm 0.7$ & -- & $8.01 \pm 0.20$ & $92.1 \pm 1.2$ & $7.6 \pm 0.7$ & $-1321.3$ & $0.3$\\
J183931$+$001447 & $31.79$ & $+2.76$ & 6.9 & 1 & 1T & $+216.9 \pm 1.3$ & -- & -- & $9.35 \pm 0.20$ & $145.9 \pm 2.6$ & $0.0$ & $-861.3$ & --\\
J183935$+$001547 & $31.81$ & $+2.76$ & 3.4 & 1 & 2T & $+172.7 \pm 2.2$ & -- & -- & $6.71 \pm 0.23$ & $148.6 \pm 4.1$ & $62.7 \pm 3.7$ & $-898.8$ & $61.3$\\
&&&&&& $+47.3 \pm 7.1$ & -- & -- & $2.13 \pm 0.21$ & $22.1 \pm 13.9$ && \\
J183307$+$011535 & $31.97$ & $+4.65$ & 2.8 & 1 & 2T & $+373.2 \pm 0.9$ & -- & -- & $1.67 \pm 0.03$ & $94.8 \pm 1.7$ & $139.2 \pm 2.6$ & $-1715.1$ & $153.7$\\
&&&&&& $+94.8 \pm 5.1$ & -- & -- & $0.32 \pm 0.03$ & $68.1 \pm 9.9$ && \\
J183437$+$010519 & $31.98$ & $+4.24$ & 2.8 & 1 & 1S & $+55.1 \pm 3.1$ & -- & $31.6 \pm 4.6$ & $6.71 \pm 0.36$ & $83.5 \pm 3.2$ & $31.6 \pm 4.6$ & $-1074.1$ & $0.3$\\
J185822$-$013654 & $32.28$ & $-2.28$ & 3.1 & 1 & 2Ed-c & $+556.8 \pm 2.0$ & $16.5 \pm 0.8$ & -- & $2.26 \pm 0.14$ & $1.3 \pm 3.1$ & $271.5 \pm 3.7$ & $-1552.1$ & $26.8$\\
&&&&&& $+15.7 \pm 7.1$ & $16.5 \pm 0.8$ & -- & $0.46 \pm 0.06$ & $154.9 \pm 15.2$ && \\
J184704$-$000446 & $32.36$ & $+0.93$ & 42.3 & c & 2Ed-c & $+140.0 \pm 7.4$ & $17.5 \pm 1.5$ & -- & $3.23 \pm 0.59$ & $64.5 \pm 11.8$ & $57.8 \pm 4.1$ & $-1397.2$ & $8.9$\\
&&&&&& $+35.6 \pm 4.9$ & $17.5 \pm 1.5$ & -- & $4.64 \pm 0.72$ & $86.7 \pm 9.4$ && \\
J190833$-$023000 & $32.65$ & $-4.95$ & 5.2 & 2 & 1T & $+122.7 \pm 1.8$ & -- & -- & $3.11 \pm 0.11$ & $45.6 \pm 3.4$ & $0.0$ & $-1238.5$ & --\\
J183511$+$014620 & $32.66$ & $+4.42$ & 53.2 & c & 2T & $+207.5 \pm 0.5$ & -- & -- & $9.69 \pm 0.07$ & $14.6 \pm 1.0$ & $59.1 \pm 3.7$ & $-1505.3$ & $90.1$\\
&&&&&& $+89.3 \pm 7.5$ & -- & -- & $0.68 \pm 0.07$ & $157.2 \pm 15.3$ && \\
J183337$+$020355 & $32.74$ & $+4.91$ & 2.7 & 1 & 2T & $+241.5 \pm 8.0$ & -- & -- & $0.37 \pm 0.04$ & $143.6 \pm 16.7$ & $68.9 \pm 6.3$ & $-1538.3$ & $42.4$\\
&&&&&& $+103.8 \pm 9.8$ & -- & -- & $0.28 \pm 0.04$ & $81.5 \pm 21.2$ && \\
J184821$+$001108 & $32.75$ & $+0.77$ & 84.2 & 2 & 1T & $-140.9 \pm 1.2$ & -- & -- & $5.78 \pm 0.14$ & $51.5 \pm 2.9$ & $0.0$ & $-1127.8$ & --\\
J185351$-$002508 & $32.84$ & $-0.73$ & 17.6 & 2 & 2T & $+374.2 \pm 0.8$ & -- & -- & $6.03 \pm 0.10$ & $33.4 \pm 1.6$ & $93.9 \pm 7.7$ & $-1267.8$ & $31.4$\\
&&&&&& $+186.4 \pm 15.4$ & -- & -- & $0.62 \pm 0.11$ & $106.4 \pm 16.3$ && \\
J185751$-$004817 & $32.95$ & $-1.80$ & 22.1 & 3 & 2Ed-c & $+729.3 \pm 3.8$ & $19.5 \pm 1.2$ & -- & $0.99 \pm 0.11$ & $69.2 \pm 6.1$ & $349.6 \pm 5.1$ & $-1682.9$ & $9.9$\\
&&&&&& $+32.3 \pm 9.4$ & $19.5 \pm 1.2$ & -- & $0.31 \pm 0.05$ & $126.0 \pm 15.7$ && \\
J190042$-$005151 & $33.22$ & $-2.46$ & 13.2 & 2 & 1T & $+410.1 \pm 1.2$ & -- & -- & $3.38 \pm 0.07$ & $91.4 \pm 2.5$ & $0.0$ & $-1277.8$ & --\\
J190407$-$011342 & $33.29$ & $-3.38$ & 12.3 & c & 1T & $+280.4 \pm 1.2$ & -- & -- & $3.58 \pm 0.08$ & $16.1 \pm 2.5$ & $0.0$ & $-1324.8$ & --\\
J185146$+$003532 & $33.50$ & $+0.19$ & 4.7 & 1 & 2S & $-164.6 \pm 7.0$ & -- & $94.6 \pm 3.1$ & $0.99 \pm 0.13$ & $143.5 \pm 11.2$ & $112.2 \pm 3.2$ & $-1721.9$ & $36.9$\\
&&&&&& $-282.4 \pm 1.7$ & -- & $13.3 \pm 3.6$ & $2.97 \pm 0.06$ & $32.2 \pm 0.9$ && \\
J190832$-$011929 & $33.70$ & $-4.41$ & 10.4 & 2 & 1T & $+170.4 \pm 3.2$ & -- & -- & $4.46 \pm 0.24$ & $112.0 \pm 6.2$ & $0.0$ & $-892.7$ & --\\
J184755$+$012221 & $33.75$ & $+1.41$ & 5.0 & 1 & 1S & $+13.0 \pm 2.6$ & -- & $95.8 \pm 1.6$ & $25.29 \pm 1.59$ & $132.8 \pm 5.4$ & $95.8 \pm 1.6$ & $-852.5$ & $5.0$\\
J185857$+$000727 & $33.90$ & $-1.61$ & 46.2 & c & 1Ed & $+267.8 \pm 1.3$ & $6.2 \pm 1.8$ & -- & $5.21 \pm 0.31$ & $110.5 \pm 2.6$ & $6.2 \pm 1.8$ & $-1196.7$ & $0.0$\\
J190017$+$000355 & $34.00$ & $-1.94$ & 4.1 & 1 & 2T & $+547.5 \pm 1.0$ & -- & -- & $2.12 \pm 0.03$ & $135.8 \pm 1.8$ & $180.9 \pm 4.2$ & $-1634.4$ & $34.3$\\
&&&&&& $+185.7 \pm 8.3$ & -- & -- & $0.24 \pm 0.03$ & $10.3 \pm 14.3$ && \\
J184435$+$020933 & $34.07$ & $+2.51$ & 16.8 & 2 & 1Ed & $+42.4 \pm 1.6$ & $12.9 \pm 1.0$ & -- & $11.03 \pm 0.67$ & $148.1 \pm 2.8$ & $12.9 \pm 0.9$ & $-759.3$ & $0.5$\\
J190832$-$005319 & $34.09$ & $-4.21$ & 86.9 & c & 1T & $-2.2 \pm 6.5$ & -- & -- & $10.53 \pm 1.23$ & $105.2 \pm 13.1$ & $0.0$ & $-220.6$ & --\\
J190831$-$004855 & $34.16$ & $-4.17$ & 20.0 & 2 & 1T & $-9.3 \pm 3.8$ & -- & -- & $8.13 \pm 0.46$ & $22.1 \pm 6.9$ & $0.0$ & $-622.7$ & --\\
J191010$-$005622 & $34.23$ & $-4.60$ & 6.0 & c & 1T & $-53.8 \pm 2.7$ & -- & -- & $7.89 \pm 0.37$ & $39.3 \pm 5.7$ & $0.0$ & $-564.8$ & --\\
J190532$-$000941 & $34.40$ & $-3.21$ & 15.6 & 2 & 1T & $+127.8 \pm 1.1$ & -- & -- & $4.39 \pm 0.10$ & $55.8 \pm 2.0$ & $0.0$ & $-1200.5$ & --\\
J190559$+$000721 & $34.70$ & $-3.18$ & 10.0 & 2 & 2T & $+62.7 \pm 14.5$ & -- & -- & $2.13 \pm 0.37$ & $179.2 \pm 8.7$ & $27.5 \pm 8.3$ & $-1242.5$ & $293.7$\\
&&&&&& $+7.7 \pm 8.2$ & -- & -- & $2.94 \pm 0.31$ & $163.1 \pm 15.8$ && \\
J190655$+$000339 & $34.75$ & $-3.42$ & 167.5 & c & 1T & $-19.3 \pm 7.5$ & -- & -- & $3.08 \pm 0.41$ & $22.8 \pm 14.5$ & $0.0$ & $-692.8$ & --\\
J190741$+$000038 & $34.80$ & $-3.61$ & 6.9 & 2 & 1S & $-85.1 \pm 1.1$ & -- & $25.8 \pm 1.9$ & $6.33 \pm 0.16$ & $149.5 \pm 1.3$ & $25.8 \pm 1.9$ & $-1365.3$ & $0.2$\\
\hline
\end{tabular}
\end{table}
\end{landscape}

\begin{landscape}
\begin{table}
\ContinuedFloat
\caption{\textit{continued}}
\begin{tabular}{lccccccccccccc}
\hline
Target source & $\ell$ & $b$ & Angular & Morphology & Best-fit & $\phi$ & $\sigma_\phi$ & $\Delta\phi$ & $p_0$ & ${\rm PA}_0$ & FD spread & BIC & $\Delta {\rm BIC}$ \\
(NVSS) & ($^\circ$) & ($^\circ$) & size ($^{\prime\prime}$) && model & (${\rm rad\,m}^{-2}$) & (${\rm rad\,m}^{-2}$) & (${\rm rad\,m}^{-2}$) & (per cent) & ($^\circ$) & (${\rm rad\,m}^{-2}$) && \\
\hline
J183848$+$040424 & $35.13$ & $+4.66$ & 3.1 & 1 & 1T & $+146.7 \pm 1.0$ & -- & -- & $2.67 \pm 0.05$ & $58.4 \pm 1.9$ & $0.0$ & $-1484.6$ & --\\
J185515$+$021054 & $35.31$ & $+0.15$ & 19.5 & 2 & 1Ed & $+93.7 \pm 1.7$ & $7.0 \pm 2.0$ & -- & $11.22 \pm 0.61$ & $33.0 \pm 3.1$ & $7.0 \pm 2.0$ & $-903.0$ & $0.1$\\
J191133$+$001449 & $35.45$ & $-4.36$ & 22.3 & c & 1T & $-29.4 \pm 3.4$ & -- & -- & $4.88 \pm 0.31$ & $137.4 \pm 7.1$ & $0.0$ & $-878.0$ & --\\
J190426$+$011036 & $35.46$ & $-2.36$ & 12.2 & 2 & 1T & $-65.5 \pm 1.8$ & -- & -- & $4.82 \pm 0.13$ & $4.6 \pm 2.9$ & $0.0$ & $-901.0$ & --\\
J185114$+$025939 & $35.57$ & $+1.41$ & 3.6 & 1 & 1Ed & $+177.2 \pm 0.8$ & $5.0 \pm 1.2$ & -- & $1.90 \pm 0.07$ & $169.0 \pm 1.7$ & $5.0 \pm 1.2$ & $-1637.6$ & $0.1$\\
J184320$+$040256 & $35.62$ & $+3.65$ & 4.7 & 1 & 1T & $+31.3 \pm 1.6$ & -- & -- & $4.28 \pm 0.12$ & $164.5 \pm 3.1$ & $0.0$ & $-1000.2$ & --\\
J190944$+$005558 & $35.85$ & $-3.65$ & 108.0 & 2 & 1S & $-44.6 \pm 3.3$ & -- & $98.7 \pm 1.3$ & $13.98 \pm 0.83$ & $126.5 \pm 6.9$ & $98.7 \pm 1.3$ & $-1029.3$ & $14.3$\\
J191417$+$002421 & $35.91$ & $-4.90$ & 17.8 & 3 & 1T & $-33.3 \pm 3.8$ & -- & -- & $8.14 \pm 0.52$ & $87.8 \pm 8.0$ & $0.0$ & $-562.4$ & --\\
J190721$+$012341 & $35.99$ & $-2.90$ & 3.2 & 1 & 1T & $+118.9 \pm 9.2$ & -- & -- & $2.23 \pm 0.32$ & $150.6 \pm 19.2$ & $0.0$ & $-563.3$ & $0.4$\\
J190712$+$012709 & $36.02$ & $-2.84$ & 4.3 & 1 & 1Ed & $+193.0 \pm 0.5$ & $10.7 \pm 0.3$ & -- & $3.71 \pm 0.07$ & $56.4 \pm 0.9$ & $10.8 \pm 0.3$ & $-1375.9$ & $4.2$\\
J185213$+$033255 & $36.18$ & $+1.44$ & 6.2 & 2 & 1S & $+176.6 \pm 3.0$ & -- & $19.5 \pm 5.5$ & $7.32 \pm 0.35$ & $125.1 \pm 2.1$ & $19.5 \pm 5.5$ & $-1129.6$ & $0.0$\\
J185837$+$024518 & $36.20$ & $-0.34$ & 2.8 & 1 & 1Ed & $+162.2 \pm 1.7$ & $14.0 \pm 0.9$ & -- & $8.51 \pm 0.49$ & $179.2 \pm 3.0$ & $14.0 \pm 0.9$ & $-941.3$ & $2.0$\\
J185222$+$033347 & $36.21$ & $+1.42$ & 17.3 & 3 & 1T & $+133.1 \pm 11.5$ & -- & -- & $5.05 \pm 0.81$ & $14.7 \pm 17.9$ & $0.0$ & $-401.4$ & --\\
J184500$+$043812 & $36.33$ & $+3.54$ & 23.3 & 3 & 1Ed & $+3.4 \pm 2.2$ & $10.5 \pm 1.6$ & -- & $19.10 \pm 1.67$ & $141.1 \pm 3.9$ & $10.6 \pm 1.6$ & $-502.4$ & $0.2$\\
J184604$+$043450 & $36.40$ & $+3.28$ & 9.7 & 2 & 1T & $+33.6 \pm 3.7$ & -- & -- & $4.80 \pm 0.32$ & $100.3 \pm 6.8$ & $0.0$ & $-779.5$ & --\\
J185802$+$031316 & $36.55$ & $+0.00$ & 4.1 & 1 & 2T & $+423.8 \pm 0.2$ & -- & -- & $4.73 \pm 0.02$ & $79.0 \pm 0.4$ & $127.9 \pm 5.2$ & $-1809.5$ & $15.0$\\
&&&&&& $+168.0 \pm 10.5$ & -- & -- & $0.09 \pm 0.02$ & $87.4 \pm 22.4$ && \\
J185306$+$044052 & $37.29$ & $+1.76$ & 6.2 & 1 & 1S & $+280.8 \pm 1.2$ & -- & $27.5 \pm 1.9$ & $17.33 \pm 0.50$ & $113.7 \pm 1.3$ & $27.5 \pm 1.9$ & $-860.8$ & $0.3$\\
J184718$+$055022 & $37.67$ & $+3.57$ & 134.8 & 2 & 1T & $+105.4 \pm 4.6$ & -- & -- & $3.22 \pm 0.29$ & $96.4 \pm 9.8$ & $0.0$ & $-685.7$ & --\\
J184438$+$062651 & $37.91$ & $+4.44$ & 11.1 & 2 & 2T & $+278.4 \pm 12.6$ & -- & -- & $1.98 \pm 0.23$ & $166.5 \pm 18.1$ & $46.4 \pm 6.6$ & $-947.2$ & $81.0$\\
&&&&&& $+185.6 \pm 4.1$ & -- & -- & $4.12 \pm 0.24$ & $30.1 \pm 8.8$ && \\
J191406$+$025549 & $38.13$ & $-3.70$ & 8.6 & 2 & 2S & $+523.7 \pm 2.7$ & -- & $35.2 \pm 3.7$ & $7.89 \pm 0.57$ & $123.5 \pm 3.8$ & $157.3 \pm 4.9$ & $-965.2$ & $14.9$\\
&&&&&& $+245.7 \pm 9.9$ & -- & $64.6 \pm 4.0$ & $5.05 \pm 0.73$ & $102.9 \pm 15.9$ && \\
J184432$+$064257 & $38.14$ & $+4.58$ & 3.2 & 1 & 1T & $+214.0 \pm 0.4$ & -- & -- & $2.75 \pm 0.02$ & $97.4 \pm 0.7$ & $0.0$ & $-1796.6$ & --\\
J185513$+$052158 & $38.14$ & $+1.60$ & 3.0 & 1 & 1T & $+369.9 \pm 0.8$ & -- & -- & $7.87 \pm 0.10$ & $37.5 \pm 1.7$ & $0.0$ & $-1150.0$ & $0.5$\\
J184919$+$063211 & $38.52$ & $+3.44$ & 18.4 & c & 2T & $+22.3 \pm 4.6$ & -- & -- & $12.80 \pm 0.61$ & $26.7 \pm 9.1$ & $46.1 \pm 9.0$ & $-538.8$ & $25.7$\\
&&&&&& $-69.8 \pm 17.3$ & -- & -- & $3.44 \pm 0.68$ & $84.7 \pm 32.1$ && \\
J191325$+$034308 & $38.76$ & $-3.18$ & 128.1 & 2 & 1S & $+333.0 \pm 1.6$ & -- & $30.7 \pm 2.6$ & $6.20 \pm 0.24$ & $10.3 \pm 2.1$ & $30.7 \pm 2.6$ & $-1329.0$ & $0.1$\\
J191849$+$030442 & $38.81$ & $-4.67$ & 14.8 & 2 & 1S & $+49.3 \pm 2.3$ & -- & $32.0 \pm 2.9$ & $1.86 \pm 0.09$ & $120.8 \pm 2.7$ & $32.0 \pm 2.9$ & $-1577.6$ & $0.4$\\
J184753$+$071538 & $39.00$ & $+4.09$ & 70.3 & c & 2T & $+168.3 \pm 4.8$ & -- & -- & $4.13 \pm 0.23$ & $47.5 \pm 9.9$ & $58.7 \pm 4.8$ & $-1046.4$ & $78.5$\\
&&&&&& $+51.0 \pm 8.3$ & -- & -- & $2.17 \pm 0.26$ & $31.8 \pm 18.2$ && \\
J190343$+$055256 & $39.56$ & $-0.04$ & 3.3 & 1 & 1T & $+445.2 \pm 2.9$ & -- & -- & $1.07 \pm 0.06$ & $55.9 \pm 5.7$ & $0.0$ & $-1524.9$ & $4.2$\\
J190043$+$064546 & $40.01$ & $+1.03$ & 3.1 & 1 & 2Ed-c & $+384.9 \pm 0.8$ & $7.6 \pm 0.8$ & -- & $8.90 \pm 0.28$ & $104.3 \pm 1.6$ & $224.6 \pm 5.9$ & $-1171.6$ & $16.5$\\
&&&&&& $-63.8 \pm 11.7$ & $7.6 \pm 0.8$ & -- & $0.71 \pm 0.13$ & $38.9 \pm 21.3$ && \\
J191725$+$044236 & $40.10$ & $-3.61$ & 11.3 & c & 2T & $+231.7 \pm 6.2$ & -- & -- & $1.24 \pm 0.13$ & $123.3 \pm 10.4$ & $71.9 \pm 3.3$ & $-1290.1$ & $100.0$\\
&&&&&& $+88.0 \pm 2.4$ & -- & -- & $2.82 \pm 0.12$ & $125.4 \pm 4.8$ && \\
J192049$+$042052 & $40.17$ & $-4.52$ & 11.4 & 2 & 2T & $+129.3 \pm 9.9$ & -- & -- & $2.60 \pm 0.45$ & $127.2 \pm 17.3$ & $42.1 \pm 5.7$ & $-1277.9$ & $138.6$\\
&&&&&& $+45.0 \pm 5.8$ & -- & -- & $4.61 \pm 0.42$ & $122.7 \pm 10.2$ && \\
J190734$+$060446 & $40.18$ & $-0.80$ & 9.9 & 2 & 1Ed & $+408.8 \pm 0.7$ & $11.0 \pm 0.4$ & -- & $5.77 \pm 0.17$ & $54.9 \pm 1.5$ & $11.1 \pm 0.4$ & $-1451.1$ & $0.7$\\
J191833$+$043928 & $40.18$ & $-3.88$ & 7.4 & 1 & 1T & $+65.9 \pm 8.1$ & -- & -- & $3.89 \pm 0.54$ & $13.9 \pm 15.5$ & $0.0$ & $-566.6$ & --\\
J191840$+$043932 & $40.20$ & $-3.91$ & 6.3 & 1 & 1S & $+153.9 \pm 9.4$ & -- & $68.2 \pm 3.2$ & $21.59 \pm 2.59$ & $82.8 \pm 12.2$ & $68.2 \pm 3.2$ & $-355.7$ & $40.3$\\
J184731$+$090047 & $40.53$ & $+4.96$ & 34.5 & 2 & 1Ed & $+230.8 \pm 1.1$ & $11.2 \pm 0.8$ & -- & $14.58 \pm 0.74$ & $134.2 \pm 2.2$ & $11.2 \pm 0.8$ & $-981.6$ & $0.1$\\
\hline
\end{tabular}
\end{table}
\end{landscape}

\begin{landscape}
\begin{table}
\ContinuedFloat
\caption{\textit{continued}}
\begin{tabular}{lccccccccccccc}
\hline
Target source & $\ell$ & $b$ & Angular & Morphology & Best-fit & $\phi$ & $\sigma_\phi$ & $\Delta\phi$ & $p_0$ & ${\rm PA}_0$ & FD spread & BIC & $\Delta {\rm BIC}$ \\
(NVSS) & ($^\circ$) & ($^\circ$) & size ($^{\prime\prime}$) && model & (${\rm rad\,m}^{-2}$) & (${\rm rad\,m}^{-2}$) & (${\rm rad\,m}^{-2}$) & (per cent) & ($^\circ$) & (${\rm rad\,m}^{-2}$) && \\
\hline
J192258$+$044354 & $40.76$ & $-4.82$ & 11.0 & 2 & 1Ed & $+126.6 \pm 2.7$ & $14.3 \pm 1.6$ & -- & $12.35 \pm 1.29$ & $49.6 \pm 4.4$ & $14.3 \pm 1.6$ & $-795.9$ & $0.3$\\
J192243$+$045126 & $40.84$ & $-4.71$ & 3.8 & 1 & 2T & $+131.9 \pm 2.4$ & -- & -- & $0.77 \pm 0.02$ & $96.0 \pm 4.8$ & $51.8 \pm 3.0$ & $-1831.0$ & $101.4$\\
&&&&&& $+28.3 \pm 5.5$ & -- & -- & $0.34 \pm 0.03$ & $83.8 \pm 12.1$ && \\
J191310$+$064158 & $41.37$ & $-1.75$ & 8.9 & 2 & 1S & $-13.8 \pm 3.1$ & -- & $44.5 \pm 2.9$ & $6.89 \pm 0.44$ & $25.9 \pm 4.5$ & $44.5 \pm 2.9$ & $-974.6$ & $1.7$\\
J184951$+$094850 & $41.51$ & $+4.81$ & 11.0 & 2 & 1Ed & $+568.1 \pm 1.1$ & $7.1 \pm 1.2$ & -- & $8.23 \pm 0.40$ & $88.1 \pm 1.9$ & $7.1 \pm 1.2$ & $-1046.8$ & $0.3$\\
J191917$+$061942 & $41.75$ & $-3.27$ & 3.4 & 1 & 1T & $+43.3 \pm 2.6$ & -- & -- & $1.51 \pm 0.06$ & $6.2 \pm 4.9$ & $0.0$ & $-1404.8$ & --\\
J190614$+$084226 & $42.36$ & $+0.70$ & 5.2 & 1 & 1T & $+127.6 \pm 2.3$ & -- & -- & $7.27 \pm 0.28$ & $107.5 \pm 4.2$ & $0.0$ & $-652.6$ & --\\
J185557$+$102011 & $42.66$ & $+3.70$ & 7.2 & 1 & 1T & $+569.9 \pm 1.9$ & -- & -- & $1.10 \pm 0.04$ & $143.5 \pm 4.0$ & $0.0$ & $-1584.3$ & --\\
J192233$+$071048 & $42.88$ & $-3.59$ & 2.9 & 1 & 2T & $+73.1 \pm 8.7$ & -- & -- & $1.11 \pm 0.15$ & $35.3 \pm 14.5$ & $58.1 \pm 9.9$ & $-1174.7$ & $0.3$\\
&&&&&& $-43.2 \pm 17.9$ & -- & -- & $0.81 \pm 0.15$ & $46.6 \pm 26.5$ && \\
J190741$+$090717 & $42.90$ & $+0.57$ & 18.2 & 3 & 2Ed-c & $+705.0 \pm 0.6$ & $8.0 \pm 0.5$ & -- & $3.99 \pm 0.08$ & $71.5 \pm 1.0$ & $257.8 \pm 4.7$ & $-1588.3$ & $17.5$\\
&&&&&& $+190.0 \pm 9.4$ & $8.0 \pm 0.5$ & -- & $0.22 \pm 0.04$ & $112.6 \pm 18.6$ && \\
J192245$+$073933 & $43.33$ & $-3.40$ & 64.5 & 2 & 1S & $+194.8 \pm 2.8$ & -- & $48.2 \pm 2.6$ & $9.53 \pm 0.64$ & $84.0 \pm 4.7$ & $48.2 \pm 2.6$ & $-954.5$ & $2.1$\\
J192802$+$070219 & $43.40$ & $-4.85$ & 2.8 & 1 & 1T & $+130.3 \pm 4.0$ & -- & -- & $6.78 \pm 0.45$ & $88.4 \pm 8.7$ & $0.0$ & $-615.1$ & --\\
J192820$+$070355 & $43.46$ & $-4.91$ & 22.8 & 2 & 1T & $+44.7 \pm 3.5$ & -- & -- & $2.27 \pm 0.14$ & $158.7 \pm 8.0$ & $0.0$ & $-1118.6$ & $12.6$\\
J191906$+$081920 & $43.49$ & $-2.30$ & 5.0 & 2 & 1T & $+226.0 \pm 1.0$ & -- & -- & $8.75 \pm 0.14$ & $65.9 \pm 2.0$ & $0.0$ & $-1008.5$ & --\\
J185728$+$111021 & $43.57$ & $+3.75$ & 65.4 & c & 1T & $+522.2 \pm 0.8$ & -- & -- & $10.55 \pm 0.16$ & $125.2 \pm 1.7$ & $0.0$ & $-1034.2$ & --\\
J191630$+$090223 & $43.83$ & $-1.39$ & 3.6 & 1 & 1T & $+545.2 \pm 2.0$ & -- & -- & $6.92 \pm 0.25$ & $99.8 \pm 4.2$ & $0.0$ & $-714.0$ & --\\
J191641$+$090147 & $43.84$ & $-1.44$ & 5.3 & 1 & 1T & $+509.8 \pm 1.5$ & -- & -- & $14.63 \pm 0.39$ & $107.2 \pm 3.1$ & $0.0$ & $-541.4$ & --\\
J185952$+$112514 & $44.06$ & $+3.34$ & 26.1 & 2 & 2Ed-c & $+658.4 \pm 2.2$ & $10.6 \pm 1.7$ & -- & $8.68 \pm 0.80$ & $169.2 \pm 4.5$ & $151.2 \pm 7.1$ & $-847.1$ & $5.8$\\
&&&&&& $+357.5 \pm 14.0$ & $10.6 \pm 1.7$ & -- & $1.86 \pm 0.35$ & $170.6 \pm 27.4$ && \\
J190323$+$112905 & $44.51$ & $+2.60$ & 15.8 & 1 & 2Ed-c & $+833.5 \pm 0.2$ & $5.4 \pm 0.2$ & -- & $18.82 \pm 0.16$ & $52.7 \pm 0.4$ & $181.6 \pm 2.9$ & $-1362.5$ & $1.6$\\
&&&&&& $+470.7 \pm 5.7$ & $5.4 \pm 0.2$ & -- & $0.63 \pm 0.06$ & $172.1 \pm 11.1$ && \\
J190319$+$112950 & $44.51$ & $+2.62$ & 9.3 & 1 & 1Ed & $+787.7 \pm 1.6$ & $11.0 \pm 1.2$ & -- & $3.02 \pm 0.18$ & $63.9 \pm 3.2$ & $11.0 \pm 1.2$ & $-1350.7$ & $0.1$\\
J192840$+$084849 & $45.04$ & $-4.15$ & 13.0 & c & 1Ed & $+528.0 \pm 0.7$ & $8.9 \pm 0.6$ & -- & $3.28 \pm 0.09$ & $76.3 \pm 1.1$ & $8.9 \pm 0.6$ & $-1575.7$ & $1.4$\\
J192355$+$094424 & $45.31$ & $-2.68$ & 26.2 & 2 & 1T & $+297.4 \pm 1.5$ & -- & -- & $4.12 \pm 0.10$ & $149.3 \pm 2.6$ & $0.0$ & $-1178.2$ & --\\
J185923$+$125912 & $45.41$ & $+4.15$ & 10.3 & c & 2Ed-c & $+258.1 \pm 0.7$ & $14.4 \pm 0.1$ & -- & $3.59 \pm 0.10$ & $75.7 \pm 0.9$ & $29.4 \pm 1.4$ & $-1636.0$ & $122.5$\\
&&&&&& $+215.7 \pm 3.9$ & $14.4 \pm 0.1$ & -- & $0.52 \pm 0.09$ & $63.0 \pm 4.3$ && \\
J191005$+$114748 & $45.54$ & $+1.28$ & 43.6 & 3 & 1Ed & $+845.4 \pm 0.7$ & $11.9 \pm 0.4$ & -- & $6.50 \pm 0.18$ & $56.3 \pm 1.4$ & $11.8 \pm 0.4$ & $-1488.3$ & $1.5$\\
J191000$+$122524 & $46.09$ & $+1.59$ & 19.1 & 3 & 1S & $+751.4 \pm 2.7$ & -- & $43.5 \pm 2.9$ & $5.52 \pm 0.38$ & $141.1 \pm 4.5$ & $43.5 \pm 2.9$ & $-1123.7$ & $0.5$\\
J192922$+$095808 & $46.14$ & $-3.76$ & 40.1 & 2 & 1Ed & $+20.6 \pm 0.5$ & $7.6 \pm 0.5$ & -- & $5.55 \pm 0.11$ & $113.4 \pm 1.0$ & $7.6 \pm 0.5$ & $-1582.4$ & $0.1$\\
J191733$+$114215 & $46.31$ & $-0.38$ & 17.7 & 2 & 1T & $-127.0 \pm 2.3$ & -- & -- & $6.58 \pm 0.23$ & $71.6 \pm 4.1$ & $0.0$ & $-903.1$ & --\\
J190501$+$132047 & $46.35$ & $+3.09$ & 19.4 & 2 & 1T & $+589.5 \pm 1.4$ & -- & -- & $4.87 \pm 0.13$ & $79.0 \pm 2.9$ & $0.0$ & $-1187.1$ & --\\
J190235$+$145023 & $47.41$ & $+4.30$ & 36.2 & 2 & 1S & $+421.0 \pm 4.3$ & -- & $90.5 \pm 1.9$ & $20.21 \pm 1.59$ & $89.7 \pm 8.5$ & $90.5 \pm 1.9$ & $-749.0$ & $4.1$\\
J193434$+$104340 & $47.43$ & $-4.52$ & 4.1 & 1 & 2T & $-115.2 \pm 6.1$ & -- & -- & $1.09 \pm 0.12$ & $113.4 \pm 12.2$ & $89.1 \pm 3.8$ & $-1193.1$ & $70.0$\\
&&&&&& $-293.3 \pm 4.6$ & -- & -- & $1.30 \pm 0.13$ & $75.2 \pm 8.3$ && \\
J190247$+$145137 & $47.46$ & $+4.26$ & 4.8 & 1 & 1T & $+538.8 \pm 1.7$ & -- & -- & $4.48 \pm 0.14$ & $163.9 \pm 3.7$ & $0.0$ & $-1075.2$ & --\\
J193357$+$105642 & $47.54$ & $-4.28$ & 9.7 & 2 & 2T & $-64.3 \pm 6.7$ & -- & -- & $1.39 \pm 0.12$ & $58.3 \pm 12.8$ & $44.7 \pm 3.5$ & $-1308.1$ & $253.8$\\
&&&&&& $-153.8 \pm 2.0$ & -- & -- & $4.54 \pm 0.11$ & $124.7 \pm 3.8$ && \\
J191025$+$140125 & $47.56$ & $+2.24$ & 329.8 & 2 & 1S & $+515.8 \pm 3.4$ & -- & $31.3 \pm 4.9$ & $4.22 \pm 0.35$ & $120.3 \pm 4.2$ & $31.3 \pm 4.8$ & $-1086.0$ & $0.4$\\
J192540$+$122738 & $47.91$ & $-1.77$ & 3.4 & 1 & 2T & $+92.7 \pm 5.7$ & -- & -- & $1.54 \pm 0.09$ & $35.1 \pm 9.8$ & $44.8 \pm 4.8$ & $-1343.6$ & $20.2$\\
&&&&&& $+3.2 \pm 7.8$ & -- & -- & $1.32 \pm 0.08$ & $179.4 \pm 14.7$ && \\
\hline
\end{tabular}
\end{table}
\end{landscape}

\begin{landscape}
\begin{table}
\ContinuedFloat
\caption{\textit{continued}}
\begin{tabular}{lccccccccccccc}
\hline
Target source & $\ell$ & $b$ & Angular & Morphology & Best-fit & $\phi$ & $\sigma_\phi$ & $\Delta\phi$ & $p_0$ & ${\rm PA}_0$ & FD spread & BIC & $\Delta {\rm BIC}$ \\
(NVSS) & ($^\circ$) & ($^\circ$) & size ($^{\prime\prime}$) && model & (${\rm rad\,m}^{-2}$) & (${\rm rad\,m}^{-2}$) & (${\rm rad\,m}^{-2}$) & (per cent) & ($^\circ$) & (${\rm rad\,m}^{-2}$) && \\
\hline
J190451$+$152148 & $48.13$ & $+4.05$ & 31.8 & 2 & 1Ed & $+545.1 \pm 0.4$ & $7.3 \pm 0.4$ & -- & $4.75 \pm 0.09$ & $99.7 \pm 0.9$ & $7.3 \pm 0.4$ & $-1707.5$ & $0.1$\\
J190414$+$153638 & $48.28$ & $+4.29$ & 4.3 & 1 & 2T & $+669.7 \pm 1.6$ & -- & -- & $2.34 \pm 0.06$ & $52.3 \pm 3.1$ & $60.4 \pm 2.5$ & $-1527.4$ & $299.3$\\
&&&&&& $+549.0 \pm 4.8$ & -- & -- & $0.88 \pm 0.05$ & $141.6 \pm 9.3$ && \\
J192458$+$130033 & $48.31$ & $-1.36$ & 106.0 & c & 1S & $+386.1 \pm 5.2$ & -- & $81.4 \pm 2.1$ & $8.21 \pm 0.79$ & $118.2 \pm 9.8$ & $81.4 \pm 2.1$ & $-985.3$ & $15.3$\\
J190655$+$152342 & $48.39$ & $+3.62$ & 8.8 & c & 2T & $+636.4 \pm 0.4$ & -- & -- & $12.61 \pm 0.12$ & $79.0 \pm 0.8$ & $84.0 \pm 3.1$ & $-1217.3$ & $40.1$\\
&&&&&& $+468.5 \pm 6.1$ & -- & -- & $0.81 \pm 0.13$ & $112.3 \pm 12.4$ && \\
J190653$+$152650 & $48.43$ & $+3.65$ & 3.6 & 1 & 1T & $+512.5 \pm 4.2$ & -- & -- & $2.17 \pm 0.17$ & $128.7 \pm 8.5$ & $0.0$ & $-1009.1$ & $9.8$\\
J193335$+$120844 & $48.56$ & $-3.62$ & 13.1 & 2 & 1S & $-111.3 \pm 4.1$ & -- & $59.8 \pm 1.9$ & $21.51 \pm 1.28$ & $59.7 \pm 6.6$ & $59.8 \pm 1.9$ & $-728.6$ & $4.0$\\
J193328$+$120953 & $48.56$ & $-3.59$ & 3.7 & 1 & 1S & $-95.9 \pm 8.5$ & -- & $99.5 \pm 0.9$ & $49.53 \pm 5.40$ & $101.6 \pm 17.1$ & $99.5 \pm 0.9$ & $-319.8$ & $3.2$\\
J190355$+$160147 & $48.62$ & $+4.55$ & 28.7 & c & 1S & $+424.8 \pm 0.7$ & -- & $30.3 \pm 1.1$ & $7.14 \pm 0.13$ & $89.1 \pm 0.9$ & $30.4 \pm 1.1$ & $-1481.9$ & $0.1$\\
J191644$+$150349 & $49.19$ & $+1.37$ & 14.2 & 2 & 1T & $+535.4 \pm 0.6$ & -- & -- & $3.15 \pm 0.03$ & $104.2 \pm 1.2$ & $0.0$ & $-1036.2$ & --\\
J192517$+$135919 & $49.21$ & $-0.97$ & 28.2 & 2 & 2Ed-c & $+506.0 \pm 1.8$ & $16.0 \pm 1.0$ & -- & $4.82 \pm 0.76$ & $2.2 \pm 1.7$ & $40.1 \pm 1.2$ & $-1263.2$ & $31.3$\\
&&&&&& $+439.6 \pm 1.2$ & $16.0 \pm 1.0$ & -- & $7.80 \pm 0.85$ & $1.6 \pm 1.1$ && \\
J190516$+$163706 & $49.30$ & $+4.53$ & 33.1 & 2 & 1S & $+478.1 \pm 1.8$ & -- & $23.0 \pm 3.0$ & $4.36 \pm 0.15$ & $38.6 \pm 1.9$ & $23.0 \pm 3.0$ & $-1498.3$ & $0.1$\\
J193302$+$131335 & $49.44$ & $-2.98$ & 67.2 & c & 1T & $-74.9 \pm 0.7$ & -- & -- & $3.66 \pm 0.05$ & $133.1 \pm 1.4$ & $0.0$ & $-1553.5$ & --\\
J191133$+$161431 & $49.65$ & $+3.02$ & 7.8 & c & 1Ed & $+611.8 \pm 1.1$ & $23.1 \pm 0.3$ & -- & $6.59 \pm 0.19$ & $41.9 \pm 1.6$ & $23.1 \pm 0.3$ & $-1695.1$ & $24.0$\\
J191158$+$161147 & $49.66$ & $+2.91$ & 3.5 & 1 & 1T & $+712.5 \pm 1.1$ & -- & -- & $1.24 \pm 0.03$ & $142.0 \pm 2.3$ & $0.0$ & $-1750.5$ & --\\
J190901$+$163944 & $49.75$ & $+3.75$ & 19.3 & 2 & 2T & $+446.3 \pm 0.6$ & -- & -- & $3.56 \pm 0.02$ & $31.1 \pm 1.1$ & $52.4 \pm 4.8$ & $-1728.2$ & $96.2$\\
&&&&&& $+341.4 \pm 9.5$ & -- & -- & $0.26 \pm 0.02$ & $120.2 \pm 17.8$ && \\
J191219$+$161628 & $49.77$ & $+2.87$ & 10.6 & 2 & 2Ed-c & $+749.5 \pm 0.3$ & $9.8 \pm 0.2$ & -- & $7.23 \pm 0.10$ & $121.6 \pm 0.6$ & $117.9 \pm 2.7$ & $-1656.7$ & $109.7$\\
&&&&&& $+515.2 \pm 5.5$ & $9.8 \pm 0.2$ & -- & $0.48 \pm 0.05$ & $67.5 \pm 9.9$ && \\
J192835$+$142156 & $49.92$ & $-1.49$ & 3.2 & 1 & 1Ed & $+128.3 \pm 3.5$ & $10.9 \pm 3.0$ & -- & $18.13 \pm 2.87$ & $105.4 \pm 7.4$ & $10.9 \pm 3.0$ & $-323.2$ & $0.3$\\
J192910$+$141952 & $49.96$ & $-1.63$ & 13.8 & c & 1T & $+125.2 \pm 1.2$ & -- & -- & $8.63 \pm 0.20$ & $21.5 \pm 2.3$ & $0.0$ & $-943.0$ & --\\
J191649$+$155836 & $50.00$ & $+1.77$ & 31.9 & c & 1T & $+394.2 \pm 1.1$ & -- & -- & $5.84 \pm 0.11$ & $37.2 \pm 2.1$ & $0.0$ & $-1243.5$ & --\\
J191549$+$160834 & $50.04$ & $+2.06$ & 17.4 & 2 & 2T & $+471.3 \pm 1.0$ & -- & -- & $5.95 \pm 0.11$ & $103.9 \pm 2.1$ & $127.5 \pm 4.6$ & $-1119.2$ & $18.0$\\
&&&&&& $+216.3 \pm 9.2$ & -- & -- & $0.60 \pm 0.12$ & $161.1 \pm 19.5$ && \\
J194012$+$125809 & $50.06$ & $-4.64$ & 27.7 & 3 & 1T & $-152.3 \pm 2.6$ & -- & -- & $2.82 \pm 0.16$ & $9.5 \pm 6.1$ & $0.0$ & $-1137.1$ & --\\
J191414$+$163640 & $50.28$ & $+2.62$ & 3.2 & 1 & 2T & $+558.7 \pm 0.7$ & -- & -- & $4.74 \pm 0.06$ & $97.6 \pm 1.4$ & $121.5 \pm 6.0$ & $-1421.8$ & $20.1$\\
&&&&&& $+315.7 \pm 11.9$ & -- & -- & $0.35 \pm 0.06$ & $50.3 \pm 38.2$ && \\
J192439$+$154043 & $50.63$ & $-0.03$ & 2.6 & 1 & 2Ed-c & $+420.4 \pm 0.4$ & $8.4 \pm 0.3$ & -- & $6.34 \pm 0.08$ & $160.7 \pm 0.6$ & $179.9 \pm 5.8$ & $-1520.6$ & $33.1$\\
&&&&&& $+61.4 \pm 11.5$ & $8.4 \pm 0.3$ & -- & $0.24 \pm 0.04$ & $104.7 \pm 21.7$ && \\
J193939$+$134604 & $50.70$ & $-4.13$ & 70.2 & c & 1T & $-182.0 \pm 3.8$ & -- & -- & $3.91 \pm 0.28$ & $57.4 \pm 7.7$ & $0.0$ & $-853.8$ & --\\
J192032$+$162557 & $50.82$ & $+1.20$ & 9.6 & 2 & 2T & $+533.4 \pm 1.2$ & -- & -- & $10.23 \pm 0.23$ & $80.9 \pm 2.3$ & $83.2 \pm 3.4$ & $-937.5$ & $1.4$\\
&&&&&& $+367.0 \pm 6.8$ & -- & -- & $1.54 \pm 0.21$ & $95.8 \pm 13.4$ && \\
J192203$+$162243 & $50.95$ & $+0.85$ & 25.1 & 3 & 1T & $+457.1 \pm 1.1$ & -- & -- & $3.77 \pm 0.07$ & $111.6 \pm 1.8$ & $0.0$ & $-1384.2$ & --\\
J193306$+$145624 & $50.95$ & $-2.17$ & 3.4 & 1 & 1T & $+146.7 \pm 1.4$ & -- & -- & $2.31 \pm 0.04$ & $128.8 \pm 2.3$ & $0.0$ & $-1239.8$ & --\\
J193321$+$150446 & $51.10$ & $-2.16$ & 2.8 & 1 & 1Ed & $+352.0 \pm 2.1$ & $11.9 \pm 1.4$ & -- & $1.63 \pm 0.11$ & $78.0 \pm 3.5$ & $11.9 \pm 1.4$ & $-1632.2$ & $0.8$\\
J193052$+$153235 & $51.22$ & $-1.41$ & 2.8 & 1 & 2T & $+214.1 \pm 2.2$ & -- & -- & $0.95 \pm 0.10$ & $176.6 \pm 2.5$ & $25.8 \pm 1.6$ & $-1665.3$ & $251.5$\\
&&&&&& $+162.5 \pm 2.2$ & -- & -- & $2.03 \pm 0.09$ & $13.1 \pm 3.6$ && \\
\hline
\multicolumn{14}{l}{\textsc{NOTE} -- The $\Delta{\rm BIC}$ values are omitted for sources with all but one Stokes \textit{QU}-fitting models being rejected.} \\
\end{tabular}
\end{table}
\end{landscape}

\section{Polarised intensity of the Galactic foreground} \label{sec:fg_pi}

Following the Stokes \textit{I}, \textit{Q}, and \textit{U} foreground subtraction routine (Section~\ref{sec:fg_sub}), we investigate the actual amount of polarised emission subtracted with respect to the PI of the target sources themselves. We pay special attention to EGSs best represented by the sinc-component models (those with $\Delta \phi$ parameter; i.e.\ 1S, 2S, and 1Id models), given that if the span of FD exhibited by those sources are owing to rapid spatial changes in FD in the Galactic foreground, it can be possible for the foreground polarised emission subtraction to be erroneous (Section~\ref{sec:fds_definition}).

\begin{figure}
\includegraphics[width=0.47\textwidth]{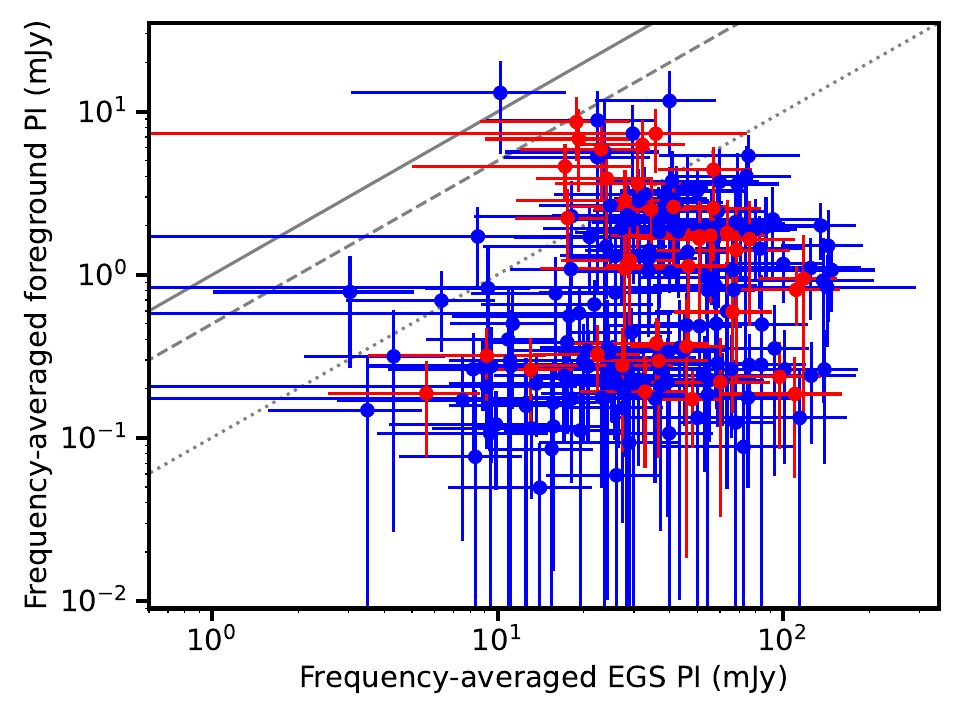}
\caption{Frequency-averaged PI of the foreground emission against that of our target EGSs, with the error bars representing the standard deviation of the pre-averaged values across frequency. The red data points correspond to sources best-fit by the sinc-component models (1S, 2S, and 1Id), while the blue data points represent the other polarised sources. The solid, dashed, and dotted lines mark where the EGS PI is 1, 2, and 10 times the foreground PI, respectively.}
\label{fig:fg_pi}
\end{figure}

For each of our 191 polarised targets, we evaluate the median foreground-subtracted EGS PI across frequency, as well as the corresponding median foreground PI across frequency, and compare them in Figure~\ref{fig:fg_pi}. The foreground PI values are the actual amounts subtracted from our EGSs (i.e.\ have been scaled by the angular size of the EGSs; see Section~\ref{sec:fg_sub}). From this, we find that the typical foreground PI is low, and is seldom significantly higher than the per-channel noise level in Stokes \textit{Q} and \textit{U} of about $1.5\,{\rm mJy/beam}$ \citep{ma20}. In addition, most of the EGSs are significantly brighter than the foreground in linear polarisation, with 29 out of the 39 sinc-component sources being more than an order of magnitude brighter than the corresponding foreground emission\footnote{The 10 EGSs that are less than an order of magnitude brighter than the foreground are NVSS~J181949$-$065524, NVSS~J182058$-$050223, NVSS~J182111$-$050219, NVSS~J182900$-$002018, NVSS~J184606$-$115808, NVSS~J185027$-$091037, NVSS~J190944$+$005558, NVSS~J191025$+$140125, NVSS~J191325$+$034308, and NVSS~J192458$+$130033.}. This suggests that the sinc-component sources may not have suffered adverse effects from erroneous foreground subtractions.

\section{The effect of foreground subtraction} \label{sec:fg_test}

\begin{figure}
\includegraphics[width=0.47\textwidth]{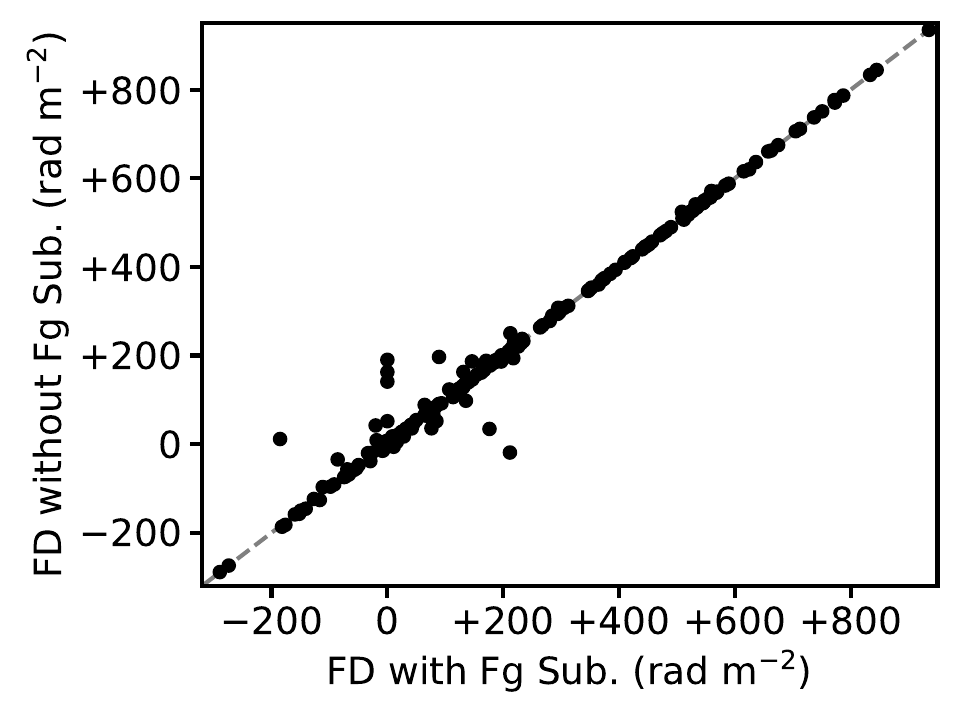}
\caption{The EGS FD values without foreground subtraction \citep{ma20} compared with those with foreground subtraction (Section~\ref{sec:fg_sub}). The error bars are plotted for both axes, but for most cases are too small to be noticeable. The grey diagonal dashed line marks where the $x$- and $y$-values are equal.}
\label{fig:fa1}
\end{figure}

In \cite{ma20}, the Stokes \textit{I}, \textit{Q}, and \textit{U} values of the target EGSs were extracted without subtraction of the potential foreground diffuse emission. Those values were then analysed using RM-Synthesis \citep{brentjens05}, with the peak FD value of each source obtained and utilised to study the Galactic-scale regular magnetic field. Here, we use our foreground-subtracted Stokes \textit{IQU} values (Section~\ref{sec:fg_sub}) to repeat the \cite{ma20} RM-Synthesis analysis, to test if their results can have been impacted by the foreground emission. The FD values with and without foreground subtraction are compared in Figure~\ref{fig:fa1}, which shows that the two sets of FD values agree well with each other in most cases. This is supported by the median FD difference of only $1.6\,{\rm rad\,m}^{-2}$. We therefore conclude that while the omission of foreground subtraction did affect the FD measurement for some EGSs, it did not impact the final conclusion of \cite{ma20}.

\begin{figure}
\includegraphics[width=0.47\textwidth]{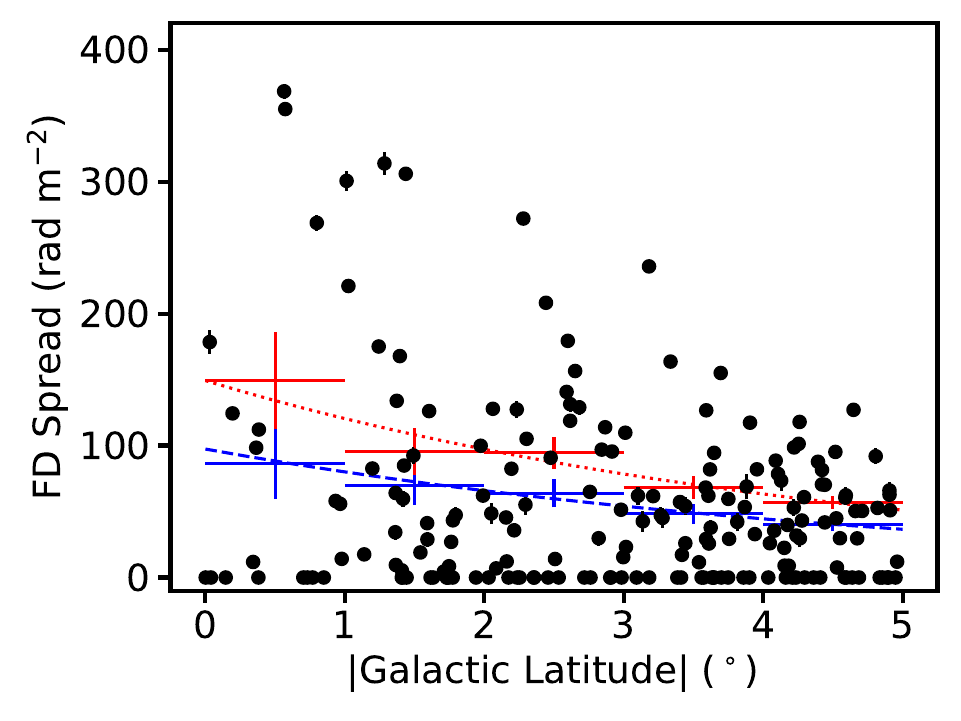}
\includegraphics[width=0.47\textwidth]{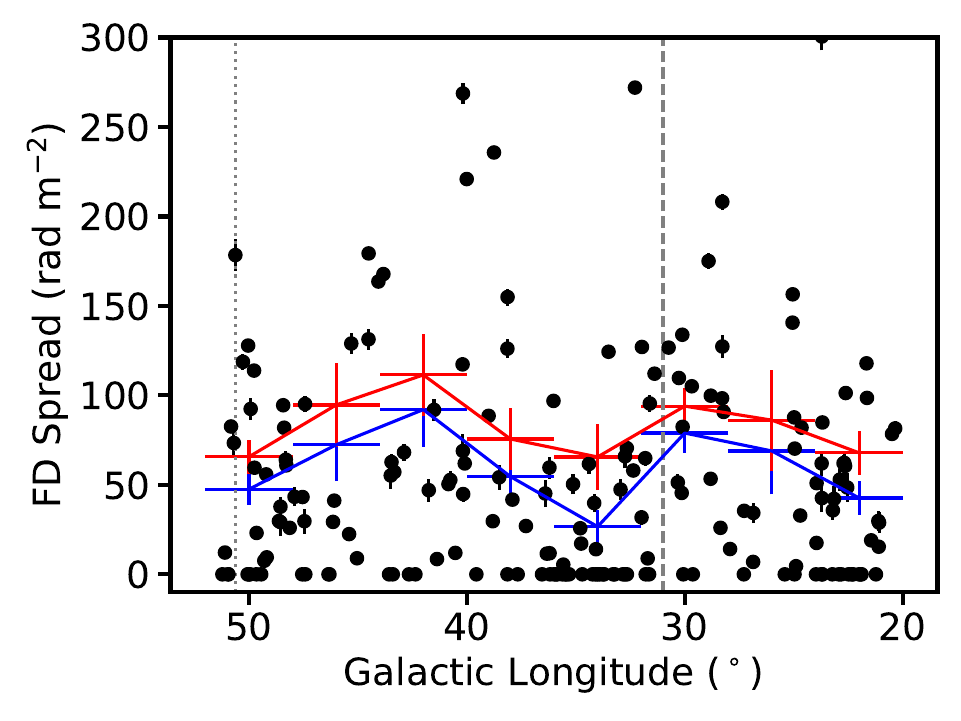}
\caption{Similar to Figure~\ref{fig:latitude}, but using FD spread values obtained without performing foreground subtraction in Stokes \textit{I}, \textit{Q}, and \textit{U} (Section~\ref{sec:fg_sub}).}
\label{fig:fa2}
\end{figure}

\begin{figure}
\includegraphics[width=0.47\textwidth]{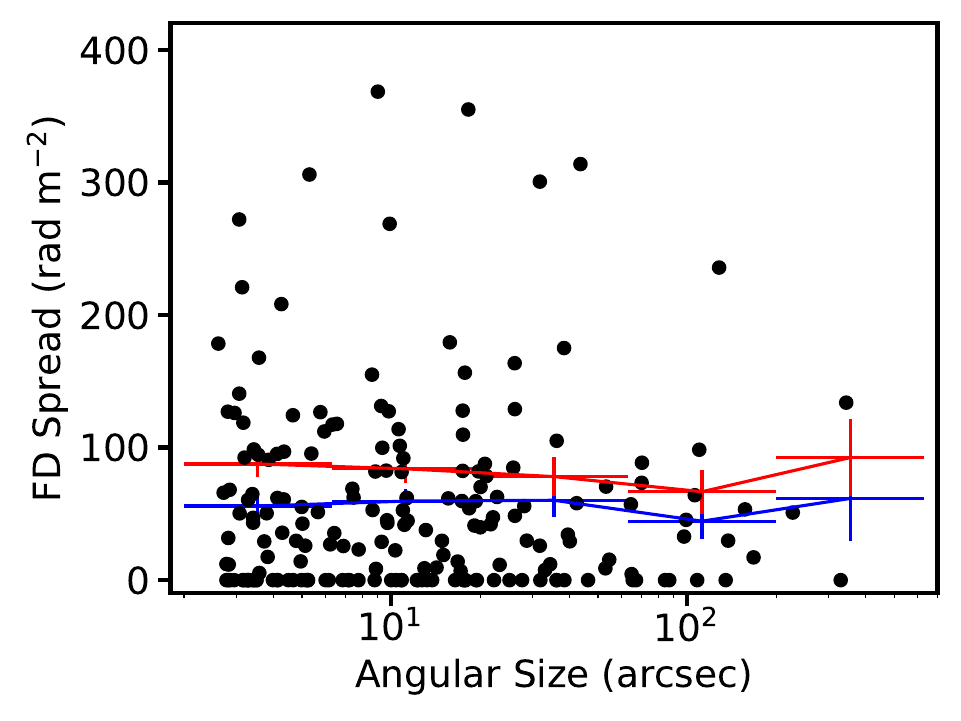}
\caption{Similar to Figure~\ref{fig:srcsize}, but using FD spread values obtained without performing foreground subtraction in Stokes \textit{I}, \textit{Q}, and \textit{U} (Section~\ref{sec:fg_sub}).}
\label{fig:fa3}
\end{figure}

Next, we turn to investigate the impact of foreground subtraction on the results of this work. We repeated all the analysis steps outlined in Section~\ref{sec:data}, but without implementing the foreground subtraction routine that was mentioned in Section~\ref{sec:fg_sub}. The FD spread profiles across both the amplitude of Galactic latitude, as well as Galactic longitude, are shown in Figure~\ref{fig:fa2}. We see a clear exponential profile of FD spread in the former, with the FD spread amplitude being about $50\,\%$ higher than that obtained with foreground subtraction in place (Section~\ref{sec:results}). Meanwhile, the modulation of FD spread across Galactic longitude can also be seen, and is more obvious than that after foreground subtraction (Section~\ref{sec:longitude}). In Figure~\ref{fig:fa3}, we see again that the FD spread is unchanged across EGS source size, with a Pearson correlation coefficient of $-9.5 \times 10^{-3}$ and a corresponding $p$-value of $0.90$. These clearly show that the omission of foreground subtraction can induce extra Faraday complexity to the background EGSs, as has been concluded by \cite{ranchod24}. In particular for our work here, while the qualitative results and interpretation will remain unchanged with or without foreground subtraction, the quantitative aspects of our work will be significantly impacted. The key conclusion here is that for future studies of the Faraday complexity exhibited by EGSs, including those using on-going and future radio polarimetric survey data, the removal of the foreground diffuse polarised emission can be essential.

\section{The effect of $\mathbf{\Delta {\rm BIC}}$} \label{sec:bictest}

As mentioned in Section~\ref{sec:qufit}, the parameter $\Delta {\rm BIC}$ is a measure of our confidence in the best-fit Stokes \textit{QU}-fitting model with respect to the second-best. By discarding sources with low $\Delta {\rm BIC}$ values ($< 10$ here; 71 sources discarded) and subsequently repeating our analysis, we can assess whether the less reliable sources have impacted our results significantly. 

We observe that the enhanced FD spread at $|b| < 3^\circ$ retains mean FD spread values of $60 \pm 10\,{\rm rad\,m}^{-2}$ and $110 \pm 10\,{\rm rad\,m}^{-2}$ for all data and non-zero data, respectively. In comparison, their counterpart values for EGSs at $|b| \geq 3^\circ$ are $27 \pm 5\,{\rm rad\,m}^{-2}$ and $62 \pm 6\,{\rm rad\,m}^{-2}$, respectively. With the FD spread distributions of the $|b| < 3^\circ$ and $|b| \geq 3^\circ$ EGSs compared by a KS-test, the resulting $p$-values remain low -- $2.9 \times 10^{-3}$ with all data and $8.1 \times 10^{-5}$ with non-zero data.

\begin{figure}
\includegraphics[width=0.47\textwidth]{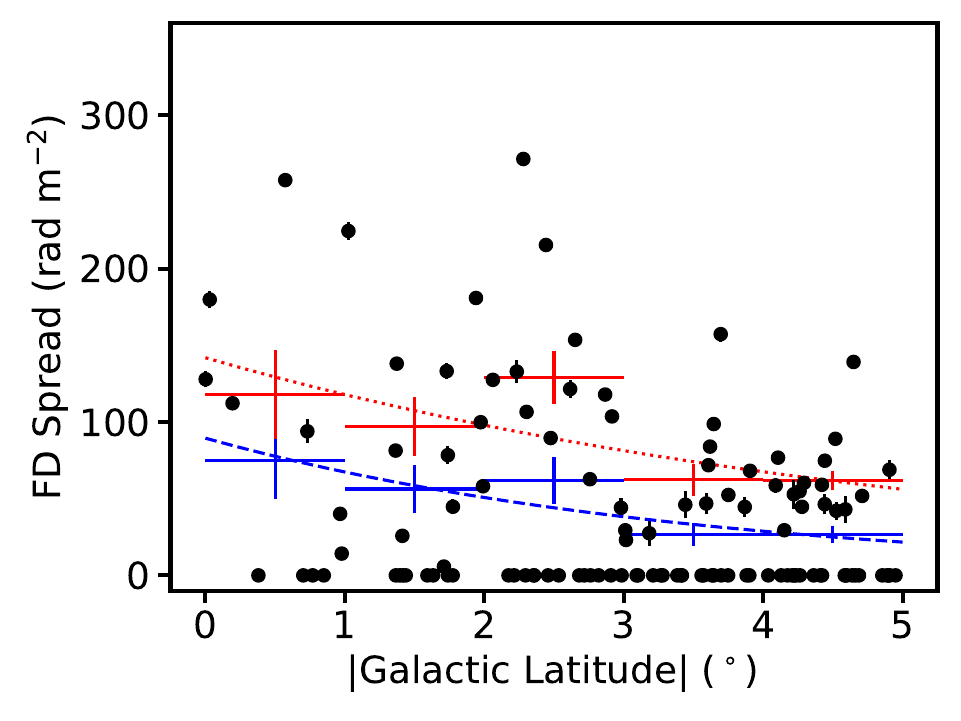}
\includegraphics[width=0.47\textwidth]{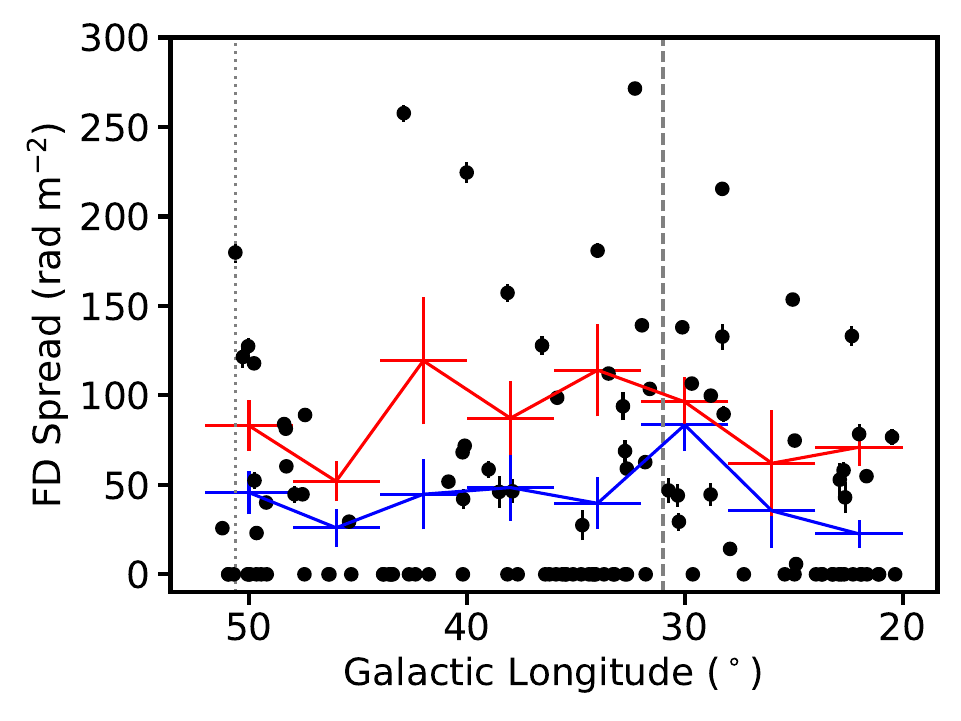}
\caption{Similar to Figure~\ref{fig:latitude}, but after having discarded EGSs that have $\Delta {\rm BIC} < 10$ from Stokes \textit{QU}-fitting.}
\label{fig:fb}
\end{figure}

We further show in Figure~\ref{fig:fb} the FD spread as a function of $|b|$ and $\ell$ after the $\Delta {\rm BIC}$ filtering. The fitted exponential functions against $|b|$ are
\begin{equation}
\text{FD Spread} = (90 \pm 20\,{\rm rad\,m}^{-2}) \cdot {\rm exp}\left( \frac{-|b|}{3.5 \pm 0.9\,{\rm deg}}\right){\rm ,}
\end{equation}
for using all of the remaining 120 sources; and
\begin{equation}
\text{FD Spread} = (140 \pm 40\,{\rm rad\,m}^{-2}) \cdot {\rm exp}\left( \frac{-|b|}{5.4 \pm 2.2\,{\rm deg}}\right){\rm ,}
\end{equation}
for using the 58 non-zero sources only. The fitted exponential functions here largely agree with our main results in Equations~\ref{eq:fds_all} and \ref{eq:fds_nonzero}, with both the amplitudes and scale heights agreeing within the uncertainties. We therefore conclude that our results presented in the main text are robust against the potential effects of $\Delta{\rm BIC}$.

\section{The effect of the $\Delta \phi$ parameter} \label{sec:remove_deltaphi}

In Section~\ref{sec:fds_definition}, we argue that the $\Delta \phi$ parameter (included in Stokes \textit{QU}-fitting models 1S, 2S, and 1Id), which has traditionally been seen as representing the magneto-ionic medium embedded within the synchrotron-emitting source (i.e.\ the Burn slab), can also be reflecting the magneto-ionic structures in the foreground ISM \citep[e.g.\ FD gradients in the plane-of-sky][]{sokoloff98}. Regardless, to ascertain the robustness of our results, we repeat our analysis in Sections~\ref{sec:fds_stats}--\ref{sec:angularsize} with EGSs best-fit by models 1S, 2S, and 1Id removed.

\begin{figure}
\includegraphics[width=0.47\textwidth]{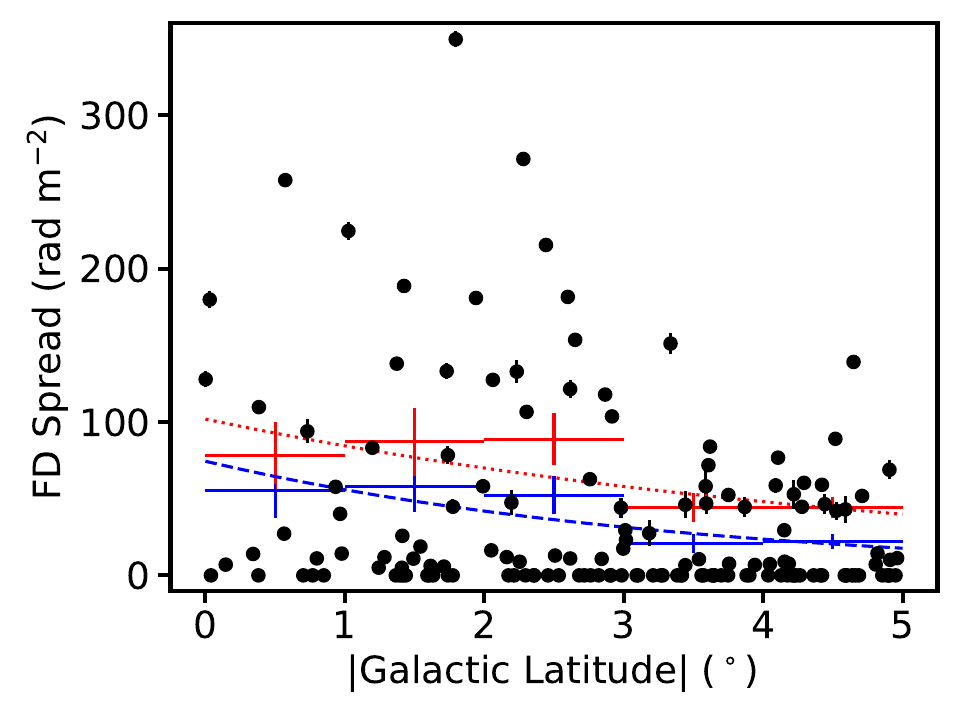}
\includegraphics[width=0.47\textwidth]{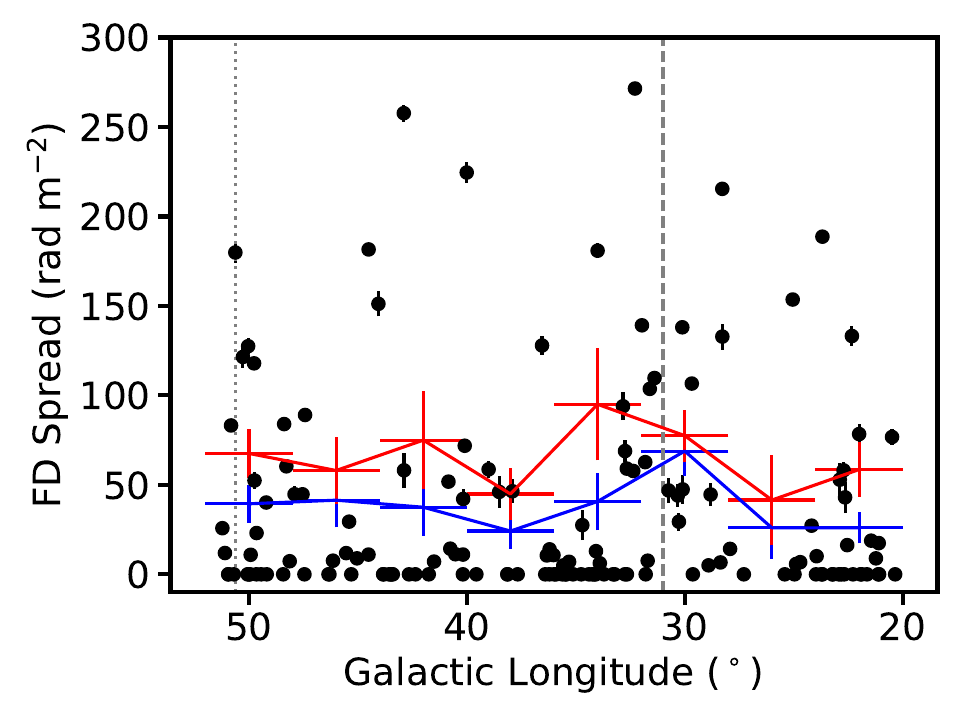}
\caption{Similar to Figure~\ref{fig:latitude}, but after having discarded the 39 EGSs best represented by the 1S, 2S, or 1Id models.}
\label{fig:reanalysis1}
\end{figure}

\begin{figure}
\includegraphics[width=0.47\textwidth]{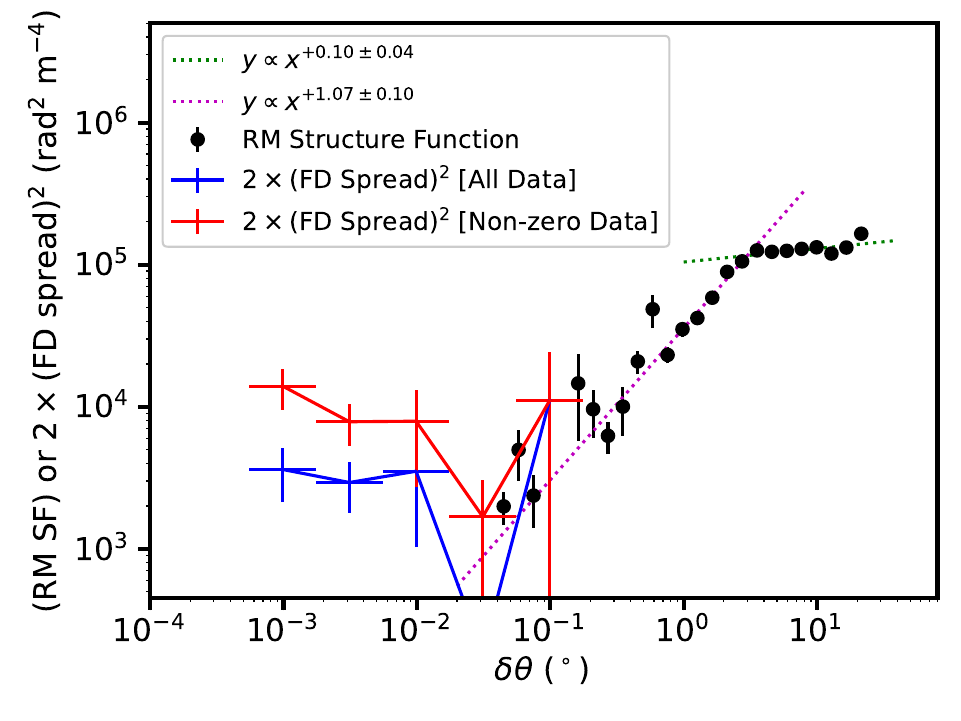}
\caption{Similar to Figure~\ref{fig:rmsf}, but with the FD spread trends (red and blue) constructed after having discarded the 39 EGSs best represented by the 1S, 2S, or 1Id models. The RM SF is calculated using the full 191 EGS sample (i.e.\ identical to Figure~\ref{fig:rmsf}).}
\label{fig:reanalysis2}
\end{figure}

We first report the statistics of our trimmed sample of 152 EGSs. The mean, median, and standard deviation of FD spread are $39$, $8$, and $63\,{\rm rad\,m}^{-2}$, respectively. Meanwhile, if we only consider the 86 EGSs that have non-zero FD spread, the mean, median, and standard deviation values are $68$, $46$, and $70\,{\rm rad\,m}^{-2}$, respectively. All these values are very similar to their counterparts listed in Section~\ref{sec:fds_stats}, and can be largely explained by the fact that we have removed a considerable sample of moderately Faraday complex sources. In particular, these 39 removed 1S, 2S, and 1Id sources have mean, median, and standard deviation in FD spread of $57$, $45$, and $32\,{\rm rad\,m}^{-2}$, respectively.

The Galactic latitude dependence of the FD spread is likewise re-examined. The Pearson correlation coefficient between FD spread and $|b|$ is $-0.25$ ($-0.25$) when considering all of the 152 EGSs (the 86 Faraday complex EGSs), with a corresponding $p$-value of $2.1 \times 10^{-3}$ ($2.3 \times 10^{-2}$). With the EGSs grouped into independent bins of $1^\circ$ along $|b|$ (Figure~\ref{fig:reanalysis1}), the enhanced FD spread at $|b| < 3^\circ$ is again apparent. With all 152 EGSs considered, the mean FD spread values are $55 \pm 9\,{\rm rad\,m}^{-2}$ and $22 \pm 4\,{\rm rad\,m}^{-2}$ at $|b| < 3^\circ$ and $|b| \geq 3^\circ$, respectively. A two-sample KS-test between the two groups yield a $p$-value of $1.8 \times 10^{-2}$. Meanwhile, considering the 86 EGSs with non-zero FD spread gives mean FD spread values of $90 \pm 10\,{\rm rad\,m}^{-2}$ and $44 \pm 6\,{\rm rad\,m}^{-2}$ for $|b| < 3^\circ$ and $|b| \geq 3^\circ$ respectively. The KS-test $p$-value is $5.3 \times 10^{-3}$. Finally, by fitting an exponential function to FD spread against $|b|$ (Equation~\ref{eq:expo}), we obtain
\begin{equation}
\text{FD Spread} = (70 \pm 20\,{\rm rad\,m}^{-2}) \cdot {\rm exp}\left( \frac{-|b|}{3 \pm 1\,{\rm deg}}\right)
\end{equation}
when considering all 152 EGSs; and
\begin{equation}
\text{FD Spread} = (100 \pm 20\,{\rm rad\,m}^{-2}) \cdot {\rm exp}\left( \frac{-|b|}{5 \pm 2\,{\rm deg}}\right)
\end{equation}
when considering the 86 Faraday complex sources only. All findings in this paragraph are quantitatively highly similar to the corresponding results presented in Section~\ref{sec:latitude}.

We end our re-analysis here by presenting the FD spread profile against Galactic longitude (Figure~\ref{fig:reanalysis1}), in addition to the RM SF and $(\textrm{FD spread})^2$ against the angular size of the EGSs (Figure~\ref{fig:reanalysis2}). Again, we find that they are qualitatively similar to our results described in the main text (i.e.\ Figures~\ref{fig:latitude} and \ref{fig:rmsf}). Overall, this re-analysis further emphasizes that inclusion of $\Delta \phi$ in the definition of FD spread do not bias/alter our results.

\label{lastpage}

\bsp
\end{document}